Thèse n° 8218

# Type-Preserving Compilation of Class-Based Languages

Présentée le 27 janvier 2023

Faculté informatique et communications
Laboratoire de méthodes de programmation 1
Programme doctoral en informatique et communications

pour l'obtention du grade de Docteur ès Sciences

par

**Guillaume André Fradji MARTRES**

Acceptée sur proposition du jury

Prof. A. Ailamaki, présidente du jury
Prof. M. Odersky, directeur de thèse
Prof. L. Parreaux, rapporteur
Prof. B. Oliveira, rapporteur
Prof. V. Kuncak, rapporteur

École
polytechnique
fédérale
de Lausanne

2023

# Acknowledgements

In 2014, I was an undergrad student at EPFL and already fascinated by programming languages when I came across an intriguing announcement[1] about a new compiler.

```
Subject: Dotty open-sourced
From: martin odersky <martin.odersky@epfl.ch>
To: <scala-internals@googlegroups.com>
Date: Feb 18 2014 19:06:07 +0200

A couple of days ago we open sourced the Dotty, a research platform
for new language concepts and compiler technologies for Scala.

    https://github.com/lampepfl/dotty

[...]

Right now, there's a (very early) compiler frontend for a subset of
Scala. We'll work on fleshing this out [...]
```

A whole new compiler that could define the future of Scala! This was enough to pique my interest, so I sent a mail to Martin asking about possible semester projects and thus began an amazing and still ongoing journey. I had no idea at the time that I would learn so much from working with Martin and I'm grateful to him for being the mentor I could hope for. Naturally, along the way I had the chance to connect with many other helpful and kind folks for which I'm equally thankful.

To Dmitry Petraskho, for showing me the ropes and taking the time to mentor me as an undergrad student.

To the generation of LAMP PhD students that started the same year as I did: Olivier Blanvillain, Liu Fengyun and Nicolas Stucki. Collaborating with them was a joy, and I had the pleasure to watch them grow into brilliant researchers and engineers. You should check out their theses!

[1] https://groups.google.com/g/scala-internals/c/6HL6lVLI3bQ/m/IY4gEyOwFhoJ

# Acknowledgements

This thesis owes a lot to Sandro Stucki and Paolo Giarrusso who listened patiently to my ideas, gave me confidence that I was on to something, and generously shared their knowledge and intuition about type theory. I cannot thank them enough for their support.

This thesis also benefited from the helpful feedback of Nada Amin, Aleksander Boruch-Gruszecki, Ondřej Lhoták, and of course my jury members: Anastasia Ailamaki, Viktor Kuncak, Bruno C. d. S. Oliveira and Lionel Parreaux. Many thanks to them!

Of all the trips I had the chance to go on during my PhD, the most memorable was certainly the journey through India to Maha & Mano's wedding. To Sébastien Doeraene, Mia Primorac, Georg Schmid and Denys Shabalin for being the best trip buddies one could ask for, and to Manohar Jonnalagedda for sharing some of his life wisdom with the rest of us and for inviting us to his wedding!

To everyone who helped make Scala 3 a reality. So many people were involved that I cannot possibly list them all, but I have fond memories of working with Martin Duhemm, Tom Grigg, Felix Mulder, Guillaume Raffin, Allan Renucci, Miles Sabin, Jamie Thompson, Dale Wijnand and many others.

To everyone I had the pleasure to interact with at LAMP and the Scala Center, including Jorge Vicente Cantero, Iulian Dragos, Philipp Haller, Vojin Jovanovic, Heather Miller, Julien Richard-Foy and Vlad Ureche. Special thanks to Darja Jovanovic for doing her best to keep me focused on finishing my PhD and to Anna Herlihy for getting me out of the office and walking dogs :).

To everyone who invested untold amount of their time and effort in the Scala community, including Fabio Labella, Adriaan Moors, Nicolas Rinaudo, Som Snytt, Daniel Spiewak, Seth Tisue, Eugene Yokota and Kenji Yoshida.

To Léonard Berney and Tim Tuuva for our weekly anime nights which I look forward to every time.

To the architect of the INR building for having the foresight to put an AC unit in what would become my office.

To the Hong Thaï Rung food truck at EPFL for keeping me fed through all these years.

Et bien sûr, je remercie et j'embrasse Maman et Papa, Finou, Tata, Tatie Karine, Tatie Gipsy, Mamie Évelyne et tous mes (petits-)cousin(e)s pour leur soutien sans faille.

*Lausanne, September 28, 2022* Guillaume Martres

# Abstract

The Dependent Object Type (DOT) calculus was designed to put Scala on a sound basis, but while DOT relies on structural subtyping, Scala is a fundamentally class-based language. This impedance mismatch means that a proof of DOT soundness by itself is not enough to declare a particular subset of the language as sound. While a few examples of Scala snippets have been manually translated into DOT, no systematic compilation scheme has been presented so far.

In this thesis we develop a series of calculi of increasing complexity to model Scala and present a type-preserving compilation scheme from each of these calculus into DOT. Along the way, we develop some necessary extensions to DOT.

# Résumé

Le calcul “Dependent Object Types” (DOT) a été conçu pour garantir la sûreté du typage de Scala. Mais alors que DOT se fonde sur le sous-typage structurel, Scala est un langage construit sur un système de classes. Dès lors, on ne peut conclure qu’un sous-ensemble particulier de Scala est sûr uniquement parce que DOT lui-même l’est. Même si quelques exemples de code Scala ont été manuellement traduits en DOT, aucun schéma de compilation systématique n’a été présenté jusqu’ici.

Dans cette thèse, nous développons une série de calculs de complexité croissante afin de modéliser Scala, et nous présentons pour chacun un schéma de compilation vers DOT préservant le typage. En chemin, nous developpons certaines extensions nécessaires à DOT.

# Contents

# Mathematical conventions

In this preliminary chapter, we briefly describe some of the notations of the meta-language we will use in the rest of this thesis to describe and analyze calculi.

As usual, terms and types that are equal up to renaming of bound variables are identified.

We write $f(a) := b$ to mean that $f$ is **defined** to map $a$ to $b$.

We write $\mathsf{dom}(f)$ for the **domain** of a function $f$.

We write $\mathsf{fv}(T)$ for the set of free variables appearing in $T$.

We write _ to denote a fresh variable we never refer to.

A **list** $X_1, \dots, X_n$ (abbreviated $\overline{X}$) is a possibly-empty ordered sequence of elements. We denote the empty list by $\varnothing$, like the empty set. The list $\overline{X}, \overline{Y}$ is the concatenation of $\overline{X}$ and $\overline{Y}$.

A **substitution** $[T_1/X_1, \dots, T_n/X_n]$ (abbreviated $[\overline{T/X}]$) simultaneously replaces every free occurence of $X_i$ by $T_i$ in the expression that appears to its right. For example, $[\overline{T/X}]\overline{X}$ is equivalent to $\overline{T}$. A substitution can be viewed as a partial function and so we define $\mathsf{dom}([\overline{T/X}]) := \overline{X}$.

By analogy with the usual **set-builder** notation $\{p \in \mathbb{P} \mid \Phi(p)\}$ we define a **list-builder** notation $[p \in \overline{P} \mid \Phi(p)]$ which preserves the order of the elements in the input list.

For convenience, we overload the usual intersection and union operators to also be defined on lists:

$$\begin{aligned} \overline{P} \cup \overline{Q} &:= \overline{P}, [q \in \overline{Q} \mid q \notin \overline{P}] \\ \overline{P} \cap \overline{Q} &:= [p \in \overline{P} \mid p \in \overline{Q}] \end{aligned}$$

For every syntactical element $\star$ such that $X_1 \star \dots \star X_n$ is valid syntax, we implicitly define a "big operator" $\bigstar$ such that $\bigstar\overline{X} := X_1 \star \dots \star X_n$.

The overline notation can be used with arbitrary syntax fragments. For example $x_1 : T_1, \dots, x_n : T_n$ can be abbreviated as $\overline{x : T}$. Note that a meta-variable might be defined outside of an overlined expression but used in such an expression which will affect its expansion.[2] For example the sentence,

> Let $\sigma = [\overline{T/X}]$ and $t = \overline{x : \sigma U}$.

expands to

> Let $\sigma = [T_1/X_1, \ \ldots, T_n/X_n]$ and $t = x_1 : \sigma U_1, \ \ldots, x_m : \sigma U_m$.

[2] This is markedly different from [Igarashi, Pierce, and Wadler 2001] (which inspired most of our notations) where $\overline{X}$ and $X$ may appear in the same context but will refer to different variables.

Overlines can be nested, although we do our best to avoid using that power for the sake of the reader. When this happens, the overlines should be expanded outside-in (because the lists represented by an inner overline might be of different lengths). For example,

$$\overline{A = \overline{X <: N}}$$

expands to

$$(A_1 = \overline{X_1 <: N_1}), \ \ldots, (A_n = \overline{X_n <: N_n})$$

which itself expands to

$$\begin{aligned} &(A_1 = X_{1_1} <: N_{1_1}, \ \ldots, X_{1_m} <: N_{1_m}), \\ &\ldots, \\ &(A_n = X_{n_1} <: N_{n_1}, \ \ldots, X_{n_z} <: N_{n_z}) \end{aligned}$$

When multiple judgments are entailed by the same context like $\Gamma \vdash X <: N$ and $\Gamma \vdash x : T$, we may "factor out" the entailement part and write $\Gamma \vdash X <: N, x : T$ instead. This can be combined with the overline notation: we write $\Gamma \vdash \overline{Y <: P}$ to mean $\Gamma \vdash Y_1 <: P_1, \ \ldots, Y_n <: P_n$.

With "postfix judgments" such as $\Gamma \vdash T$ wf, we allow $\Gamma \vdash T, S$ wf to stand for $\Gamma \vdash T$ wf, $S$ wf, this can also be combined with the overline notation: we write $\Gamma \vdash \overline{T}$ wf to mean $\Gamma \vdash T_1, \ \ldots, T_n$ wf

In a context where $\Gamma \vdash T <: S$ is a subtyping judgment, we write $\Gamma \vdash T =:= S$ as a short-hand for $\Gamma \vdash T <: S, S <: T$.

In proofs, we abbreviate "induction hypothesis" to "IH".

# 1 Introduction

## 1.1 Background

How can we reason about the behavior of our programs without running them first? Assuming our language of choice has a static type system, a type theorist might answer with the following very broad recipe:

1. Write down the rules that determine which programs are *well-typed* in our language.
2. Write down the rules that determine how a program is *evaluated*.
3. Prove that all well-typed programs will behave in a particular way when evaluated.

But modern programming languages are fiendishly complex, so much so that step 1 by itself might already prove too arduous unless the language has already been carefully specified. Even if we manage to exhaustively specify the static and operational semantics of our language, the sheer number of rules involved will likely make any interesting property too hard to prove in a reasonable amount of time. This is compounded by the fact that languages keep evolving, and what we can prove about any particular version of it might not hold for the next.

As exemplified by Featherweight Java [Igarashi, Pierce, and Wadler 2001], the pragmatic approach in this situation has been to formally specify only a tractable subset of the original language which is then carefully studied, while reasoning informally about other parts of the language.

This has been very successful in practice but the downside is that important properties established in our core language might not in fact hold in practice due to under-studied interactions with other parts of the language such as `null` in Java [Amin and Tate 2016].

Another possible way to tame complexity is to design a simpler language that can serve as a *compilation target* for our source language. Assuming well-typed programs in our source language are translated into well-typed programs of the target language, then results we prove about well-typed programs in the target language also apply to programs in our source language.

This technique was pioneered by the GHC Haskell compiler using System $F_C$ [Sulzmann et al.

2007] as an intermediate representation. It isn't completely without pitfalls either:

1. The operational semantics of a program in our source language is now determined by the operational semantics of its translation. If the translation procedure itself is complex, we'll have a hard time figuring out how our program will be executed.

2. The translation might in fact not always produce well-typed programs in the target language. To guard against this, GHC can re-typecheck the translated program as a consistency check. If it turns out not to be well-typed, then it can stop and report to the user that a compiler bug has been found.

## 1.2 Reasoning about Scala

The Dependent Object Types (DOT) calculus [Amin, Grütter, et al. 2016; Rompf and Amin 2016] was designed as a compilation target for Scala. But unlike System $F_C$, DOT isn't meant to be a practical intermediate language: Scala's primary backend is Java bytecode which can be seen as a simple class-based language. Using a class-less language such as DOT as an intermediate step when compiling to Java bytecode would be counter-productive both for performance and interoperability with other languages on the JVM as too much of the program structure would be lost.[1]

Instead, DOT should be seen as a theoretical framework for reasoning about Scala. In that respect, it has been very successful: type system features first developed in DOT such as intersection types and union types were added to the language, and type soundness holes in the language were patched based on ideas developed in DOT.

But still, we can't help but have a nagging feeling that something is missing here: can we actually compile Scala to DOT? In fact, we know that some Scala features such as higher-kinded types are not encodable in DOT [Odersky, Martres, and Petrashko 2016; Stucki and Giarrusso 2021]. So at most we may be able to compile a subset of valid Scala programs into DOT, but it isn't clear what that subset would be.

Even if we were to write down a compilation scheme from a subset of Scala into DOT, how would we know whether it is actually correct? Unlike with System $F_C$, there is no known practical algorithm for typechecking DOT [Nieto 2017], so we cannot simply check that our translation is correct in practice. This leaves us with only one clear path ahead: given a particular subset of Scala, we need to prove that well-typed programs in it can always be compiled into well-typed DOT programs. In other words, we need to develop a *type-preserving* compilation scheme. This is the approach we choose to pursue in this thesis.

## 1.3 Thesis organization

We begin our journey with a whirlwind tour of the DOT calculus family in Chapter 2. After settling on **oopslaDOT** [Rompf and Amin 2016] as our target calculus of choice, we present a series of calculi of increasing complexity, each accompanied by a type-preserving compilation proof, as summarized in Figure 1.1.

[1] Even alternative backends such as Scala.js implement JVM-like operational semantics to ease cross-platform development [Doeraene 2018, § 2.1].

**Figure 1.1: Thesis organization**

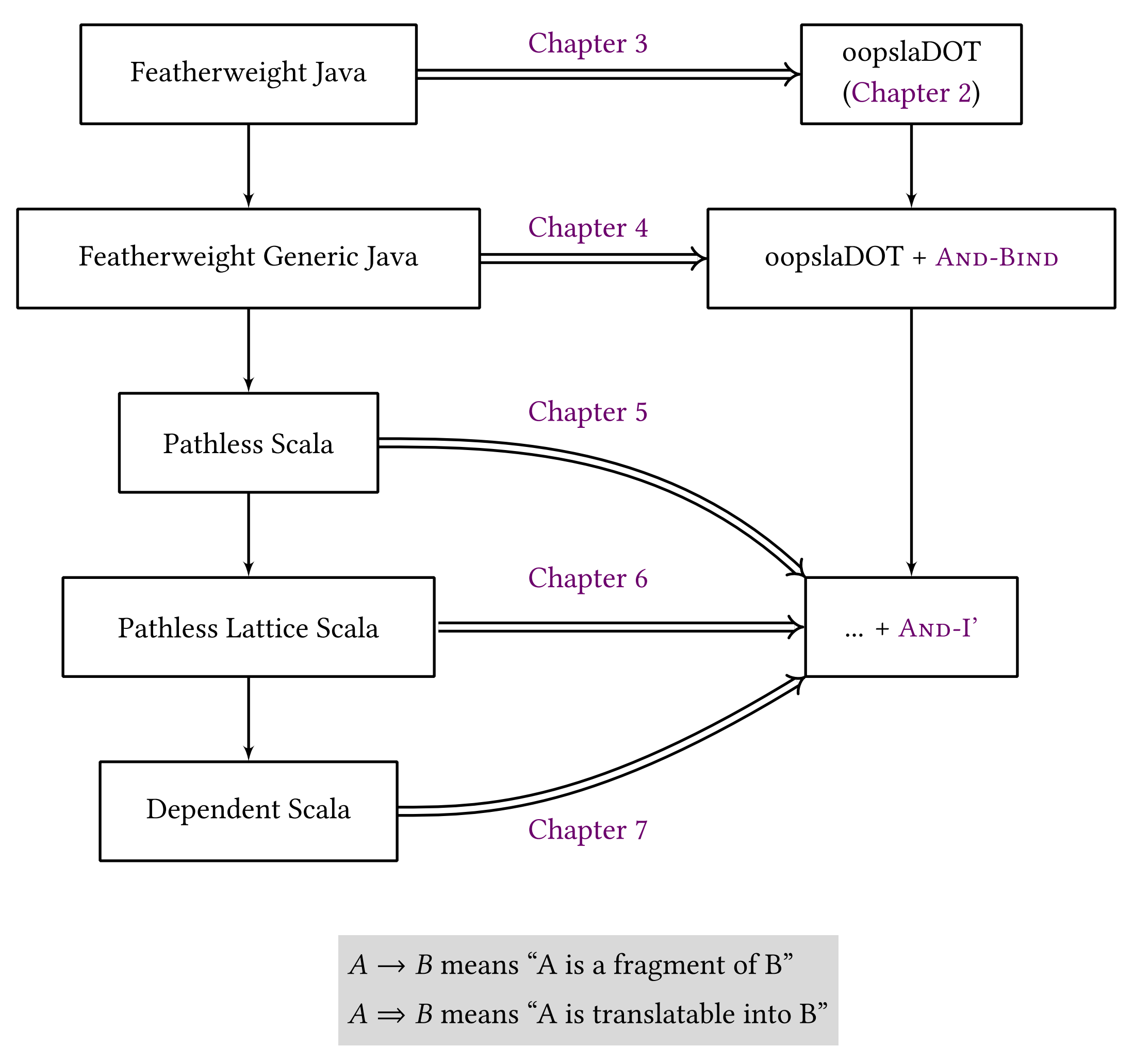

Chapter 3 reviews **Featherweight Java** which conveniently happens to already be isomorphic to a subset of Scala. We make this correpondance explicit by swapping the original syntax of Featherweight Java for a more Scala-like syntax.

In Chapter 4, we apply the same treatment to **Featherweight Generic Java**, an extension of Featherweight Java with type parameters. This makes the type-preservation proof significantly more challenging. In fact, and against all expectations, no existing version of DOT appears to be expressive enough for this task and we are forced to extend oopslaDOT with an additional subtyping rule AND-BIND. We provide a mechanized type safety proof for our extension which we base on the existing mechanization of oopslaDOT.

Having run out of Java calculi we could repurpose, we develop **Pathless Scala** in Chapter 5

which adds intersection types and multiple inheritance to Featherweight Generic Java. Here again, the existing DOT rules fall short and we end up needing an extra typing rule And-I' to complete our type-preservation proof. Proving the resulting extended DOT calculus sound requires generalizing the statement of the type soundness theorem originally presented in [Rompf and Amin 2016], this is reflected in our updated mechanized type safety proof.

**Pathless Lattice Scala** in Chapter 6 turns subtyping into a lattice by adding union types (which represent least upper bounds) and Nothing (which represents bottom). This is also the first chapter where we define algorithmic subtyping rules.

Finally, **Dependent Scala** in Chapter 7 adds type members and type selections to our source language. Besides justifying our use of DOT as a target language, this sheds a new light on DOT itself: we find that the seemingly problematic restrictions of oopslaDOT's declarative subtyping rules involving type selections do not prevent us from developing algorithmic subtyping rules for our source calculus that match the expressiveness of real Scala.

As a bonus, and to demonstrate that the calculi we develop here are useful for more than establishing soundness, Appendix A develops a translation from Pathless Scala into a superset of Featherweight Java with interfaces to model how type erasure from Scala to Java bytecode is implemented in the compiler. We believe that specifying type erasure in detail is important and cannot be left as an implementation detail because it is critical to maintaining binary-compatibility of artifacts produced by different versions of the Scala compiler.

# 2 Dependent Object Types

In this chapter we review the Dependent Object Types (DOT) family of calculi. In particular, we contrast [Rompf and Amin 2016] with [Amin, Grütter, et al. 2016] and justify why we chose the former as a basis for the target calculi we use in subsequent chapters. We then introduce some "syntactic sugar" (that is, derived syntactic forms) to improve the readability of our translations. Finally, we prove various meta-theoretic properties of DOT that will be useful in our type-preserving translation proofs.

## 2.1 A short and incomplete history of the DOT family

As Figure 2.1 attests, there is not one DOT.[1] But while each of these papers may have its own take on exactly what DOT is, the running theme among them is clear: Scala features a rich type system but its defining characteristic is its support for *path-dependent* types. $p.L$ is a path-dependent type if $p$ is a reference to an object with a type member $L$ where $p$ itself is either a variable $x$ or a reference to a term member $p_1.l$. The type of $p$ specifies both an upper- and lower-bound for $L$ that determine its place in the subtyping hierarchy. In particular, this means that depending on the context, the subtyping hierarchy can be extended in arbitrary ways which is a major source of complexity for the meta-theory of DOT.

The first publication on DOT [Amin, Moors, and Odersky 2012] did not include a soundness proof, but it served as motivation and roadmap for the development of the Scala 3 language and compiler: it argued both for replacing the non-commutative "compound types" $A$ **with** $B$ of Scala 2 with true intersection types and for adding union types to ensure that the least upper bound of a type is always defined.[2] The intuition was that each aspect of the Scala 3 type system ought to be translatable into DOT,[3] but this translation was never formally defined.

[1] Amusingly, this figure was also generated using DOT (https://en.wikipedia.org/wiki/DOT_(graph_description_language)).

[2] In Scala 2, the least upper bound of two types could have an infinite expansion. This required the compiler to rely on heuristics when typing a conditional expression for example.

[3] In fact, initial versions of the compiler implemented support for type parameters by desugaring them into type members. However, we were unable to scale this approach to support the full power of higher-kinded types that Scala 2 users were accustomed to. So type parameters were reintroduced as a first class concept in the compiler [Odersky, Martres, and Petrashko 2016]. Much theoretical work remains to be done to combine DOT with higher-kinded

**Figure 2.1: DOT: A family tree**

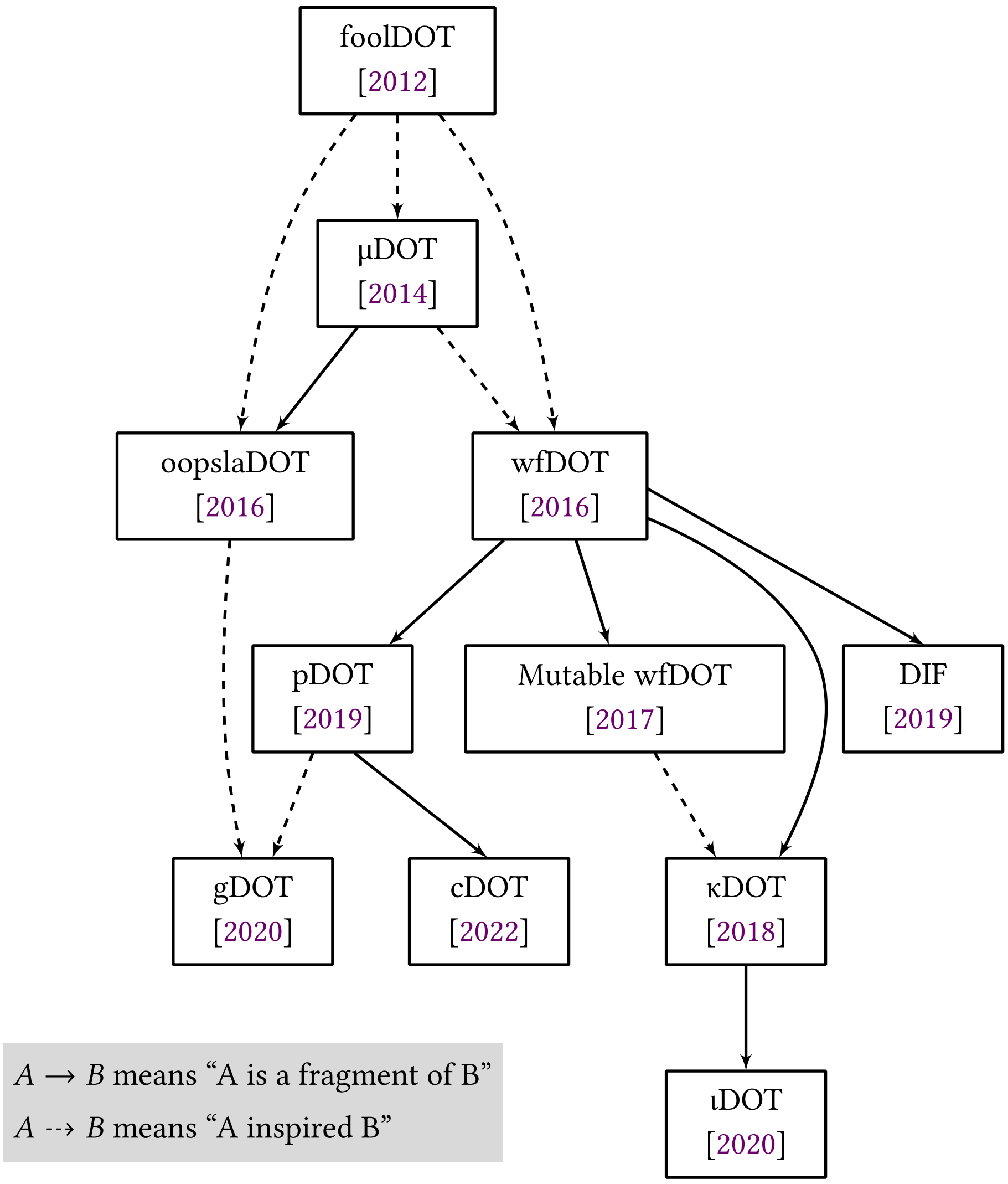

Four years later, soundness proofs were finally published[4] for two closely related calculi. At this point, we need to introduce nicknames to distinguish these calculi since they are all known as "DOT". We will refer to them respectively as "foolDOT" [Amin, Moors, and Odersky 2012], "wfDOT" [Amin, Grütter, et al. 2016] and "oopslaDOT" [Rompf and Amin 2016] based on the name of the conference they were published at (respectively FOOL'2012, WadlerFest'2016 and OOPSLA'2016).

Compared to foolDOT, both later DOTs restricted the paths in path-dependent types to just variables. Compared to oopslaDOT, wfDOT trades off some expressiveness for a simpler meta-types [Stucki and Giarrusso 2021] and in this thesis we will only consider fragments of Scala without such types.

[4] There exists an earlier soundness proof for the $\mu$DOT [Amin, Rompf, and Odersky 2014] fragment which features a minimal type system that only supports record types and path-dependent types.

theory. We will explore the exact differences and their impact on our work in the next section.

Thanks to its relative simplicity, wfDOT has since been successfully extended in multiple ways. The restriction of path-dependent types to variables was lifted in pDOT [Rapoport and Lhoták 2019]. Other variants of DOT explored mutability [Rapoport and Lhoták 2017], object initialization [Kabir and Lhoták 2018; Kabir, Li, and Lhoták 2020], implicit functions [Jeffery 2019] and pattern matching with GADT-like inferred local constraints [Boruch-Gruszecki et al. 2022]. In this thesis, we will only consider fragments of Scala with variable-dependent types and without mutability, implicits or pattern matching, so these extensions are outside the scope of our discussion.

We mention in passing [Giarrusso et al. 2020] which features a very extensive type system backed by an impressive meta-theory machinery based on the Iris framework [Jung et al. 2018]. However, the actual degree of expressiveness of gDOT is still an open question due to its reliance on annotations as described in Section 9 of the paper:

> "[Amin, Grütter, et al. 2016] prove that all $F_{<:}$ programs can be translated into DOT. Due to the presence of the ◂ operator and the **coerce** annotations, it is unclear how to create a translation from either (p)DOT or $F_{<:}$ into gDOT. However, we have been able to translate many given $F_{<:}$ and DOT examples into gDOT by hand by adding a sufficient number of ◂ and **coerce** annotations. We thus conjecture that there exists a whole-program encoding of $F_{<:}$ programs into gDOT."

We now turn our attention towards evaluating which of wfDOT and oopslaDOT is a better target calculus in our quest towards establishing soundness for a significant subset of Scala. We first present oopslaDOT in detail and then contrast it with wfDOT, before ultimately settling on oopslaDOT due to its support for subtyping between recursive types which our type-preserving compilation proofs will critically rely on.

## 2.2 Syntax and semantics of oopslaDOT

Figures 2.2 to 2.4 are adapted from [Rompf and Amin 2016]. The notation $t^x$ emphasizes that $x$ may appear free in $t$ and $\Gamma_{[x]}$ is a truncated context where all bindings to the right of $x$ in $\Gamma$ are dropped. Figure 2.4 simultaneously defines a regular typing judgment $\Gamma \vdash t : T$ and a less powerful "strict typing" judgment $\Gamma \vdash t :_! T$ using the syntax $:_{(!)}$ to denote the rules which are applicable to both judgments. Unlike the original presentation, our syntax definition allows optional type ascriptions on method arguments and result types (the paper notes that their mechanized proof supports both variants). We denote optional syntax elements with wavy underlines.

### 2.2.1 Well-formedness

Although [Rompf and Amin 2016] does not formally define a well-formedness judgment, it does implicitly rely on one as stated in Section 3 of the paper:

> "For readability, we omit well-formedness requirements from the rules, and assume

**Figure 2.2: oopslaDOT: Syntax**

| | |
|---|---|
| $x, y, z$ | Variable |
| $L$ | Type label |
| $m$ | Method label |
| | |
| $s, t, u$ ::= | Term |
| $x$ | variable reference |
| $\{z \Rightarrow \overline{d}\}$ | object |
| $s.m(t)$ | method invocation |
| $\Gamma ::= \overline{x : T}$ | Context |
| | |
| $\sigma, \tau ::= [\overline{S/T}]$ | Type substitution |
| $\theta ::= [\overline{y/x}]$ | Variable substitution |
| $d$ ::= | Declaration |
| $L = T$ | type tag |
| $m(x : \utilde{S}) : \utilde{U^x} = t$ | method member |
| $S, T, U$ ::= | Type |
| $\top$ | top type |
| $\bot$ | bottom type |
| $L : S \mathinner{..} U$ | type member |
| $m(x : S) : U^x$ | method member |
| $x.L$ | type selection |
| $\{z \Rightarrow T^z\}$ | recursive self type |
| $T \wedge T$ | intersection type |
| $T \vee T$ | union type |

all types to be syntactically well-formed in the given environment."

The type-preserving proofs we present in later chapters will require us to pay close attention to well-formedness (intuitively, we'd like our translation to preserve some notion of well-formedness), so we explicitly define it in Figure 2.5. Note that $\Gamma \vdash x.L$ wf doesn't require $x$ to have a type member $L$.

### 2.2.2 Evaluating wfDOT and oopslaDOT as compilation targets

So how does oopslaDOT measure up against wfDOT? In this comparison we will only consider the static semantics of both calculi. While the operational semantics of oopslaDOT described in [Rompf and Amin 2016, Figure 2] are more complex due to the use of a store, there exists an alternative store-less presentation in [Amin 2016, § 3.5] which relies on augmenting the syntax to allow values and not just variables as paths.

We can safely ignore some syntactic differences which do not significantly affect expressiveness:

- wfDOT syntax directly supports let bindings, but oopslaDOT can encode them (see Definition 2.3.4).
- wfDOT does not have methods, but it can encode them using fields that return lambdas.
- wfDOT only allows function applications where both the function and the argument are variables, but arbitrary applications can be translated into that form using let bindings.

The only significant syntactic difference between the two system is the lack of union types in wfDOT. This is concerning since, as we described in Section 2.1, unions are an important aspect of the Scala 3 type system which we model in Chapter 6. While this omission hasn't yet been rectified by subsequent work, there are no known meta-theoretical difficulties unique to the interaction of union types with the rest of DOT. So this is likely more of a practical than

**Figure 2.3: oopslaDOT: Subtyping rules**

$\boxed{\Gamma \vdash S <: U}$

Lattice structure

$$\Gamma \vdash \bot <: T \quad \textsc{(Bot)}$$

$$\Gamma \vdash T <: \top \quad \textsc{(Top)}$$

$$\frac{\Gamma \vdash T_1 <: T}{\Gamma \vdash T_1 \wedge T_2 <: T} \quad \textsc{(And11)}$$

$$\frac{\Gamma \vdash T <: T_1}{\Gamma \vdash T <: T_1 \vee T_2} \quad \textsc{(Or21)}$$

$$\frac{\Gamma \vdash T_2 <: T}{\Gamma \vdash T_1 \wedge T_2 <: T} \quad \textsc{(And12)}$$

$$\frac{\Gamma \vdash T <: T_2}{\Gamma \vdash T <: T_1 \vee T_2} \quad \textsc{(Or22)}$$

$$\frac{\Gamma \vdash T <: T_1,\ T <: T_2}{\Gamma \vdash T <: T_1 \wedge T_2} \quad \textsc{(And2)}$$

$$\frac{\Gamma \vdash T_1 <: T,\ T_2 <: T}{\Gamma \vdash T_1 \vee T_2 <: T} \quad \textsc{(Or1)}$$

Type and method members

$$\frac{\Gamma \vdash S_2 <: S_1,\ U_1 <: U_2}{\Gamma \vdash L : S_1 \,..\, U_1 <: L : S_2 \,..\, U_2} \quad \textsc{(Typ)}$$

$$\frac{\Gamma \vdash S_2 <: S_1 \quad \Gamma, x : S_2 \vdash U_1^x <: U_2^x}{\Gamma \vdash m(x : S_1) : U_1^x <: m(x : S_2) : U_2^x} \quad \textsc{(Fun)}$$

Path selections

$$\frac{\Gamma_{[x]} \vdash x :_! (L : \bot \,..\, T)}{\Gamma \vdash x.L <: T} \quad \textsc{(Sel1)}$$

$$\frac{\Gamma_{[x]} \vdash x :_! (L : S \,..\, \top)}{\Gamma \vdash S <: x.L} \quad \textsc{(Sel2)}$$

$$\Gamma \vdash x.L <: x.L \quad \textsc{(SelX)}$$

Recursive self types

$$\frac{\Gamma, z : T_1^z \vdash T_1^z <: T_2^z}{\Gamma \vdash \{z \Rightarrow T_1^z\} <: \{z \Rightarrow T_2^z\}} \quad \textsc{(BindX)}$$

$$\frac{\Gamma, z : T_1^z \vdash T_1^z <: T_2}{\Gamma \vdash \{z \Rightarrow T_1^z\} <: T_2} \quad \textsc{(Bind1)}$$

Transitivity

$$\frac{\Gamma \vdash T_1 <: T_2,\ T_2 <: T_3}{\Gamma \vdash T_1 <: T_3} \quad \textsc{(Trans)}$$

**Figure 2.4: oopslaDOT: Typing rules**

**Type assignment** $\boxed{\Gamma \vdash t :_{(!)} T}$

Variables, self packing/unpacking

$$\frac{\Gamma(x) = T^x}{\Gamma \vdash x :_{(!)} T^x} \quad \text{(Var)}$$

$$\frac{\Gamma \vdash x : T^x}{\Gamma \vdash x : \{z \Rightarrow T^z\}} \quad \text{(VarPack)}$$

$$\frac{\Gamma \vdash x :_{(!)} \{z \Rightarrow T^z\}}{\Gamma \vdash x :_{(!)} T^x} \quad \text{(VarUnpack)}$$

Subsumption

$$\frac{\Gamma \vdash t :_{(!)} T_1,\ T_1 <: T_2}{\Gamma \vdash t :_{(!)} T_2} \quad \text{(Sub)}$$

Method invocation

$$\frac{\Gamma \vdash t : (m(x : T_1) : T_2^x),\ y : T_1}{\Gamma \vdash t.m(y) : T_2^y} \quad \text{(TAppVar)}$$

$$\frac{\Gamma \vdash t : (m(x : T_1) : T_2),\ t_2 : T_1 \quad x \notin \mathsf{fv}(T_2)}{\Gamma \vdash t.m(t_2) : T_2} \quad \text{(TApp)}$$

Object creation

$$\frac{\overset{\text{(labels disjoint)}}{\Gamma, z : T_1^z \wedge \ldots \wedge T_n^z \vdash d_i : T_i^z \quad \forall i.1 \leq i \leq n}}{\Gamma \vdash \{z \Rightarrow d_1 \ldots d_n\} : \{z \Rightarrow T_1^z \wedge \ldots \wedge T_n^z\}} \quad \text{(TNew)}$$

**Member initialization** $\boxed{\Gamma \vdash d : T}$

$$\frac{\Gamma \vdash T <: T}{\Gamma \vdash (L = T) : (L : T\ ..\ T)} \quad \text{(DTyp)}$$

$$\frac{\Gamma, x : T_1 \vdash t : T_2^x}{\Gamma \vdash (m(x : T_1) : T_2^x = t) : (m(x : T_1) : T_2^x)} \quad \text{(DFun)}$$

**Figure 2.5: oopslaDOT: Free variables and well-formedness**

**Well-formed type** $\boxed{\Gamma \vdash T \text{ wf}}$

$$\frac{\mathsf{fv}(T) \subseteq \mathsf{dom}(\Gamma)}{\Gamma \vdash T \text{ wf}} \quad (\text{WTyp})$$

**Well-formed term** $\boxed{\Gamma \vdash t \text{ wf}}$

$$\frac{\mathsf{fv}(t) \subseteq \mathsf{dom}(\Gamma)}{\Gamma \vdash t \text{ wf}} \quad (\text{WTerm})$$

**Well-formed environment** $\boxed{\Gamma \text{ wf}}$

$$\varnothing \text{ wf} \quad (\text{WEmpty})$$

$$\frac{\Gamma \text{ wf} \quad \Gamma, x : T^x \vdash T^x \text{ wf}}{\Gamma, x : T^x \text{ wf}} \quad (\text{WEnv})$$

theoretical problem and does not by itself disqualify wfDOT as a target calculus assuming we are willing to add back unions ourselves.

wfDOT does have one typing rule that has no counterpart in oopslaDOT:

$$\frac{\Gamma \vdash x : T \quad \Gamma \vdash x : U}{\Gamma \vdash x : T \wedge U} \quad (\text{And-I})$$

However, we will show in Subsection 5.5.1 that oopslaDOT can be extended with rules that generalize And-I.

In the end, the only fundamental differences between wfDOT and oopslaDOT lie in their subtyping rules, as summarized in Figure 2.6. oopslaDOT supports subtyping between recursive types via rules BindX and Bind1 and these rules have no equivalent in wfDOT. The price oopslaDOT pays for this is a significantly more complex soundness proof and some seemingly arbitrary restrictions on subtyping involving type selections (in rules Sel1 and Sel2): the variable containing the type member being selected must be typed in a truncated context using the "strict typing" judgment $\Gamma \vdash x :_! T$ which prohibits use of VarPack.[5,6]

Having described the differences between these two calculi, it is now time to determine which one we shall use as the target of our type-preserving compilation schemes. At first glance, wfDOT looks like the better candidate: Scala does not have a direct equivalent to the recursive subtyping rules wfDOT lacks, and the Scala compiler never performs context truncation in subtyping, so the restrictions imposed by oopslaDOT seem like potential impediments. In fact,

[5] See [Hu 2019] for an example illustrating the effect of context truncation on expressiveness.

[6] In addition, Sel1 requires $x$ to have type $(L : \bot .. T)$ whereas Sel-<: uses type $(L : S .. T)$ for some arbitrary $S$ instead, but the more general rule can be recovered via Typ, Bot and Trans.

**Figure 2.6: Comparison of oopslaDOT and wfDOT subtyping rules**

**oopslaDOT subtyping**

$$\frac{\Gamma_{[x]} \vdash x :_{!} (L : \bot \mathinner{..} T)}{\Gamma \vdash x.L <: T} \quad (\textsc{Sel1})$$

$$\frac{\Gamma_{[x]} \vdash x :_{!} (L : S \mathinner{..} \top)}{\Gamma \vdash S <: x.L} \quad (\textsc{Sel2})$$

$$\frac{\Gamma, z : T_1^z \vdash T_1^z <: T_2^z}{\Gamma \vdash \{z \Rightarrow T_1^z\} <: \{z \Rightarrow T_2^z\}} \quad (\textsc{BindX})$$

$$\frac{\Gamma, z : T_1^z \vdash T_1^z <: T_2}{\Gamma \vdash \{z \Rightarrow T_1^z\} <: T_2} \quad (\textsc{Bind1})$$

**wfDOT subtyping**

$$\frac{\Gamma \vdash x : (L : S \mathinner{..} T)}{\Gamma \vdash x.L <: T} \quad (\textsc{Sel-<:})$$

$$\frac{\Gamma \vdash x : (L : S \mathinner{..} T)}{\Gamma \vdash S <: x.L} \quad (\textsc{<:-Sel})$$

Amin expressed a similar sentiment in her thesis [Amin 2016, § 3.5.2]:

> "Let's first consider the stepping-stone option pursued by pragmatism in prior work [Amin, Grütter, et al. 2016] of omitting recursive types from subtyping, making them second-class types. This option has the big advantage of simplicity: typing can be used without caveats in subtyping type selections. Furthermore, this option is a decent match for Dotty / Scala which already has several restrictions on structural recursive types."

The most surprising result of this thesis is that wfDOT is in fact not a good target calculus for Scala, but oopslaDOT is! Both calculi would likely work equally well as compilation targets for Featherweight Java (Chapter 3), but as soon as we extend our source calculus to Featherweight Generic Java in Chapter 4, our proofs of type-preserving compilation end up critically relying on the subtyping rules involving recursive types.[7] These rules let us establish *subtyping preservation*[8]: if $\Gamma \vdash S <: T$ holds in our source language, then given the function $|\cdot|$ that translates types and environments into our target language, we should be able to prove $|\Gamma| \vdash |S| <: |T|$.

What about the restrictions present in Sel1 and Sel2? The use of "strict typing" does not cause any issue in our proofs in practice, but the context truncation restriction from Sel1 and Sel2 do need to be reflected in the declarative subtyping rules DS-SelOther1 and DS-SelOther2 in Chapter 7. However, we find that we can define sound algorithmic subtyping rules AS-Sel1 and AS-Sel2 that do not require context truncation and match the behavior of the Scala 3 compiler.

[7] Subsection 4.3.1 presents an alternative translation scheme which does not require subtyping between recursive types but forces us to restrict the set of valid class hierarchies.

[8] If our translation didn't have this property we would have to inserts coercions to emulate subtyping, but this would likely make the expression translation much more complex and defeat the point of relying on the DOT type system to encode and reason about core Scala semantics.

In other words, the restrictions imposed by oopslaDOT do not prevent us from translating Scala programs that the compiler would accept, which is great news!

Having established oopslaDOT as the most appropriate target calculus for our purposes, we will spend the rest of this chapter studying it but will now refer to it simply as “DOT”. Note however that the DOT we discuss here will still need to be extended in subsequent chapters. In Chapter 4, we introduce applied class types which require augmenting oopslaDOT with an extra subtyping rule (And-Bind in subsection 4.3.1) for the subtyping preservation proof to go through. In Chapter 5, we introduce intersection types which require an extra typing rule (And-I' in subsection 5.5.1). In both cases, we prove the resulting extended calculus sound by updating the existing Coq mechanization of oopslaDOT. In the latter case, this requires generalizing the original type soundness theorem [Rompf and Amin 2016, Theorem 1] to imply the usual property of *preservation* in Theorem 5.5.4.

## 2.3 Syntactic sugar

The following derived syntactic forms will come in handy in our translations.

**Definition 2.3.1: Type alias**

$$(X = T) \rightsquigarrow (X : T \,..\, T)$$

**Definition 2.3.2: List in recursive type**

$$\{z \Rightarrow \overline{T}\} \rightsquigarrow \{z \Rightarrow \bigwedge \overline{T}\}$$

**Definition 2.3.3: Anonymous function**

**Derived type**

$$(x : S) \Rightarrow U \rightsquigarrow \{_ \Rightarrow \mathsf{apply}(x : S) : U\}$$

**Derived term**

$$\lambda x.\, u \rightsquigarrow \{_ \Rightarrow \mathsf{apply}(x) = u\}$$

**Definition 2.3.4: Let bindings**

$$\textbf{let } x = s \textbf{ in } u \rightsquigarrow (\lambda x.\, u).\mathsf{apply}(s)$$

$$\textbf{let } x = s,\ \overline{y = t} \textbf{ in } u \rightsquigarrow \textbf{let } x = s \textbf{ in } (\textbf{let } \overline{y = t} \textbf{ in } u)$$

**Lemma 2.3.5**

$$\frac{\Gamma \vdash t : T \qquad \Gamma, x : T \vdash s : S \qquad x \notin \mathsf{fv}(S)}{\Gamma \vdash \textbf{let } x = t \textbf{ in } s : S} \quad (\textsc{Let})$$

*Proof.*

$$
\dfrac{\Gamma \vdash t : T \quad x \notin \mathsf{fv}(S) \quad \dfrac{\dfrac{\dfrac{\Gamma, x : T \vdash s : S}{\Gamma \vdash (\mathsf{apply}(x) = s) : (\mathsf{apply}(x : T) : S)}\ (\text{DFun})}{\Gamma \vdash \{_ \Rightarrow \mathsf{apply}(x) = s\} : \{_ \Rightarrow \mathsf{apply}(x : T) : S\}}\ (\text{TNew, WeakenTp})}{\Gamma \vdash \{_ \Rightarrow \mathsf{apply}(x) = s\} : (\mathsf{apply}(x : T) : S)}\ (\text{Sub, Bind1})}{\Gamma \vdash \textbf{let } x = t \textbf{ in } s : S}\ (\text{TApp})
$$

∎

### Definition 2.3.6: Methods with variable number of parameters

In a parameter list, we allow each parameter type to refer to all previous parameters and the result type to refer to all parameters.

**Derived types**

$$m() : U_0 \rightsquigarrow m(_ : \top) : U_0$$

$$m(\overline{x : S}, y : T) : U_0 \rightsquigarrow m(\overline{x : S}) : ((y : T) \Rightarrow U_0)$$

**Derived declarations** (all type ascriptions are optional)

$$m() : U_0 = t \rightsquigarrow m(_ : \top) : U_0 = t$$

$$m(\overline{x : S}, y : T) : U_0 = t \rightsquigarrow m(\overline{x : S}) : ((y : T) \Rightarrow U_0) = t$$

**Derived terms**

$$t.m() \rightsquigarrow t.m(\{_ \Rightarrow\})$$

$$t.m(\overline{x}, y) \rightsquigarrow t.m(\overline{x}).\mathsf{apply}(y)$$

### Lemma 2.3.7

We can generalize DFun, Fun, TApp and TAppVar to methods with variable number of parameters. Note that TApp' generalizes both TApp and TAppVar since it lets the result type depend on a subset of the method arguments. We intentionally make the names of each parameter coincide with the name of the corresponding argument to avoid having to write down all the variable substitutions that could be involved.

$$
\dfrac{\begin{array}{c}\Delta_0 = \Gamma \quad \Delta_{i+1} = \Delta_i, x_{i+1} : S_{i+1} \\ \Delta_i \vdash S_{i+1}\ \mathsf{wf} \quad \Delta_n \vdash u : U\end{array}}{\Gamma \vdash (m(\overline{x : S}) : U = u) : (m(\overline{x : S}) : U)} \quad (\text{DFun'})
$$

$$
\dfrac{\begin{array}{c}\Delta_0 = \Gamma \quad \Delta_{i+1} = \Delta_i, x_{i+1} : S_{i+1} \\ \Delta_{i+1} \vdash T_{i+1} <: S_{i+1} \quad \Delta_n \vdash U_1 <: U_2\end{array}}{\Gamma \vdash m(\overline{x : S}) : U_1 <: m(\overline{x : T}) : U_2} \quad (\text{Fun'})
$$

$$
\dfrac{\begin{array}{c}\Gamma \vdash t : (m(\overline{x : S}, \overline{y : T}) : U) \\ \Delta_0 = \Gamma \quad \Delta_{i+1} = \Gamma, x_{i+1} : S_{i+1} \\ \Delta_n \vdash \overline{t : T}\end{array}}{\Gamma \vdash t.m(\overline{x}, \overline{t}) : U} \quad (\text{TApp'})
$$

## 2.4 Meta-theory

The following derived subtyping rules are defined in [Rompf and Amin 2016]:

$$\Gamma \vdash T <: T \qquad \text{(Refl)}$$

$$\frac{\Gamma_1 \vdash T_1 <: T_2 \quad \Gamma_2(x) = T_2 \quad \Gamma_1 = \Gamma_2(x \to T_1) \quad \Gamma_2 \vdash S <: U}{\Gamma_1 \vdash S <: U} \qquad \text{(Narrow)}$$

where $\Gamma_1 = \Gamma_2(x \to T_1)$ means that $\Gamma_1$ is equal to $\Gamma_2$ for all inputs except $x$ which it maps to $T_1$.

**Lemma 2.4.1: Weakening**

$$\frac{\Gamma_1, \Gamma_2 \vdash T_1 <: T_2 \quad y \notin \mathsf{dom}(\Gamma_1)}{\Gamma_1, y : U, \Gamma_2 \vdash T_1 <: T_2} \qquad \text{(Weaken)}$$

$$\frac{\Gamma_1, \Gamma_2 \vdash t :_{(!)} T \quad y \notin \mathsf{dom}(\Gamma_1)}{\Gamma_1, y : U, \Gamma_2 \vdash t :_{(!)} T} \qquad \text{(WeakenTp)}$$

*Proof.* Both rules are proved together by simultaneous induction on the size of the subtyping and typing derivations, we only show a few representative cases:

**Case** $\frac{(\Gamma_1, \Gamma_2)_{[x]} \vdash x :_! (L : T \mathrel{..} \top)}{\Gamma_1, \Gamma_2 \vdash T <: x.L}$ (Sel2)

> We can distinguish two sub-cases:
> - If $x \in \Gamma_1$, then $(\Gamma_1, y : U, \Gamma_2)_{[x]} = (\Gamma_1, \Gamma_2)_{[x]}$ and Sel2 finishes the case.
> - If $x \in \Gamma_2$, then $(\Gamma_1, \Gamma_2)_{[x]} = \Gamma_1, \Gamma_{2_{[x]}}$ and $(\Gamma_1, y : U, \Gamma_2)_{[x]} = \Gamma_1, y : U, \Gamma_{2_{[x]}}$, therefore by the IH we have $(\Gamma_1, y : U, \Gamma_2)_{[x]} \vdash x :_! (L : T \mathrel{..} \top)$ and Sel2 finishes the case again.

**Case** $\frac{\Gamma_1, \Gamma_2, z : T_1^z \vdash T_1^z <: T_2^z}{\Gamma_1, \Gamma_2 \vdash \{z \Rightarrow T_1^z\} <: \{z \Rightarrow T_2^z\}}$ (BindX)

> By the IH we have $\Gamma_1, y : U, \Gamma_2, z : T_1^z \vdash T_1^z <: T_2^z$ and BindX finishes the case.

■

**Lemma 2.4.2: Narrowing of types**

$$\frac{\Gamma_1 \vdash T_1 <: T_2 \quad \Gamma_2(x) = T_2 \quad \Gamma_1 = \Gamma_2(x \to T_1) \quad \Gamma_2 \vdash s :_{(!)} S}{\Gamma_1 \vdash s :_{(!)} S} \qquad \text{(NarrowTp)}$$

*Proof.* By induction on the derivation of $\Gamma_2 \vdash s :_{(!)} S$, with a case analysis on the final rule. We only show Var and TNew as all other cases follow directly from the IH and Narrow.

**Case** $\dfrac{\Gamma_2(s) = S}{\Gamma_2 \vdash s :_{(!)} S}$ (Var)

> We can distinguish two sub-cases:
> - If $s = x$, then $S = T_2$, $\Gamma_1(s) = T_1$ and Sub finishes the case.
> - Otherwise, $\Gamma_1(s) = S$ and Var finishes the case.

**Case** $\dfrac{\begin{array}{c} U = U_1^z \wedge \ldots \wedge U_n^z \\ \Gamma_2, z : U \vdash d_i : U_i^z \quad \forall i.\, 1 \leq i \leq n \end{array}}{\Gamma_2 \vdash \{z \Rightarrow d_1 \ldots d_n\} : \{z \Rightarrow U\}}$ (TNew)

> By Weaken we have $\Gamma_1, z : U \vdash T_1 <: T_2$ so by the IH $\Gamma_1, z : U \vdash d_i : U_i^z$ and TNew finishes the case.

■

**Lemma 2.4.3**

$$\frac{\Gamma_2(x) = T^x \quad \Gamma_1 = \Gamma_2(x \to \{z \Rightarrow T^z\}) \quad \Gamma_2 \vdash S <: U}{\Gamma_1 \vdash S <: U} \quad (\textsc{EnvPack})$$

$$\frac{\Gamma_2(x) = T^x \quad \Gamma_1 = \Gamma_2(x \to \{z \Rightarrow T^z\}) \quad \Gamma_2 \vdash s :_{(!)} S}{\Gamma_1 \vdash s :_{(!)} S} \quad (\textsc{EnvPackTp})$$

*Proof.* By simultaneous induction on the size of the subtyping and typing derivations, we only show the Var case as all others follow by the IH:

**Case** $\dfrac{\Gamma_2(s) = S}{\Gamma_2 \vdash s :_{(!)} S}$ (Var)

> We can distinguish two sub-cases:
> - If $s = x$, then $S = T^x$ and $\Gamma_1 \vdash s :_{(!)} \{z \Rightarrow T^z\}$ by Var. VarUnpack finishes the case.
> - Otherwise, $\Gamma_1(s) = S$ and Var finishes the case.

■

**Lemma 2.4.4: Commutativity and associativity of intersection**

$\Gamma \vdash T_1 \wedge T_2 <: T_2 \wedge T_1$, $\Gamma \vdash T_1 \wedge (T_2 \wedge T_3) <: (T_1 \wedge T_2) \wedge T_3$ and $\Gamma \vdash (T_1 \wedge T_2) \wedge T_3 <: T_1 \wedge (T_2 \wedge T_3)$

*Proof.* By And2, And11, And12 and Refl. ■

**Lemma 2.4.5: Width and depth subtyping**

1. $\Gamma \vdash \overline{T_0} \wedge \overline{T_1} \wedge \overline{T_2} <: \overline{T_1}$
2. $\Gamma \vdash \{z \Rightarrow \overline{T_0} \wedge \overline{T_1} \wedge \overline{T_2}\} <: \{z \Rightarrow \overline{T_1}\}$
3. If $\Gamma \vdash \overline{S <: T}$ then $\Gamma \vdash \bigwedge \overline{S} <: \bigwedge \overline{T}$
4. If $\Gamma, z : \bigwedge \overline{S} \vdash \overline{S <: T}$ then $\Gamma \vdash \{z \Rightarrow \overline{S}\} <: \{z \Rightarrow \overline{T}\}$

**Lemma 2.4.6: Substituting type selection by equal type preserves type equality**

Given $\sigma = [\overline{T/x.L}]$ and $\Gamma \vdash \overline{T =:= x.L}$, if $\Gamma \vdash U$ wf then $\Gamma \vdash \sigma U =:= U$

*Proof.* By structural induction on $U$. Cases $U = \bot$ and $U = \top$ are trivial since in those cases $\sigma U = U$.

**Case** $U = y.L'$

If $U \notin \mathrm{dom}(\sigma)$ then $\sigma U = U$. Otherwise, $\sigma U = T_i$ for some $i$ and we know that $\Gamma \vdash T_i =:= U$.

**Case** $U = (L : U_1 .. U_2)$

We have $\sigma U = (L : \sigma U_1 .. \sigma U_2)$. By the IH, $\Gamma \vdash \sigma U_1 =:= U_1$ and $\Gamma \vdash \sigma U_2 =:= U_2$. Typ finishes the case.

**Case** $U = m(x : U_1) : U_2^x$

We have $\sigma U = m(x : \sigma U_1) : \sigma U_2^x$.

$$\dfrac{\dfrac{}{\Gamma \vdash U_1 <: \sigma U_1}\,(\text{IH}) \qquad \dfrac{\dfrac{}{\Gamma, x : \sigma U_1 \vdash \overline{T =:= x.L}}\,(\textsc{Weaken})}{\Gamma, x : \sigma U_1 \vdash U_2^x <: \sigma U_2^x}\,(\text{IH})}{\Gamma \vdash m(x : U_1) : U_2^x <: m(x : \sigma U_1) : \sigma U_2^x}\,(\textsc{Fun})$$

$$\dfrac{\dfrac{}{\Gamma \vdash \sigma U_1 <: U_1}\,(\text{IH}) \qquad \dfrac{\dfrac{}{\Gamma, x : U_1 \vdash \overline{T =:= x.L}}\,(\textsc{Weaken})}{\Gamma, x : U_1 \vdash \sigma U_2^x <: U_2^x}\,(\text{IH})}{\Gamma \vdash m(x : \sigma U_1) : \sigma U_2^x <: m(x : U_1) : U_2^x}\,(\textsc{Fun})$$

**Case** $U = \{z \Rightarrow U_1^z\}$

We have $\sigma U = \{z \Rightarrow \sigma U_1^z\}$, we only show one direction since the other proceeds similarly.

$$\dfrac{\dfrac{\dfrac{}{\Gamma, z : U_1^z \vdash \overline{T =:= x.L}}\,(\textsc{Weaken})}{\Gamma, z : U_1^z \vdash U_1^z <: \sigma U_1^z}\,(\text{IH})}{\Gamma \vdash \{z \Rightarrow U_1^z\} <: \{z \Rightarrow \sigma U_1^z\}}\,(\textsc{BindX})$$

**Case** $U = U_1 \wedge U_2$

> We have $\sigma U = \sigma U_1 \wedge \sigma U_2$ and again we only show one direction.
>
> $$\dfrac{\dfrac{\dfrac{}{\Gamma \vdash U_1 <: \sigma U_1}\text{(IH)}}{\Gamma \vdash U_1 \wedge U_2 <: \sigma U_1}\text{(And12)} \quad \dfrac{\dfrac{}{\Gamma \vdash U_2 <: \sigma U_2}\text{(IH)}}{\Gamma \vdash U_1 \wedge U_2 <: \sigma U_2}\text{(And12)}}{\Gamma \vdash U_1 \wedge U_2 <: \sigma U_1 \wedge \sigma U_2}\text{(And2)}$$

**Case** $U = U_1 \vee U_2$

> We have $\sigma U = \sigma U_1 \vee \sigma U_2$ and we only show one direction here too.
>
> $$\dfrac{\dfrac{\dfrac{}{\Gamma \vdash U_1 <: \sigma U_1}\text{(IH)}}{\Gamma \vdash U_1 <: \sigma U_1 \vee \sigma U_2}\text{(Or21)} \quad \dfrac{\dfrac{}{\Gamma \vdash U_2 <: \sigma U_2}\text{(IH)}}{\Gamma \vdash U_2 <: \sigma U_1 \vee \sigma U_2}\text{(Or22)}}{\Gamma \vdash U_1 \vee U_2 <: \sigma U_1 \vee \sigma U_2}\text{(Or1)}$$

■

**Lemma 2.4.7: Substituting type selection by equal type preserves typing**

Given $\sigma = [\overline{T/x.L}]$ and $\Gamma \vdash \overline{T =:= x.L}$, then

1. $\Gamma \vdash d : S$ implies $\Gamma \vdash \sigma d : \sigma S$
2. $\Gamma \vdash t : T$ implies $\Gamma \vdash \sigma t : \sigma T$

*Proof.* By simultaneous induction on the derivations of $\Gamma \vdash d : S$ and $\Gamma \vdash t : T$ using Lemma 2.4.6. We only show a few representative cases.

**Case** $\dfrac{\Gamma(x) = T^x}{\Gamma \vdash x : T^x}\text{(Var)}$

> We have $\sigma x = x$. By Lemma 2.4.6, $\Gamma \vdash T^x <: \sigma T^x$ and Sub finishes the case.

**Case** $\dfrac{\Gamma \vdash S' <: S'}{\Gamma \vdash (L = S') : (L : S' \,..\, S')}\text{(DTyp)}$

> By Refl, $\Gamma \vdash \sigma S' <: \sigma S'$ and DTyp finishes the case.

**Case** $\dfrac{\Gamma, x : T_1 \vdash t : T_2^x}{\Gamma \vdash (m(x : T_1) : T_2^x = t) : (m(x : T_1) : T_2^x)}\text{(DFun)}$

$$\dfrac{\dfrac{}{\Gamma, x : T_1 \vdash \sigma t : \sigma T_2^x}\text{(IH 2.)} \qquad \dfrac{}{\Gamma, x : \sigma T_1 \vdash \sigma T_1 <: T_1}\text{(Lemma 2.4.6)}}{\dfrac{\Gamma, x : \sigma T_1 \vdash \sigma t : \sigma T_2^x}{\vdash (m(x : \sigma T_1) : \sigma T_2^x = \sigma t) : (m(x : \sigma T_1) : \sigma T_2^x)}\text{(DFun)}}\text{(NarrowTp)}$$

■

# 3 Featherweight Java (Scala-flavored)

In this chapter, we review the Featherweight Java (FJ) calculus [Igarashi, Pierce, and Wadler 2001]. We then develop a translation scheme from FJ into DOT and prove that it is type-preserving.

## 3.1 Syntax and semantics

**Figure 3.1: FJ: Syntax**

| | | | |
|---|---|---|---|
| | | $L$ ::= | Class declaration |
| $x, y, z$ | Variable | **class** $C(\overline{f : D}) \lhd B(\overline{g})\ \{\overline{M}\}$ | |
| $B, C, D, E$ | Class type | $M$ ::= | Method declaration |
| $f, g$ | Class parameter | **def** $m(\overline{x : D}) : D_0 = e_0$ | |
| $m$ | Method name | $e$ ::= | Expression |
| | | $x$ | variable |
| $\Gamma$ ::= | Context | $e.f$ | parameter access |
| $\varnothing \mid \Gamma, x : C$ | | $e_0.m(\overline{e})$ | method call |
| | | **new** $C(\overline{e})$ | object |

FJ models a single-class inheritance language where subtyping is defined by subclassing. It was originally designed to be a proper subset of Java but it also happens to be a good match for the semantics of Scala. To make this more obvious, we alter its syntax to resemble Scala.

Besides the syntax changes, the version of FJ we present here lacks support for casts. In principle, they should be translatable into DOT using an approach similar to [League, Shao, and Trifonov 2002] but we consider them out of scope for this thesis.

An FJ program is a pair $(CT, e)$ composed of a class table $CT$ and an expression $e$. The class table maps class names $C$ to class declarations **class** $C(\overline{f : D}) \lhd B(\overline{g})\ \{\overline{M}\}$ where,

- $C$ is the name of the class,
- $\overline{f : D}$ declares the names and types of the parameters accepted by the class constructor,

**Figure 3.2: FJ: Comparison of the original and Scala-flavored syntax**

**Original**

```
class C ◁ Object {
  A a;
  C(A a) {
    super(); this.a = a
  }
}
class D ◁ C {
  B b;
  A(A a, B b) {
    super(a); this.b = b;
  }
  F foo(E e) {
    return new G(e).bar(this.b).f;
  }
}
```

**Scala-flavored**

```
class C(a: A) ◁ Object {}
class D(a: A, b: B) ◁ C(a) {
  def foo(e: E): F =
    new G(e).bar(b).f
}
```

- $B$ is the parent class that $C$ extends (the special class name Object can be used here and denotes the root of the class hierarchy),
- $\overline{g}$ is the subset of the class parameters which are passed to the constructor of the parent class,
- and $\overline{M}$ is the list of methods defined in the class.

In turn, method declarations have the form **def** $m(\overline{x : D}) : D_0 = e_0$ where,

- $m$ is the name of the method,
- $\overline{x : D}$ declares the names and types of the parameters accepted by the method,
- $D_0$ is the result type of the method,
- and $e_0$ is the body of the method.

A valid expression is either,

- a reference to a variable $x$ in the environment,
- a constructor call **new** $C(\overline{e})$ which returns an object of type $C$ instantiated using class parameters $\overline{e}$,
- a method call $e_0.m(\overline{e})$ where the class type of the receiver $e_0$ has a method $m$ which

accepts arguments $\overline{e}$,

- or a parameter access $e.f$ where the class type of $e$ has a constructor parameter $f$.

A well-typed program written in our calculus is almost, but not quite, valid Scala. For the sake of brevity, we omit the `val` keyword in front of constructor parameters which is normally needed to allow access to class parameters via the $e.f$ syntax. We also write $\triangleleft$ as a short-hand for **extends** as in the original paper.

Figure 3.2 informally defines the mapping between the original syntax and our Scala-flavored version. Although it may not look like it, all well-formed cast-less FJ programs can be expressed in our syntax due to the restrictions imposed on well-formed classes by FJ. The subtyping and typing rules in Figures 3.3 to 3.5 are adapted from the original paper to fit our syntax. As in the original definitions, every class $C$ mentioned in a rule is assumed to be defined in the global class table $CT$.

We intentionally omit the definition of evaluation rules. Instead, we give meaning to a well-typed FJ program via a type-preserving translation into DOT defined in the next section.[1] Since DOT is sound [Rompf and Amin 2016, Definition 1], this indirectly establishes soundness for our source calculus.

**Figure 3.3: FJ: Subtyping rules and lookup functions**

**Subtyping** $\boxed{C <: D}$

$$C <: C \qquad \text{(S-Refl)}$$

$$\frac{C <: D \quad D <: B}{C <: B} \qquad \text{(S-Trans)}$$

$$\frac{\textbf{class}\, C \ldots \triangleleft B \ldots}{C <: B} \qquad \text{(S-Class)}$$

[1] It would be interesting to formally relate the traditional way FJ evaluation proceeds with the evaluation of an FJ program translated into DOT, but the use of a store in the operational semantics of oopslaDOT makes this non-trivial. The store-less version of DOT from [Amin 2016, § 3.5] might be more appropriate for this task.

**Figure 3.4: FJ: lookup functions**

**Value parameters lookup**

$$\boxed{\mathsf{vparams}(C) \mathrel{:=} \overline{f : D}}$$

$$\mathsf{vparams}(\mathsf{Object}) \mathrel{:=} \varnothing$$

$$\frac{\textbf{class}\ C(\overline{f : D})\ \ldots}{\mathsf{vparams}(C) \mathrel{:=} \overline{f : D}}$$

**Method names lookup**

$$\boxed{\mathsf{mnames}(C) \mathrel{:=} \overline{m}}$$

$$\mathsf{mnames}(\mathsf{Object}) \mathrel{:=} \varnothing$$

$$\frac{\begin{array}{c}\textbf{class}\ C\ \ldots \lhd B\ \{\overline{\textbf{def}\ m_C\ \ldots}\} \\ \mathsf{mnames}(B) = \overline{m_B} \\ \overline{n} = \left[m \in \overline{m_C}\ \middle|\ m \notin \overline{n}\right]\end{array}}{\mathsf{mnames}(C) \mathrel{:=} \overline{m_B}, \overline{n}}$$

**Method type and body lookup**

$$\boxed{\begin{array}{l}\mathsf{mtype}(m, C) \mathrel{:=} (\overline{x : D}) \to D_0 \\ \mathsf{mbody}(m, C) \mathrel{:=} e_0\end{array}}$$

$$\frac{\begin{array}{c}\textbf{class}\ C\ \ldots\ \{\overline{M}\} \\ \textbf{def}\ m(\overline{x : D}) : D_0 = e_0 \in \overline{M}\end{array}}{\begin{array}{l}\mathsf{mtype}(m, C) \mathrel{:=} (\overline{x : D}) \to D_0 \\ \mathsf{mbody}(m, C) \mathrel{:=} e_0\end{array}} \quad \text{(M-Class)}$$

$$\frac{\textbf{class}\ C\ \ldots \lhd B\ \{\overline{M}\} \qquad m\ \ldots \notin \overline{M}}{\begin{array}{l}\mathsf{mtype}(m, C) \mathrel{:=} \mathsf{mtype}(m, B) \\ \mathsf{mbody}(m, C) \mathrel{:=} \mathsf{mbody}(m, B)\end{array}} \quad \text{(M-Super)}$$

**Figure 3.5: FJ: Typing rules**

**Expression typing** $\boxed{\Gamma \vdash e : C}$

$$\frac{\Gamma(x) = C}{\Gamma \vdash x : C} \quad \text{(T-VAR)}$$

$$\frac{\Gamma \vdash e_0 : C \quad \mathsf{vparams}(C) = \overline{f : D}}{\Gamma \vdash e_0.f_i : D_i} \quad \text{(T-GETTER)}$$

$$\frac{\Gamma \vdash e_0 : C \quad \mathsf{mtype}(m, C) = (\overline{x : D}) \longrightarrow D_0 \quad \Gamma \vdash \overline{e : E} \quad \overline{E <: D}}{\Gamma \vdash e_0.m(\overline{e}) : D_0} \quad \text{(T-INVK)}$$

$$\frac{\textbf{class}\, C(\overline{f : D}) \quad \Gamma \vdash \overline{e : E} \quad \overline{E <: D}}{\Gamma \vdash \textbf{new}\, C(\overline{e}) : C} \quad \text{(T-NEW)}$$

**Method typing** $\boxed{\Gamma \vdash m \text{ ok}}$

$$\frac{\begin{array}{c} C = \Gamma(\mathsf{this}) \quad \textbf{class}\, C \lhd B \ldots \\ \mathsf{mtype}(m, C) = (\overline{x : D}) \longrightarrow D_0 \\ \mathsf{mbody}(m, C) = e_0 \\ \Gamma, \overline{x : D} \vdash e_0 : E_0 \quad E_0 <: D_0 \\ \mathsf{mtype}(m, B) \text{ defined implies } \mathsf{mtype}(m, B) = \mathsf{mtype}(m, C) \end{array}}{\Gamma \vdash m \text{ ok}} \quad \text{(T-METHOD)}$$

**Class typing** $\boxed{\vdash C \text{ ok}}$

$$\frac{\begin{array}{c} \textbf{class}\, C(\overline{g : E}, \overline{f : D}) \lhd B(\overline{g}) \ \{\overline{\textbf{def}\, m \ldots}\} \\ \mathsf{vparams}(B) = \overline{g : E} \quad \overline{\mathsf{this} : C \vdash m \text{ ok}} \end{array}}{\vdash C \text{ ok}} \quad \text{(T-CLASS)}$$

**Class table typing** $\boxed{\vdash CT \text{ ok}}$

$$\frac{\begin{array}{c} C \in \mathsf{dom}(CT) \text{ implies } \vdash C \text{ ok} \\ \text{No inheritance cycle between the classes in } CT \end{array}}{\vdash CT \text{ ok}} \quad \text{(T-CT)}$$

## 3.2 Translation

Our translation scheme is defined using three operators defined in Figures 3.6 and 3.7:

- $|\cdot|$ translates FJ types into DOT types and FJ terms into DOT terms.
- $(\!|\cdot|\!)$ translates lists of FJ declarations into one or more DOT declarations.
- If $(\!|\cdot|\!)$ returns a DOT declaration, then $[\![\cdot]\!]$ is defined to return the type of the declaration, for example since $(\!| f : D |\!) := (f() : |D| = f_{\mathsf{param}})$ we have $[\![ f : D ]\!] = (f : |D|)$. If $(\!|\cdot|\!)$ returns multiple DOT declarations, then $[\![\cdot]\!]$ returns the intersection of their types. For convenience, we additionally define $[\![\mathsf{Object}]\!] := \top$.

**Figure 3.6: Translating FJ types and expressions to DOT**

**Type Translation** $\quad |C| := T_{\text{DOT}}$

$$|\mathsf{Object}| := \mathsf{ct.Object}$$

$$\frac{\textbf{class}\ C \ldots \lhd D \ldots}{|C| := \mathsf{ct}.C}$$

**Expression Translation** $\quad |e| := t_{\text{DOT}}$

$$|x| := x$$

$$|e.f| := |e|.f()$$

$$|e_0.m(\overline{e})| := |e_0|.m(\overline{|e|})$$

$$|\textbf{new}\ \mathsf{Object}| := \{_ \Rightarrow\}$$

$$|\textbf{new}\ C(\overline{e})| := \mathsf{ct.new}_C(\overline{|e|})$$

We illustrate our translation scheme with an example. Given the class table $CT$,

```
class B(obj: Object) ◁ Object {}
class C() ◁ Object {
  def foo(): C = this
}
class D() ◁ C() {
  def bar(b: B): Object = b.obj
}
```

we translate it to the object $\{\mathsf{ct} \Rightarrow (\!| CT |\!)\}$ which expands to

**Figure 3.7: Translating FJ definitions to DOT**

**Getter Translation** $\boxed{(\!|\, \overline{f : D} \,|\!) \mathrel{::=} d_{\text{DOT}}}$

$$(\!|\, f : D \,|\!) \mathrel{::=} f() : |D| = f_{\mathsf{param}}$$

$$(\!|\, \overline{f : D} \,|\!) \mathrel{::=} \overline{(\!|\, f : D \,|\!)}$$

**Method Translation** $\boxed{(\!|\, \overline{m} \,|\!)_C \mathrel{::=} d_{\text{DOT}}}$

$$\frac{\mathsf{mtype}(m, C) = (\overline{x : D}) \longrightarrow D_0 \qquad \mathsf{mbody}(m, C) = e_0}{(\!|\, m \,|\!)_C \mathrel{::=} m(\overline{x : |D|}) : |D_0| = |e_0|}$$

$$(\!|\, \overline{m} \,|\!)_C \mathrel{::=} \overline{(\!|\, m \,|\!)_C}$$

**Class Translation** $\boxed{(\!|\, C \,|\!) \mathrel{::=} \overline{d_{\text{DOT}}}}$

$$(\!|\, C \,|\!) \mathrel{::=} (\!|\, \mathsf{vparams}(C) \,|\!),\ (\!|\, \mathsf{mnames}(C) \,|\!)_C$$

**Class Table Translation** $\boxed{(\!|\, CT \,|\!) \mathrel{::=} \overline{d_{\text{DOT}}}}$

$$(\!|\, \varnothing \,|\!) \mathrel{::=} (\mathsf{Object} = \top)$$

$$\frac{L_C = \textbf{class}\, C[\overline{X_C <: N}](\overline{f : U}) \lhd B \ldots}{(\!|\, \overline{L}, L_C \,|\!) \mathrel{::=} (\!|\, \overline{L} \,|\!),\ \Big(C = \mathsf{ct}.B \wedge \{\mathsf{this} \Rightarrow [\![C]\!]\}\Big),\ \Big(\mathsf{new}_C(\overline{f_{\mathsf{param}} : |D|}) : |C| = \{\mathsf{this} \Rightarrow (\!|\, C \,|\!)\}\Big)}$$

**Environment Translation** $\boxed{|\Gamma| \mathrel{::=} \Gamma_{\text{DOT}}}$

$$|\varnothing| \mathrel{::=} \mathsf{ct} : [\![CT]\!]$$

$$\frac{\mathsf{vparams}(C) = \overline{f : D}}{|\Gamma, \mathsf{this} : C| \mathrel{::=} |\Gamma|, \overline{f_{\mathsf{param}} : |D|}, \mathsf{this} : [\![C]\!]}$$

$$\frac{x \neq \mathsf{this}}{|\Gamma, x : C| \mathrel{::=} |\Gamma|, x : |C|}$$

$$
\begin{array}{l}
\{\mathsf{ct} \Rightarrow \\
\quad \mathsf{Object} = \top, \\
\quad \mathtt{B} = \mathsf{ct}.\mathtt{Object} \wedge \{\mathsf{this} \Rightarrow (\mathtt{obj}() : \mathsf{ct}.\mathtt{Object})\}, \\
\quad \mathsf{new}_{\mathtt{B}}(\mathtt{obj}_{\mathsf{param}} : \mathsf{ct}.\mathtt{Object}) : \mathsf{ct}.\mathtt{B} = \{\mathsf{this} \Rightarrow \\
\quad\quad \mathtt{obj}() : \mathsf{ct}.\mathtt{Object} = \mathtt{obj}_{\mathsf{param}} \\
\quad \}, \\
\quad \mathtt{C} = \mathsf{ct}.\mathtt{Object} \wedge \{\mathsf{this} \Rightarrow (\mathtt{foo}() : \mathsf{ct}.\mathtt{C})\}, \\
\quad \mathsf{new}_{\mathtt{C}}() : \mathsf{ct}.\mathtt{C} = \{\mathsf{this} \Rightarrow \\
\quad\quad \mathtt{foo}() : \mathsf{ct}.\mathtt{C} = \mathsf{this} \\
\quad \}, \\
\quad \mathtt{D} = \mathsf{ct}.\mathtt{C} \wedge \{\mathsf{this} \Rightarrow (\mathtt{foo}() : \mathsf{ct}.\mathtt{C}) \wedge (\mathtt{bar}(\mathtt{b} : \mathsf{ct}.\mathtt{Object}) : \mathsf{ct}.\mathtt{C})\}, \\
\quad \mathsf{new}_{\mathtt{D}}() : \mathsf{ct}.\mathtt{D} = \{\mathsf{this} \Rightarrow \\
\quad\quad \mathtt{foo}() : \mathsf{ct}.\mathtt{C} = \mathsf{this}, \mathit{ch} \\
\quad\quad \mathtt{bar}(\mathtt{b} : \mathsf{ct}.\mathtt{B}) : \mathsf{ct}.\mathtt{Object} = \mathtt{b}.\mathtt{obj}() \\
\quad \} \\
\}
\end{array}
$$

When possible, a translated definition reuses the name of the original definition, but a class parameters like `obj` above must be translated both into a parameter to the constructor method $\mathsf{new}_{\mathtt{B}}$ and a method in the translated class body, so we name the constructor method parameter $\mathtt{obj}_{\mathsf{param}}$ to avoid any ambiguity.

A well-typed FJ program,

$$(CT, e)$$

can be translated into a DOT expression well-typed in the empty context,

$$\textbf{let } \mathsf{ct} = \{\mathsf{ct} \Rightarrow (\!| CT |\!)\} \textbf{ in } |e|$$

as established by Theorem 3.2.13.

### 3.2.1 Meta-theory

Every class $C$ we refer to is implicitly required to be defined in the global class table $CT$ such that $\vdash CT$ ok.

**Lemma 3.2.1: Well-formed translation**

1. $|\Gamma| \vdash |C|, [\![C]\!]$ wf
2. If $(\mathsf{this} : C) \in \Gamma$ then $|\Gamma| \vdash (\!| C |\!)$ wf

*Proof.*

1. The only free variable that can appear in $|C|$ or $[\![C]\!]$ is ct which is always present in $|\Gamma|$.
2. Let $\mathsf{vparams}(C) = \overline{f : D}$, then we have $\{\overline{f_{\mathsf{param}}}, \mathsf{this}\} \subseteq \mathsf{dom}(|\Gamma|)$ which covers all addi-tional free variables that can appear in $(\!| C |\!)$ by inspection.

■

**Theorem 3.2.2: Subtyping preservation**

If $C <: B$ and $|\Gamma|$ defined then $|\Gamma| \vdash |C| <: |B|$.

*Proof.* By induction on the derivation of $C <: B$.

**Case** $C <: C$ (S-Refl)

By Refl.

**Case** $\dfrac{\textbf{class}\ C \ldots \lhd B \ldots}{C <: B}$ (S-Class)

$$\dfrac{\dfrac{}{|\varnothing| \vdash \mathsf{ct} :_! [\![CT]\!]}\ (\textsc{Var}) \qquad \dfrac{\dfrac{\dfrac{\dfrac{}{|\varnothing| \vdash \mathsf{ct}.B <: \mathsf{ct}.B}\ (\textsc{Refl})}{|\varnothing| \vdash \mathsf{ct}.B \wedge \{\mathsf{this} \Rightarrow [\![C]\!]\} <: \mathsf{ct}.B}\ (\textsc{And11})}{|\varnothing| \vdash (C = \mathsf{ct}.B \wedge \{\mathsf{this} \Rightarrow [\![C]\!]\}) <: (C : \bot \mathinner{..} \mathsf{ct}.B)}\ (\textsc{Typ})}{|\varnothing| \vdash [\![CT]\!] <: (C : \bot \mathinner{..} \mathsf{ct}.B)}\ (2.4.5)}{\dfrac{|\varnothing| \vdash \mathsf{ct} :_! (C : \bot \mathinner{..} \mathsf{ct}.B)}{|\Gamma| \vdash \mathsf{ct}.C <: \mathsf{ct}.B}\ (\textsc{Sel1})}\ (\textsc{Sub})$$

**Case** $\dfrac{C <: D \quad D <: B}{C <: B}$ (S-Trans)

By the IH, $|\Gamma| \vdash |C| <: |D|$ and $|\Gamma| \vdash |D| <: |B|$. Trans finishes the case.

■

**Lemma 3.2.3**

If $|\Gamma|$ defined then $|\Gamma| \vdash |C| <: [\![C]\!]$

*Proof.* $C = \mathsf{Object}$ follows by Top, otherwise we have $|C| = \mathsf{ct}.C$ and

$$\dfrac{\dfrac{\dfrac{}{|\Gamma| \vdash \{\mathsf{this} \Rightarrow [\![C]\!]\} <: [\![C]\!]}\ (\textsc{Bind1})}{|\Gamma| \vdash \mathsf{ct}.B \wedge \{\mathsf{this} \Rightarrow [\![C]\!]\} <: [\![C]\!]}\ (\textsc{And2})}{|\Gamma| \vdash \mathsf{ct}.C <: [\![C]\!]}\ (\textsc{Trans}, \textsc{Sel1})$$

■

**Corollary 3.2.4: Class translation preserves value parameters and methods**

- If $\mathsf{vparams}(C) = \overline{f : D}$ then $|\Gamma| \vdash \overline{|C| <: (f() : |D|)}$.
- If $\mathsf{mtype}(m, C) = (\overline{x : D}) \longrightarrow D_0$ then $|\Gamma| \vdash |C| <: (m(\overline{x : |D|}) : |D_0|)$.

*Proof.* By definition, $[\![C]\!] = [\![\mathsf{vparams}(C)]\!] \wedge [\![\mathsf{mnames}(C)]\!]_C$ so this follows from the previous lemma, transitivity and width subtyping. ■

**Lemma 3.2.5**

Given **class** $C \ldots \lhd B \ldots$ and $|\Gamma|$ defined then $|\Gamma| \vdash [\![C]\!] <: [\![B]\!]$.

*Proof.* By definition, we want to show:

$$|\Gamma| \vdash [\![\mathsf{vparams}(C)]\!] \wedge [\![\mathsf{mnames}(C)]\!]_C <: [\![\mathsf{vparams}(B)]\!] \wedge [\![\mathsf{mnames}(B)]\!]_B$$

After proving the following claims, we can finish the case by depth subtyping.

**Claim 1:** $|\Gamma| \vdash [\![\mathsf{vparams}(C)]\!] <: [\![\mathsf{vparams}(B)]\!]$

> $\vdash C\ \mathsf{ok}$ implies that $\mathsf{vparams}(C) = (\mathsf{vparams}(B), \ldots)$ so by definition, $[\![\mathsf{vparams}(C)]\!] = [\![\mathsf{vparams}(B)]\!] \wedge T$ for some $T$ and width subtyping finishes the claim.

**Claim 2:** $|\Gamma| \vdash [\![\mathsf{mnames}(C)]\!]_C <: [\![\mathsf{mnames}(B)]\!]_B$

> By definition, $\mathsf{mnames}(C) = (\mathsf{mnames}(B), \ldots)$ so $[\![\mathsf{mnames}(C)]\!]_C = [\![\mathsf{mnames}(B)]\!]_C \wedge T$ for some $T$ and we only need to prove that $|\Gamma| \vdash [\![m]\!]_C <: [\![m]\!]_B$ for all $m \in \mathsf{mnames}(B)$. If $m \in \mathsf{mnames}(B)$ then either $m \notin M$ or $\Gamma \vdash m\ \mathsf{ok}$, in both cases this implies $[\![m]\!]_C = [\![m]\!]_B$.

■

**Lemma 3.2.6**

If $|\Gamma|$ defined then $|\Gamma| \vdash \{\mathsf{this} \Rightarrow [\![C]\!]\} <: |C|$

*Proof.* Since $\vdash C\ \mathsf{ok}$, we can have a sequence of class $\overline{D}$ such that $D_1 = C$, $D_n = \mathsf{Object}$ and $D_i <: D_{i+1}$ derived by the rules S-Refl and S-Class for any $i$. We prove by induction on the length $n$ ($\geq 1$) of the sequence.

**Case** $(n = 1)$

> By Top.

**Case** $\textsf{class}\, C \ldots \lhd B \ldots \quad (n \geq 2)$

By Sel1, $|\Gamma| \vdash \mathsf{ct}.B \wedge \{\mathsf{this} \Rightarrow [\![C]\!]\} <: \mathsf{ct}.C$. Hence,

$$\dfrac{\dfrac{\dfrac{\dfrac{}{|\Gamma|, _ : [\![C]\!] \vdash [\![C]\!] <: [\![B]\!]}\ (3.2.5,\ \textsc{Weaken})}{|\Gamma| \vdash \{\mathsf{this} \Rightarrow [\![C]\!]\} <: \{\mathsf{this} \Rightarrow [\![B]\!]\}}\ (\textsc{BindX}) \qquad \dfrac{}{|\Gamma| \vdash \{\mathsf{this} \Rightarrow [\![B]\!]\} <: \mathsf{ct}.B}\ (\mathrm{IH})}{|\Gamma| \vdash \{\mathsf{this} \Rightarrow [\![C]\!]\} <: \mathsf{ct}.B}\ (\textsc{Trans})}{\dfrac{|\Gamma| \vdash \{\mathsf{this} \Rightarrow [\![C]\!]\} <: \mathsf{ct}.B \wedge \{\mathsf{this} \Rightarrow [\![C]\!]\}}{|\Gamma| \vdash \{\mathsf{this} \Rightarrow [\![C]\!]\} <: \mathsf{ct}.C}\ (\textsc{Trans})}\ (\textsc{And2},\ \textsc{Refl})$$

■

At this point in our proof, it would be convenient if we could establish that $|\Gamma| \vdash [\![C]\!] <: |C|$ to show that $|\Gamma| \vdash \mathsf{this} : |C|$ by subsumption. This would follow from Lemma 3.2.6 if we had a Bind2 rule symmetric to the existing Bind1 to prove $|\Gamma| \vdash [\![C]\!] <: \{\mathsf{this} \Rightarrow [\![C]\!]\}$, but this rule is missing from [Rompf and Amin 2016] as mentioned in Section 3 of the paper[2]:

> "[...] Note as well that there is no Bind2 rule, symmetric to Bind1, which is another kind of contractiveness restriction. We conjecture that these contractiveness restrictions could be lifted without breaking soundness, since we can always construct explicit conversion functions that use rules VarPack and VarUnpack on proper term bindings. However, removing these contractiveness restrictions would likely require different and harder to mechanize proof techniques such as a coinductive interpretation of subtyping."

For our purposes, VarPack is indeed enough:

**Lemma 3.2.7: this translation is type-preserving**

If $\Gamma \vdash \mathsf{this} : C$ and $|\Gamma|$ defined then $|\Gamma| \vdash \mathsf{this} : |C|$

*Proof.* By inversion of $\Gamma \vdash \mathsf{this} : C$ via T-Var, we must have $\Gamma(\mathsf{this}) = C$ and so $|\Gamma|(\mathsf{this}) = [\![C]\!]$ by definition. Hence,

$$\dfrac{\dfrac{\dfrac{}{|\Gamma| \vdash \mathsf{this} : [\![C]\!]}\ (\textsc{Var})}{|\Gamma| \vdash \mathsf{this} : \{\mathsf{this} \Rightarrow [\![C]\!]\}}\ (\textsc{VarPack}) \qquad \dfrac{}{|\Gamma| \vdash \{\mathsf{this} \Rightarrow [\![C]\!]\} <: |C|}\ (3.2.6)}{|\Gamma| \vdash \mathsf{this} : |C|}\ (\textsc{Sub})$$

■

**Theorem 3.2.8: Typing translation is type-preserving**

If $\Gamma \vdash e : C$ and $|\Gamma|$ defined then $|\Gamma| \vdash |e| : |C|$.

[2] Interestingly, this rule is derivable in gDOT ([Giarrusso et al. 2020, Figure 7]).

*Proof.* By induction on the derivation of $\Gamma \vdash e : C$.

**Case** $\dfrac{\Gamma(x) = C}{\Gamma \vdash x : C}$ (T-Var)

> By definition, $|\Gamma|(|x|) = |\Gamma|(x)$, and we can distinguish two sub-cases:
> - If $x = \mathsf{this}$, then $|\Gamma|(\mathsf{this}) = [\![C]\!]$ by definition and Lemma 3.2.7 finishes the case.
> - Otherwise, $|\Gamma|(x) = |C|$ and Var finishes the case.

**Case** $\dfrac{\Gamma \vdash e_0 : C \quad \mathsf{vparams}(C) = \overline{f : D}}{\Gamma \vdash e_0.f_i : D_i}$ (T-Getter)

> By the IH, $|\Gamma| \vdash |e_0| : |C|$. By Corollary 3.2.4 and Sub, $|\Gamma| \vdash |e_0| : (f_i() : |D_i|)$. TApp finishes the case.

**Case** $\dfrac{\Gamma \vdash e_0 : C \quad \mathsf{mtype}(m, C) = (\overline{x : D}) \to D_0 \quad \Gamma \vdash \overline{e : E} \quad \overline{E <: D}}{\Gamma \vdash e_0.m(\overline{e}) : D_0}$ (T-Invk)

> By the IH, $|\Gamma| \vdash |e_0| : |C|$ and $|\Gamma| \vdash \overline{|e| : |E|}$. By Corollary 3.2.4 and Sub, $|\Gamma| \vdash |e_0| : (m(\overline{x : |D|}) : |D_0|)$. TApp' finishes the case since by Theorem 3.2.2, $|\Gamma| \vdash \overline{|E| <: |D|}$ and so by Sub, $|\Gamma| \vdash \overline{|e| : |D|}$.

**Case** $\dfrac{\mathsf{class}\, C(\overline{f : D}) \quad \Gamma \vdash \overline{e : E} \quad \overline{E <: D}}{\Gamma \vdash \mathsf{new}\, C(\overline{e}) : C}$ (T-New)

> By the IH, $|\Gamma| \vdash \overline{|e| : |E|}$. By Lemma 2.4.5, Sub and Var, $|\Gamma| \vdash \mathsf{ct} : (\mathsf{new}_C(\overline{f_{\mathsf{param}} : |D|}) : |C|)$. We can finish using TApp' like in the previous case.

∎

**Lemma 3.2.9**

> Given $\mathsf{class}\, C(\overline{f : D}) \lhd B \ldots$, $\Gamma_C = \mathsf{this} : C$, and $\Gamma_B = \mathsf{this} : B$, then $|\Gamma_B| \vdash t : T$ implies $|\Gamma_C| \vdash t : T$.

*Proof.* Let $\mathsf{vparams}(B) = \overline{g : E}$. Then,

$$|\Gamma_B| = |\varnothing|, \overline{g_{\mathsf{param}} : |E|}, \mathsf{this} : [\![B]\!]$$

By inversion of $\vdash C$ ok via T-Class we must have $\overline{f : D} = \overline{g : E}, \overline{f' : D'}$ and so

$$|\Gamma_C| = |\varnothing|, \overline{g_{\mathsf{param}} : |E|}, \overline{f'_{\mathsf{param}} : |D'|}, \mathsf{this} : [\![C]\!]$$

Therefore,

$$\dfrac{|\Gamma| \vdash t : T \quad \dfrac{\dfrac{}{|\varnothing| \vdash [\![C]\!] <: [\![B]\!]}\ (3.2.5)}{|\varnothing|, \overline{g_{\text{param}} : |E|}, \text{this} : [\![C]\!] \vdash [\![C]\!] <: [\![B]\!]}\ (\textsc{Weaken})}{\dfrac{|\varnothing|, \overline{g_{\text{param}} : |E|}, \text{this} : [\![C]\!] \vdash t : T}{|\Gamma_C| \vdash t : T}\ (\textsc{WeakenTp})}\ (\textsc{NarrowTp})$$

■

**Lemma 3.2.10: Method translation is well-typed**

Given $\textbf{class}\, C(\ldots) \lhd B \ldots \{\overline{M}\}$, $\Gamma = \text{this} : C$, $\text{mtype}(m, C) = (\overline{x : D}) \to D_0$ and $\text{mbody}(m, C) = e_0$, then $|\Gamma| \vdash (\!| m |\!)_C : [\![m]\!]_C$.

*Proof.* By induction on the derivation of $\text{mtype}(m, C)$ and $\text{mbody}(m, C)$.

**Case** $\dfrac{\textbf{def}\ m(\overline{x : D}) : D_0 = e_0 \in \overline{M}}{\text{mtype}(m, C) \coloneqq (\overline{x : D}) \to D_0}$ (M-Class)
$\text{mbody}(m, C) \coloneqq e_0$

$\vdash CT$ ok implies $\vdash C$ ok which implies $\Gamma \vdash m$ ok which in turn can be inverted to reveal,

$$\Gamma, \overline{x : D} \vdash e_0 : E_0$$
$$E_0 <: D_0$$

Hence,

$$\dfrac{\dfrac{\dfrac{\Gamma, \overline{x : D} \vdash e_0 : E_0}{|\Gamma, \overline{x : D}| \vdash |e_0| : |E_0|}\ (3.2.8) \quad \dfrac{E_0 <: D_0}{|\Gamma, \overline{x : D}| \vdash |E_0| <: |D_0|}\ (3.2.2)}{|\Gamma, \overline{x : D}| \vdash |e_0| : |D_0|}\ (\textsc{Sub}, 3.2.2)}{|\Gamma| \vdash (\!| m |\!)_C : [\![m]\!]_C}\ (\text{DF}\textsc{un}')$$

**Case** $\dfrac{m \ldots \notin \overline{M}}{\text{mtype}(m, C) \coloneqq \text{mtype}(m, B)}$ (M-Super)
$\text{mbody}(m, C) \coloneqq \text{mbody}(m, B)$

By definition we have $(\!| m |\!)_C = (\!| m |\!)_B$ and $[\![m]\!]_C = [\![m]\!]_B$. By the IH, $|\text{this} : B| \vdash (\!| m |\!)_B : [\![m]\!]_B$ so by Lemma 3.2.9 we have $|\Gamma| \vdash (\!| m |\!)_B : [\![m]\!]_B$ which finishes the case since $(\!| m |\!)_C = (\!| m |\!)_B$ and $[\![m]\!]_C = [\![m]\!]_B$ by definition.

■

**Lemma 3.2.11: Class translation is well-typed**

Given $\textbf{class}\, C(\overline{f : D}) \ldots$ then $|\varnothing|, \overline{f_{\text{param}} : |D|} \vdash \{\text{this} \Rightarrow (\!| C |\!)\} : \{\text{this} \Rightarrow [\![ C ]\!]\}$

*Proof.* Let $\Gamma = \text{this} : C$ and note that $|\Gamma| = |\varnothing|, \overline{f_{\text{param}} : |D|}, \text{this} : [\![C]\!]$ by definition. By TNew, we only need to prove the following claims.

**Claim 1:** $|\Gamma| \vdash (\!| f : D |\!) : [\![ f : D ]\!] \quad \forall (f : D) \in \text{vparams}(C)$

By Var.

**Claim 2:** $|\Gamma| \vdash (\!| m |\!)_C : [\![ m ]\!]_C \quad \forall m \in \text{mnames}(C)$

By Lemma 3.2.10.

∎

**Lemma 3.2.12: Class table translation is well-typed**

$\varnothing \vdash \{\text{ct} \Rightarrow (\!| CT |\!)\} : \{\text{ct} \Rightarrow [\![ CT ]\!]\}$.

*Proof.* After proving the following claims for each $\textbf{class}\, C(\overline{f : D})$ in $CT$, we can finish the proof by TNew.

**Claim 1:** $|\varnothing| \vdash (C = \{\text{this} \Rightarrow [\![C]\!]\}) : (C = \{\text{this} \Rightarrow [\![C]\!]\})$

By DTyp.

**Claim 2:** $|\varnothing| \vdash (\text{new}_C(\overline{f_{\text{param}} : |D|}) : |C| = \{\text{this} \Rightarrow (\!| C |\!)\}) : (\text{new}_C(\overline{f_{\text{param}} : |D|}) : |C|)$

Let $\Gamma_0 = |\varnothing|, \overline{f_{\text{param}} : |D|}$. Then,

$$\dfrac{\dfrac{}{\Gamma_0 \vdash \{\text{this} \Rightarrow (\!| C |\!)\} : \{\text{this} \Rightarrow [\![C]\!]\}}\,(3.2.11) \qquad \dfrac{\dfrac{}{|\varnothing| \vdash \{\text{this} \Rightarrow [\![C]\!]\} <: |C|}\,(3.2.6)}{\Gamma_0 \vdash \{\text{this} \Rightarrow [\![C]\!]\} <: |C|}\,(\text{Weaken})}{\Gamma_0 \vdash \{\text{this} \Rightarrow (\!| C |\!)\} : |C|}\,(\text{Sub})$$

and DFun' finishes the claim.

∎

**Theorem 3.2.13: Program translation is type-preserving**

If $\varnothing \vdash_{\text{FJ}} e : C$ then $\varnothing \vdash_{\text{DOT}} \textbf{let}\ \text{ct} = \{\text{ct} \Rightarrow (\!| CT |\!)\}\ \textbf{in}\ |e| : |C|$.

*Proof.*

$$\dfrac{\dfrac{}{\varnothing \vdash \{\mathsf{ct} \Rightarrow (\!| CT |\!)\} : \{\mathsf{ct} \Rightarrow [\![ CT ]\!]\}}\ (3.2.12) \qquad \dfrac{\dfrac{\varnothing \vdash e : C}{\mathsf{ct} : [\![ CT ]\!] \vdash |e| : |C|}\ (3.2.8)}{\mathsf{ct} : \{\mathsf{ct} \Rightarrow [\![ CT ]\!]\} \vdash |e| : |C|}\ (\textsc{EnvPackTp})}{\varnothing \vdash \textbf{let}\ \mathsf{ct} = \{\mathsf{ct} \Rightarrow (\!| CT |\!)\}\ \textbf{in}\ |e| : |C|}\ (\textsc{Let})$$

■

# 4 Featherweight Generic Java (Scala-flavored)

In this chapter, we review the Featherweight Generic Java (FGJ) calculus [Igarashi, Pierce, and Wadler 2001] which extends FJ by adding support for type parameters as they exist in Java. As in the previous chapter, we develop a type-preserving translation scheme to DOT which requires extending DOT with an extra subtyping rule AND-BIND.

## 4.1 Syntax and semantics

**Figure 4.1: FGJ: Syntax**

| | |
|---|---|
| $x, y, z$ | Variable |
| $B, C, D, E$ | Class name |
| $f, g$ | Class parameter |
| $m$ | Method name |
| $X_C$ | Class variable |
| $X_m$ | Method variable |
| $X, Y, Z ::= X_C \mid X_m$ | Type variable |
| $N, P, Q ::= C[\overline{T}]$ | Non-variable |
| $S, T, U, V ::= X \mid N$ | Type |
| $\Gamma ::=$ | Context |
| $\varnothing \mid \Gamma, x : T \mid \Gamma, \overline{X <: N}$ | |
| $L ::=$ | Class declaration |
| $\textbf{class}\ C\,[\overline{X_C <: N}]\,(\overline{f : T}) \triangleleft P(\overline{f})\ \{\overline{M}\}$ | |
| $M ::=$ | Method declaration |
| $\textbf{def}\ m\,[\overline{X_m <: N}]\,(\overline{x : T}) : T_0 = e_0$ | |
| $e ::=$ | Expression |
| $x$ | variable |
| $e.f$ | parameter access |
| $e_0.m\,[\overline{T}]\,(\overline{e})$ | method call |
| $\textbf{new}\ C\,[\overline{T}]\,(\overline{e})$ | object |
| $\sigma, \tau ::= [\overline{T/X}]$ | Type substitution |

Compared to FJ, an FGJ class or method declaration takes an additional type parameter clause $[\overline{X <: N}]$, where $\overline{X}$ is a list of type variable names that are accessible in the scope of the definition. The only thing known about each type variable $X_i$ is its upper-bound $N_i$, note that forward references to type parameters such as $[X <: C[Y], Y <: \text{Object}]$ are allowed.

Constructor and method call syntax is similarly extended to pass a type argument clause $[\overline{T}]$ where each $T_i$ must be a subtype of the subtituted upper-bound $[\overline{T/X}]N_i$. Constructors now return applied class types $C[\overline{T}]$.

FGJ also relaxes the definition of overriding to allow *covariant overriding* where the result type of the overriding method can be a subtype of the result type of the overridden method.

The version of FGJ we present in Figures 4.1 to 4.3, 4.5 and 4.6 differs from [Igarashi, Pierce, and Wadler 2001] in a few ways:

- As in Chapter 3, we drop casts and use Scala-like syntax.
- We introduce an additional lookup function $\mathsf{tparams}(C)$ that returns the type parameters of $C$ to reduce the amount of changes we will need to make when we extend the calculus in Chapter 5.
- We distinguish between class type variables $X_C$ and method type variables $X_m$ in the syntax so we can translate them differently in Figure 4.7.
- We use a single context $\Gamma$ to store both term and type variables whereas the original presentation used a separate context $\Delta$ for type variables instead. This simplifies our translation since DOT only has one context.
- Our definition of method overriding in Figure 4.4 is more expressive than the original one[1] as it takes into account the environment $\Gamma$ containing the class type variables. This is needed to typecheck the following class table:

  ```
  class A
  class Base { def foo(): A = … }
  class Sub[S <: A] ◁ Base { def foo(): S = … }
  ```

  The equivalent Java code is valid and yet Sub is not well-formed in [Igarashi, Pierce, and Wadler 2001, Figure 6] because the type parameter S <: A is not part of the environment when the override check is done.

[1] However, unlike the original definition, we require that the names of the parameters of the overriding method match the names used in the overridden method to simplify the translation.

**Figure 4.2: FGJ: Subtyping**

$\Gamma \vdash S <: T$

$$\Gamma \vdash S <: S \qquad \text{(GS-Refl)}$$

$$\frac{\Gamma(X) = N}{\Gamma \vdash X <: N} \qquad \text{(GS-Var)}$$

$$\frac{\textbf{class}\, C[\overline{X <: N}](\ldots) \lhd P \ldots}{\Gamma \vdash C[\overline{T}] <: [\overline{T/X}]P} \qquad \text{(GS-Class)}$$

$$\frac{\Gamma \vdash S <: U \quad \Gamma \vdash U <: T}{\Gamma \vdash S <: T} \qquad \text{(GS-Trans)}$$

**Figure 4.3: FGJ: Well-formedness**

**Well-formed type** $\Gamma \vdash T \text{ wf}$

$$\Gamma \vdash \mathsf{Object}\ \mathsf{wf} \qquad \text{(WF-Object)}$$

$$\frac{X \in \mathsf{dom}(\Gamma)}{\Gamma \vdash X\ \mathsf{wf}} \qquad \text{(WF-Var)}$$

$$\frac{\mathsf{tparams}(C) = \overline{X <: N} \quad \sigma = [\overline{T/X}] \quad \Gamma \vdash \overline{T}\ \mathsf{wf} \quad \Gamma \vdash \overline{T <: \sigma N}}{\Gamma \vdash C[\overline{T}]\ \mathsf{wf}} \qquad \text{(WF-Class)}$$

**Well-formed environment** $\Gamma \text{ wf}$

$$\varnothing\ \mathsf{wf}$$

$$\frac{\Gamma, \overline{X <: N} \vdash \overline{N}\ \mathsf{wf}}{\Gamma, \overline{X <: N}\ \mathsf{wf}}$$

$$\frac{\Gamma \vdash T\ \mathsf{wf}}{\Gamma, x : T\ \mathsf{wf}}$$

**Figure 4.4: FGJ: Overriding**

***m* in *N* overrides *m* in *P***  $\boxed{\mathsf{override}_{\Gamma}(m, N, P)}$

$$\frac{\begin{array}{c}\mathsf{mtype}(m, N) = [\overline{Y <: P}] \to (\overline{x : U}) \to U \\ \mathsf{mtype}(m, P) = [\overline{Y <: P}] \to (\overline{x : U}) \to V \\ \Gamma, \overline{Y <: P} \vdash U <: V\end{array}}{\mathsf{override}_{\Gamma}(m, N, P)} \quad \text{(OV-Present)}$$

$$\frac{\begin{array}{c}\mathsf{mtype}(m, N) \text{ defined} \\ \mathsf{mtype}(m, P) \text{ undefined}\end{array}}{\mathsf{override}_{\Gamma}(m, N, P)} \quad \text{(OV-Absent)}$$

**Figure 4.5: FGJ: Lookup functions**

**Non-variable upper bound of type** $\quad \boxed{\mathsf{bound}_\Gamma(T) \coloneqq N}$

$$\mathsf{bound}_\Gamma(X) \coloneqq \Gamma(X) \qquad \text{(B-Var)}$$

$$\mathsf{bound}_\Gamma(N) \coloneqq N \qquad \text{(B-Class)}$$

**Type parameters lookup** $\quad \boxed{\mathsf{tparams}(C) \coloneqq \overline{X <: N}}$

$$\frac{\textbf{class}\ C[\overline{X <: N}]\ \dots}{\mathsf{tparams}(C) \coloneqq \overline{X <: N}}$$

**Value parameters lookup** $\quad \boxed{\mathsf{vparams}(N) \coloneqq \overline{f : T}}$

$$\mathsf{vparams}(\mathsf{Object}) \coloneqq \varnothing \qquad \text{(G-Object)}$$

$$\frac{\textbf{class}\ C[\overline{X <: N}](\overline{f : U})\ \dots \qquad \sigma = [\overline{S/X}]}{\mathsf{vparams}(C[\overline{T}]) \coloneqq \overline{f : \sigma U}} \qquad \text{(G-Class)}$$

**Method names lookup** $\quad \boxed{\mathsf{mnames}(C) \coloneqq \overline{m}}$

$$\mathsf{mnames}(\mathsf{Object}) \coloneqq \varnothing$$

$$\frac{\textbf{class}\ C\ \dots \lhd B\ \{\overline{\textbf{def}\ m_C\ \dots}\} \quad \mathsf{mnames}(B) = \overline{m_B} \quad \overline{n} = \left[m \in \overline{m_C} \,\middle|\, m \notin \overline{n}\right]}{\mathsf{mnames}(C) \coloneqq \overline{m_B}, \overline{n}}$$

**Method type and body lookup** $\quad \boxed{\begin{array}{l}\mathsf{mtype}(m, N) \coloneqq [\overline{Y <: P}] \to (\overline{x : T}) \to T_0 \\ \mathsf{mbody}(m, N) \coloneqq e_0\end{array}}$

$$\frac{\textbf{class}\ C[\overline{X <: N}]\ \dots\ \{\overline{M}\} \quad \sigma = [\overline{T/X}] \quad (\textbf{def}\ m[\overline{Y <: P}](\overline{x : T}) : T_0 = e_0) \in \overline{M}}{\begin{array}{l}\mathsf{mtype}(m, C[\overline{T}]) \coloneqq [\overline{Y <: \sigma P}] \to (\overline{x : \sigma T}) \to \sigma T_0) \\ \mathsf{mbody}(m, C[\overline{T}]) \coloneqq \sigma e_0\end{array}} \qquad \text{(GM-Class)}$$

$$\frac{\textbf{class}\ C[\overline{X <: N}](\dots) \lhd P\ \{\overline{M}\} \quad \sigma = [\overline{T/X}] \quad (\textbf{def}\ m\ \dots) \notin \overline{M}}{\begin{array}{l}\mathsf{mtype}(m, C[\overline{T}]) \coloneqq \mathsf{mtype}(m, \sigma P) \\ \mathsf{mbody}(m, C[\overline{T}]) \coloneqq \mathsf{mbody}(m, \sigma P)\end{array}} \qquad \text{(GM-Super)}$$

**Figure 4.6: FGJ: Typing rules**

**Expression typing** $\boxed{\Gamma \vdash e : T}$

$$\frac{\Gamma(x) = T}{\Gamma \vdash x : T} \quad \text{(GT-Var)}$$

$$\frac{\Gamma \vdash e_0 : T_0 \quad \mathsf{vparams}(\mathsf{bound}_\Gamma(T_0)) = \overline{f : T}}{\Gamma \vdash e_0.f_i : T_i} \quad \text{(GT-Getter)}$$

$$\frac{\begin{array}{c}\Gamma \vdash e_0 : T_0 \quad \mathsf{mtype}(m,\ \mathsf{bound}_\Gamma(T_0)) = [\overline{Y <: P}] \to (\overline{x : U}) \to U_0 \\ \sigma = [\overline{V}/\overline{Y}] \quad \Gamma \vdash \overline{V}\ \mathsf{wf}, \overline{V <: \sigma P}, \overline{e : S}, \overline{S <: \sigma U}\end{array}}{\Gamma \vdash e_0.m[\overline{V}](\overline{e}) : T_0} \quad \text{(GT-Invk)}$$

$$\frac{\Gamma \vdash N\ \mathsf{wf} \quad \mathsf{vparams}(N) = \overline{f : U} \quad \Gamma \vdash \overline{e : S}, \overline{S <: U}}{\Gamma \vdash \mathbf{new}\ N(\overline{e}) : N} \quad \text{(GT-New)}$$

**Method typing** $\boxed{\Gamma \vdash m\ \mathsf{ok}}$

$$\frac{\begin{array}{c}\Gamma = \overline{X <: N}, \mathsf{this} : C[\overline{X}] \\ \mathbf{class}\ C \ldots \lhd Q \\ \mathsf{mtype}(m, C[\overline{X}]) = [\overline{Y <: P}] \to (\overline{x : U}) \to U_0 \quad \mathsf{mbody}(m, C[\overline{X}]) = e_0 \\ \Gamma, \overline{Y <: P} \vdash \overline{U}, U_0, \overline{P}\ \mathsf{wf} \\ \Gamma, \overline{Y <: P}, \overline{x : U} \vdash e_0 : E_0, E_0 <: U_0 \\ \mathsf{override}_\Gamma(m, C[\overline{X}], Q)\end{array}}{\Gamma \vdash m\ \mathsf{ok}} \quad \text{(GT-Method)}$$

**Class typing** $\boxed{\vdash C\ \mathsf{ok}}$

$$\frac{\begin{array}{c}\mathbf{class}\ C[\overline{X <: N}](\overline{g : U}, \overline{f : T}) \lhd P(\overline{g})\ \{\overline{\mathbf{def}\ m \ldots}\} \\ \Gamma = \overline{X <: N}, \mathsf{this} : C[\overline{X}] \\ \Gamma \vdash \overline{N}, \overline{U}, \overline{T}, P\ \mathsf{wf} \quad \mathsf{vparams}(P) = \overline{g : U} \quad \Gamma \vdash \overline{m\ \mathsf{ok}}\end{array}}{\vdash C\ \mathsf{ok}} \quad \text{(GT-Class)}$$

**Class table typing** $\boxed{\vdash CT\ \mathsf{ok}}$

$$\frac{\begin{array}{c}C \in \mathsf{dom}(CT) \text{ implies } \vdash C\ \mathsf{ok} \\ \text{No inheritance cycle between the classes in } CT\end{array}}{\vdash CT\ \mathsf{ok}} \quad \text{(GT-CT)}$$

## 4.2 Meta-theory

**Lemma 4.2.1: Correctness of** bound

If $\text{bound}_\Gamma(S) = N$, then $\Gamma \vdash S <: N$.

*Proof.* By induction on the derivation of $\text{bound}_\Gamma(S)$. ■

The two following lemmas are partially adapted from [Igarashi, Pierce, and Wadler 2001, Lemmas A.2.5 and A.2.6].

**Lemma 4.2.2: Substitution preserves subtyping**

Let $\Gamma_1 = \overline{X <: N}$. If $\Gamma_1 \vdash S <: U$ and $\Gamma_2 \vdash \overline{T <: \sigma N}$ where $\sigma = [\overline{T/X}]$ then $\Gamma_2 \vdash \sigma S <: \sigma U$.

*Proof.* By induction on the derivation of $\Gamma_1 \vdash S <: U$.

**Case** $\Gamma_1 \vdash S <: S$ (GS-Refl)

By GS-Refl, $\Gamma_2 \vdash \sigma S <: \sigma S$.

**Case** $\dfrac{\Gamma_1(Z) = P}{\Gamma_1 \vdash Z <: P}$ (GS-Var)

Since $Z \in \overline{X}$ and $\overline{\sigma X = T}$ this follows from the premise $\Gamma_2 \vdash \overline{T <: \sigma N}$.

**Case** $\dfrac{\textsf{class}\ C[\overline{Z <: Q}](\ldots) \lhd P \ldots}{\Gamma_1 \vdash C[\overline{V}] <: [\overline{V/Z}]P}$ (GS-Class)

By inversion of GT-Class, $\overline{Z <: Q} \vdash P$ wf, so $P$ does not include any $\overline{X}$ as a free variable and therefore $\sigma[\overline{V/Z}]P = [\overline{\sigma V/Z}]P$. By GT-Class, $\Gamma_2 \vdash C[\overline{\sigma V}] <: [\overline{\sigma V/Z}]P$ which completes the case.

**Case** $\dfrac{\Gamma_1 \vdash S <: V \quad \Gamma_1 \vdash V <: U}{\Gamma_1 \vdash S <: U}$ (GS-Trans)

By the IH, $\Gamma_2 \vdash \sigma S <: \sigma V$, $\sigma V <: \sigma U$ and GS-Trans completes the case. ■

**Lemma 4.2.3: Substitution preserves well-formedness**

Let $\Gamma_1 = \overline{X <: N}$. If $\Gamma_1 \vdash S$ wf, $\Gamma_2 \vdash \overline{T}$ wf and $\Gamma_2 \vdash \overline{T <: \sigma N}$ where $\sigma = [\overline{T/X}]$ then $\Gamma_2 \vdash \sigma S$ wf.

*Proof.* By induction on the derivation of $\Gamma_1 \vdash S$ wf. Case WF-Object is trivial.

**Case** $\dfrac{oZ \in \mathsf{dom}(\Gamma_1)}{\Gamma_1 \vdash Z \;\mathsf{wf}}$ (WF-Var)

> Since $Z \in \overline{X}$ and $\overline{\sigma X = T}$ this follows from the premise $\Gamma_2 \vdash \overline{T}$ wf.

**Case** $\dfrac{\mathsf{class}\, C[\overline{Z <: Q}] \lhd P \ldots \quad \sigma' = [\overline{V/Z}] \quad \Gamma_1 \vdash \overline{V} \;\mathsf{wf} \quad \Gamma_1 \vdash \overline{V <: \sigma' Q}}{\Gamma_2 \vdash C[\overline{V}] \;\mathsf{wf}}$ (WF-Class)

> By Lemma 4.2.2, $\Gamma_2 \vdash \overline{\sigma V <: \sigma(\sigma' Q)}$. By inversion of GT-Class, $\overline{Z <: Q} \vdash \overline{Q}$ wf, so none of the $\overline{Q}$ include any $\overline{X}$ as a free variable and therefore $\sigma(\sigma' Q) = (\sigma\sigma')Q$. Since $\Gamma_2 \vdash \overline{\sigma V}$ wf by the IH and $\sigma\sigma' = [\overline{\sigma V/Z}]$, we can conclude that $\Gamma_2 \vdash C[\overline{\sigma V}]$ wf by WF-Class.

■

**Lemma 4.2.4**

If Γ wf, $\Gamma \vdash S$ wf and $\Gamma \vdash S <: T$, then $\Gamma \vdash T$ wf.

*Proof.* By induction on the derivation of $\Gamma \vdash S <: T$, case GS-Refl is trivial.

**Case** $\dfrac{\Gamma(X) = N}{\Gamma \vdash X <: N}$ (GS-Var)

> Γ wf implies $\Gamma \vdash N$ wf.

**Case** $\dfrac{\mathsf{class}\, C[\overline{X <: N}](\ldots) \lhd P \ldots \quad \sigma = [\overline{T/X}]}{\Gamma \vdash C[\overline{T}] <: \sigma P}$ (GS-Class)

> By inversion of $\Gamma \vdash C[\overline{T}]$ wf via WF-Class, $\Gamma \vdash \overline{T}$ wf and $\Gamma \vdash \overline{T <: \sigma N}$. By inversion of $\vdash C$ ok via GT-Class, $\overline{X <: N} \vdash P$ wf. So by Lemma 4.2.3, $\Gamma \vdash \sigma P$ wf.

**Case** $\dfrac{\Gamma \vdash S <: U \quad \Gamma \vdash U <: T}{\Gamma \vdash S <: T}$ (GS-Trans)

> By the IH, $\Gamma \vdash U$ wf so by the IH again $\Gamma \vdash T$ wf.

■

## 4.3 Translation

As in Section 3.2, our translation scheme is defined using $|\cdot|$, $(\!|\cdot|\!)$ and $[\![\cdot]\!]$.

Expression translation is now parameterized by the context Γ, this is necessary to translate type arguments in method applications, although in practice this wouldn't be needed if we used de Bruijn indices to represent method type variables[2] like the Scala 3 compiler.

[2] But not to represent class type variables which are assumed to be globally unique by our translation.

A well-typed FJ program,

$$(CT, e)$$

can be translated into a DOT expression well-typed in the empty context,

$$\textbf{let}\ \mathsf{ct} = \{\mathsf{ct} \Rightarrow (\!| CT |\!)\}\ \textbf{in}\ |e|_\varnothing$$

but before we can establish this in Theorem 3.2.13 we'll need to augment DOT with an extra subtyping rule.

**Figure 4.7: Translating FGJ types and expressions to DOT**

**Type Translation** $\boxed{|T| \mathrel{::=} T_{\text{DOT}}}$

$$|\mathsf{Object}| \mathrel{::=} \mathsf{ct.Object} \qquad \text{(TR-OBJ)}$$

$$|X_C| \mathrel{::=} \mathsf{this}.X_C \qquad \text{(TR-CVAR)}$$

$$|X_m| \mathrel{::=} \mathsf{mtag}.X_m \qquad \text{(TR-MVAR)}$$

$$\frac{\mathsf{tparams}(C) = \overline{X <: \ldots}}{|C[\overline{T}]| \mathrel{::=} \mathsf{ct}.C \wedge \{_ \Rightarrow \overline{X = |T|}\}} \qquad \text{(TR-CLASS)}$$

**Type Parameter Clause Translation** $\boxed{|X <: N| \mathrel{::=} T_{\text{DOT}}}$

$$|\overline{X_C <: N}| \mathrel{::=} \{\mathsf{this} \Rightarrow \overline{X_C : \bot .. |N|}\}$$

$$|\overline{X_m <: N}| \mathrel{::=} \{\mathsf{mtag} \Rightarrow \overline{X_m : \bot .. |N|}\}$$

**Expression Translation** $\boxed{|e|_\Gamma \mathrel{::=} t_{\text{DOT}}}$

$$|x|_\Gamma \mathrel{::=} x$$

$$|e_0.f|_\Gamma \mathrel{::=} |e_0|_\Gamma.f()$$

$$\frac{x_{\mathsf{mtag}} \text{ is fresh} \quad \Gamma \vdash e_0 : T_0 \qquad \mathsf{mtype}(m,\ \mathsf{bound}_\Gamma(T_0)) = [\overline{Y <: P}] \to \ldots}{|e_0.m[\overline{V}](\overline{e})|_\Gamma \mathrel{::=} \textbf{let}\ x_{\mathsf{mtag}} = \{_ \Rightarrow \overline{Y = |V|}\}\ \textbf{in}\ |e_0|_\Gamma.m(x_{\mathsf{mtag}}, \overline{|e|_\Gamma})}$$

$$|\textbf{new}\ \mathsf{Object}|_\Gamma \mathrel{::=} \{_ \Rightarrow\}$$

$$\frac{x_{\mathsf{ctag}} \text{ is fresh} \quad \mathsf{tparams}(C) = \overline{X <: \ldots}}{|\textbf{new}\ C[\overline{V}](\overline{e})|_\Gamma \mathrel{::=} \textbf{let}\ x_{\mathsf{ctag}} = \{_ \Rightarrow \overline{X = |V|}\}\ \textbf{in}\ \mathsf{ct.new}_C(x_{\mathsf{ctag}}, \overline{|e|_\Gamma})}$$

**Figure 4.8: Translating FGJ definitions to DOT**

**Getter Translation** $\boxed{(\!| f : T |\!) \coloneqq d_{\text{DOT}}}$

$$(\!| f : T |\!) \coloneqq f() : |T| = f_{\text{param}}$$

**Method Translation** $\boxed{(\!| m |\!)_C \coloneqq d_{\text{DOT}}}$

$$\frac{\begin{array}{l}\textbf{class}\, C[\overline{X <: N}] \ldots \quad \Gamma = \overline{X <: N}, \text{this} : C[\overline{X}] \\ \text{mtype}(m, C[\overline{X}]) = [\overline{Y <: P}] \rightarrow (\overline{x : U}) \rightarrow U_0 \\ \text{mbody}(m, C[\overline{X}]) = e_0\end{array}}{(\!| m |\!)_C \coloneqq m(\text{mtag} : |\overline{Y <: P}|, \overline{x : |U|}) : |U_0| = |e_0|_{\Gamma, \overline{Y <: P}, \overline{x : U}}}$$

**Class Translation** $\boxed{(\!| C |\!) \coloneqq \overline{d_{\text{DOT}}}}$

$$\frac{\textbf{class}\, C[\overline{X <: N}] \ldots \quad \text{baseArgs}(C) = \bigwedge \overline{Z = S}}{\begin{array}{l}(\!| C |\!) \coloneqq (\!| \text{vparams}(C[\overline{X}]) |\!), (\!| \text{mnames}(C) |\!)_C, \overline{Z = |S|} \\ (\!| C |\!)^{\overline{T}} \coloneqq (\!| C |\!), \overline{X = T}\end{array}}$$

**Class Table Translation** $\boxed{(\!| CT |\!) \coloneqq \overline{d_{\text{DOT}}}}$

$$(\!| \varnothing |\!) \coloneqq (\text{Object} = \top)$$

$$\frac{L_C = \textbf{class}\, C[\overline{X_C <: N}](\overline{f : U}) \lhd B \ldots \quad \tau = [\overline{\text{ctag}.X_C / |X_C|}]}{\begin{array}{c}(\!| \overline{L}, L_C |\!) \coloneqq (\!| \overline{L} |\!), C = \text{ct}.B \wedge \{\text{this} \Rightarrow [\![C]\!], \overline{X_C : \bot .. |N|}\}, \\ \text{new}_C(\text{ctag} : |\overline{X_C <: N}|, \overline{f_{\text{param}} : \tau|U|}) : \tau|C[\overline{X_C}]| = \{\text{this} \Rightarrow (\!| C |\!)^{\tau|\overline{X_C}|}\}\end{array}}$$

**Environment Translation** $\boxed{|\Gamma| \coloneqq \Gamma_{\text{DOT}}}$

$$|\varnothing| \coloneqq \text{ct} : [\![CT]\!] \qquad \text{(E-Empty)}$$

$$|\Gamma, \overline{X_m <: N}| \coloneqq |\Gamma|, \text{mtag} : |\overline{X_m <: N}| \qquad \text{(E-MVar)}$$

$$\frac{\text{tparams}(C) = \overline{X_C <: N}}{|\overline{X_C <: N}, \text{this} : C[\overline{X_C}]| \coloneqq |\varnothing|, \text{ctag} : |\overline{X_C <: N}|, \text{this} : [\![C]\!]^{\overline{\text{ctag}.X}}} \qquad \text{(E-This)}$$

$$\frac{x \neq \text{this}}{|\Gamma, x : T| \coloneqq |\Gamma|, x : |T|} \qquad \text{(E-Var)}$$

**Arguments of Base Types** $\boxed{\text{baseArgs}(C) \coloneqq T_{\text{DOT}}}$

$$\text{baseArgs}(\text{Object}) \coloneqq \top$$

$$\frac{\textbf{class}\, C \ldots \lhd B[\overline{S}] \ldots \quad \text{tparams}(B) = \overline{X <: \ldots}}{\text{baseArgs}(C) \coloneqq \left(\bigwedge \overline{X = |S|}\right) \wedge \text{baseArgs}(B)}$$

### 4.3.1 Required addition to DOT

Consider the following class table:

```
class C[X] extends Object
class D[Y] extends C[Y]
```

Then the environment translation as defined by Figure 4.8 will be

$$\begin{aligned} |\varnothing| = \mathsf{ct} : (&\mathsf{Object} = \top) \land \\ &(C = \mathsf{ct.Object} \land \{\mathsf{this} \Rightarrow X : \bot \,..\, \top\}) \land \ldots \\ &(D = \mathsf{ct}.C \land \{\mathsf{this} \Rightarrow X = \mathsf{this}.Y,\ Y : \bot \,..\, \top\}) \land \ldots \end{aligned}$$

It is easy to see that $\varnothing \vdash D[\mathsf{Object}] <: C[\mathsf{Object}]$, therefore if subtyping preservation holds, we should be able to establish that $|\varnothing| \vdash |D[\mathsf{Object}]| <: |C[\mathsf{Object}]|$. While it is easy to show that $|\varnothing| \vdash \mathsf{ct}.D <: \mathsf{ct}.C \land \{\mathsf{this} \Rightarrow X = \mathsf{this}.Y\}$ via SEL1, we get stuck pretty quickly after that:

$$\dfrac{\dfrac{\dfrac{}{|\varnothing| \vdash \{\mathsf{this} \Rightarrow X = \mathsf{this}.Y\} \land \{\mathsf{this} \Rightarrow Y = \top\} <: \{\mathsf{this} \Rightarrow X = \top\}}\ (???)}{|\varnothing| \vdash \mathsf{ct}.C \land \{\mathsf{this} \Rightarrow X = \mathsf{this}.Y\} \land \{\mathsf{this} \Rightarrow Y = \top\} <: \mathsf{ct}.C \land \{\mathsf{this} \Rightarrow X = \top\}}\ (2.4.5)}{|\varnothing| \vdash \mathsf{ct}.D \land \{\mathsf{this} \Rightarrow Y = \top\} <: \mathsf{ct}.C \land \{\mathsf{this} \Rightarrow X = \top\}}\ (\textsc{Trans}, \textsc{Sel1})$$

Intuitively, this subtyping relation should be true: if $X$ is equal to $Y$ and $Y$ is equal to $\top$, then $X$ is equal to $\top$, but there is no existing subtyping rule which would let us establish that (BINDX is close but it only works at the top-level). To remedy this predicament, we propose adding the following axiom to DOT:

$$\Gamma \vdash \{z \Rightarrow S\} \land \{z \Rightarrow T\} <: \{z \Rightarrow S \land T\} \qquad (\textsc{And-Bind})$$

Combined with BINDX, this solves our problem:

$$\dfrac{\dfrac{\dfrac{\dfrac{\dfrac{}{|\varnothing|, \mathsf{this} : \ldots \land (Y = \top) \vdash \mathsf{this} :_! (Y = \top)}\ (\textsc{Sub}, \textsc{Var})}{|\varnothing|, \mathsf{this} : \ldots \land (Y = \top) \vdash \top <: \mathsf{this}.Y}\ (\textsc{Sel2})}{|\varnothing|, \mathsf{this} : \ldots \land (Y = \top) \vdash (X = \mathsf{this}.Y) \land \ldots <: (X = \top)}\ (\textsc{Trans}, \textsc{Typ})}{|\varnothing| \vdash \{\mathsf{this} \Rightarrow X = \mathsf{this}.Y,\ Y = \top\} <: \{\mathsf{this} \Rightarrow X = \top\}}\ (\textsc{BindX})}{|\varnothing| \vdash \{\mathsf{this} \Rightarrow X = \mathsf{this}.Y\} \land \{\mathsf{this} \Rightarrow Y = \top\} <: \{\mathsf{this} \Rightarrow X = \top\}}\ (\textsc{Trans}, \textsc{And-Bind})$$

**Theorem 4.3.1**

oopslaDOT extended with AND-BIND is sound.

*Proof.* The Coq mechanization of oopslaDOT is available at https://oopsla16.namin.net. The calculus is defined in `dot.v` and two soundness proofs using different techniques but proving the same theorem are provided in `dot_soundness.v` and `dot_soundness_alt.v` respectively. In [Rompf and Amin 2016], the main proof is described in Section 6.1 to 6.5 and the alternative proof is described in Section 6.6. In practice, we found the alternative proof easier to work with and we extended it with And-Bind in https://github.com/smarter/minidot/commit/527762074f74df09b0a6241bafb1202ba92a5ebf. ■

### Alternative translation scheme

Recall that when comparing oopslaDOT against wfDOT in Chapter 2, we chose oopslaDOT primarily because of its inclusion of subtyping rules involving recursive types. Indeed, in the example above we relied on BindX to establish subtyping preservation for $\varnothing \vdash D[\mathsf{Object}] <: C[\mathsf{Object}]$. But one might wonder if this is just an artifact of the translation scheme we chose in Figure 4.7. Could we design an alternative type translation function that removes the need for such rules? The answer is yes, but as usual there are trade-offs involved. If we replace TR-Class by

$$\frac{\textbf{class}\; C[\overline{X <: \ldots}](\ldots) \lhd B[\overline{U}] \ldots \;\; \sigma = [\overline{T/X}]}{|C[\overline{T}]| \coloneqq \mathsf{ct}.C \wedge \{_ \Rightarrow \overline{X = |T|}\} \wedge |B[\overline{\sigma U}]|} \qquad \text{(TR-ClassAlt)}$$

Then subtyping preservation becomes almost trivial. In our previous example, we would have $|D[\mathsf{Object}]| = \ldots \wedge |C[\mathsf{Object}]|$ and thus $|\varnothing| \vdash |D[\mathsf{Object}]| <: |C[\mathsf{Object}]|$ would simply follow by width subtyping. The catch is that TR-ClassAlt is not applicable to all valid FGJ class hierarchies. For example given,

```
class B[X] ⊲ Object
class C ⊲ B[C]
```

Then the expansion of $|C|$ using TR-ClassAlt is non-terminating:

$$\begin{aligned} \boxed{|C|} &= \mathsf{ct}.C \wedge |B[C]| \\ |B[C]| &= \mathsf{ct}.B \wedge \left(X_B = \boxed{|C|}\right) \wedge |\mathsf{Object}| \end{aligned}$$

Indirect cycles are also problematic, which rule out a simple syntactic check:

```
class B[X] ⊲ Object
class D ⊲ B[E]
class E ⊲ D
```

$$\begin{aligned} \boxed{|E|} &= |D| \wedge \mathsf{ct}.E \\ |D| &= |B[E]| \wedge \mathsf{ct}.D \\ |B[E]| &= |\mathsf{Object}| \wedge \mathsf{ct}.B \wedge \left(X_B = \boxed{|E|}\right) \end{aligned}$$

To safely use TR-ClassAlt we would need to disallow all class hierarchies where a type parameter of a base type of a class refers back to the class itself. This can be accomplished by a more strict well-formedness check for classes:

$$\frac{\textsf{class}\ C[\overline{X <: N}] \lhd P \ldots \qquad \sigma = [\overline{T/X}] \qquad \Gamma \vdash \overline{T}\ \textsf{wf} \qquad \boxed{\Gamma \vdash \sigma P\ \textsf{wf}} \qquad \Gamma \vdash \overline{T <: \sigma N}}{\Gamma \vdash C[\overline{T}]\ \textsf{wf}} \tag{WF-ClassAlt}$$

We dub FGJ$^-$ ("FGJ minus") the calculus obtained by replacing WF-Class by WF-ClassAlt in FGJ.

**Conjecture 4.3.2**

If we replace TR-Class by TR-ClassAlt then there exists a type-preserving translation from FGJ$^-$ to wfDOT.

*Proof sketch.* While there are uses of Bind1 and BindX in our proof which are unrelated to subtyping preservation, we conjecture that these uses are inessential and could be replaced by sufficiently creative uses of typing rules like And-I as in [Amin, Grütter, et al. 2016, § 5.2] (which might make the proof more complex). In particular, note that Lemma 2.4.6 relies on BindX and would have to be replaced by an alternative lemma, perhaps of the form "Given $\sigma = [\overline{T/x.L}]$ and $\Gamma \vdash \overline{T =:= x.L}$, if $\Gamma \vdash U$ wf and $\Gamma \vdash t :_{(!)} U$ then $\Gamma \vdash t :_{(!)} \sigma U$". ◊

We will not study FGJ$^-$ in more detail because it is not expressive enough to encode F-bounded polymorphism [Canning et al. 1989; Greenman, Muehlboeck, and Tate 2014] which is commonly used in the Java standard library (e.g., with `java.lang.Comparable`) and therefore important for Scala to support.

### 4.3.2 Meta-theory

Like in the previous chapter, we'd like to relate FGJ judgments in an environment $\Gamma$ with DOT judgments in the translated environment $|\Gamma|$, but $|\Gamma|$ needs to account for implementation details of our constructor translation which makes it inconvenient to work with. In particular, the FGJ equivalent of Lemma 3.2.9 does not hold because of the presence of ctag in the environment.

To remedy this, we introduce an *environment entailment* judgment $\Gamma \dashv\vdash \Delta$ such that $\Gamma \dashv\vdash |\Gamma|$ and we generalize our theorems to apply to all $\Delta$ such that $\Gamma \dashv\vdash \Delta$. This lets us use Theorem 4.3.19 in place of Lemma 3.2.9. It is possible that a different environment translation $|\Gamma|$ could alleviate the need for environment entailement but we were not able to come up with a satisfying alternative.

**Definition 4.3.3: Environment entailment**

$$\boxed{\Gamma_{\text{FGJ}} \Dashv \Delta_{\text{DOT}}}$$

$$\varnothing \Dashv \mathsf{ct} : [\![CT]\!], \Delta \qquad \text{(EE-Empty)}$$

$$\frac{\begin{array}{c}\Gamma' \Dashv \Delta \\ \Delta \vdash \overline{|X| <: |N|}\end{array}}{\Gamma', \overline{X <: N} \Dashv \Delta} \qquad \text{(EE-Typs)}$$

$$\frac{\begin{array}{c}\mathsf{tparams}(C) = \overline{X <: N} \\ \overline{X <: N} \Dashv \Delta', \mathsf{this} : T \\ \Delta', \mathsf{this} : T \vdash \mathsf{this} :_{(!)} [\![C]\!]^{\overline{|X|}}\end{array}}{\overline{X <: N}, \mathsf{this} : C[\overline{X}] \Dashv \Delta', \mathsf{this} : T, \Delta''} \qquad \text{(EE-This)}$$

$$\frac{\Gamma' \Dashv \Delta' \quad x \neq \mathsf{this}}{\Gamma', x : T \Dashv \Delta', x : |T|, \Delta''} \qquad \text{(EE-Var)}$$

**Theorem 4.3.4: Environment translation conforms to entailment**

If $|\Gamma|$ wf then $\Gamma \Dashv |\Gamma|$.

*Proof.* By structural induction on $\Gamma$.

**Case** $\Gamma = \varnothing$

By EE-Empty.

**Case** $\Gamma = \Gamma', \overline{X <: N}$

By inversion of $|\Gamma|$ via E-MVar, we must have $\overline{X = X_m}$ and $\overline{|X_m| = \mathsf{mtag}.X_m}$. Hence,

$$\frac{\dfrac{}{\Gamma' \Dashv |\Gamma'|}\,\text{(IH)} \qquad \dfrac{\dfrac{\dfrac{}{|\Gamma', \overline{X_m <: N}| \vdash \mathsf{mtag} :_! |\overline{X_m <: N}|}\,\text{(Var)}}{|\Gamma', \overline{X_m <: N}| \vdash \mathsf{mtag} :_! \overline{(X_m : \bot \,..\, |N|)}}\,\text{(Sub, VarUnpack)}}{|\Gamma', \overline{X_m <: N}| \vdash \overline{|X_m| <: |N|}}\,\text{(Sel1)}}{\Gamma', \overline{X_m <: N} \Dashv |\Gamma', \overline{X_m <: N}|}\,\text{(EE-Typs)}$$

**Case** $\Gamma = \Gamma', \mathsf{this} : T$

By inversion of $|\Gamma|$ via E-This we must have $\Gamma' = \overline{X_C <: N}$ and $T = C[\overline{X_C}]$. We have,

$$\dfrac{\dfrac{\dfrac{}{|\Gamma| \vdash \mathsf{this} :_! \llbracket C \rrbracket^{\overline{\mathsf{ctag}.X}}}\ (\textsc{Var})}{|\Gamma| \vdash \mathsf{this} :_! (X_i = \mathsf{ctag}.X_i)}\ (2.4.5)}{|\Gamma| \vdash \mathsf{ctag}.X_i =:= |X_i|}\ (\textsc{Sel1}, \textsc{Sel2})$$

Let $\tau = \overline{[\mathsf{ctag}.X/\mathsf{this}.X]}$. Then,

$$\dfrac{\dfrac{\dfrac{\dfrac{\dfrac{\dfrac{\dfrac{\dfrac{\dfrac{}{|\Gamma|_{[\mathsf{ctag}]} \vdash \mathsf{ctag} :_! |\overline{X <: N}|}\ (\textsc{Var})}{|\Gamma|_{[\mathsf{ctag}]} \vdash \mathsf{ctag} :_! (X_i : \bot \mathinner{..} \tau|N_i|)}\ (\textsc{Sub}, \textsc{VarUnpack})}{|\Gamma| \vdash \mathsf{ctag}.X_i <: \tau|N_i|}\ (\textsc{Sel1})}{|\Gamma| \vdash \mathsf{ctag}.X_i <: |N_i|}\ (2.4.6)}{|\Gamma| \vdash (X_i = \mathsf{ctag}.X_i) <: (X_i : \bot \mathinner{..} |N_i|)}\ (\textsc{Typ})}{|\Gamma| \vdash \llbracket C \rrbracket^{\overline{\mathsf{ctag}.X}} <: (X_i : \bot \mathinner{..} |N_i|)}\ (\textsc{Trans}, 2.4.5)}{|\Gamma| \vdash \mathsf{this} :_! (X_{Ci} : \bot \mathinner{..} |N_i|)}\ (\textsc{Sub}, \textsc{Var})}{|\Gamma| \vdash \overline{|X_C| <: |N|}}\ (\textsc{Sel1})}{\Gamma' \dashv |\Gamma|}\ (\textsc{EE-Typs}) \quad \dfrac{\dfrac{}{|\Gamma| \vdash \llbracket C \rrbracket^{\overline{\mathsf{ctag}.X}} <: \llbracket C \rrbracket^{\overline{|X|}}}\ (2.4.5, \textsc{Typ})}{|\Gamma| \vdash \mathsf{this} :_{(!)} \llbracket C \rrbracket^{\overline{|X|}}}\ (\textsc{Sub})}{\Gamma \vdash |\Gamma|}\ (\textsc{EE-This})$$

**Case** $\Gamma = \Gamma', x : T$ where $x \neq \mathsf{this}$

We have $|\Gamma| = |\Gamma'|, x : |T|$. By the IH, $\Gamma' \dashv |\Gamma'|$ and EE-Var finishes the case.

■

**Lemma 4.3.5: Appending on the right preserves environment entailment**

If $\Gamma \dashv \Delta$ then $\Gamma \dashv \Delta, \Delta'$.

*Proof.* By straightforward induction on the derivation of $\Gamma \dashv \Delta$. ■

**Lemma 4.3.6: Truncating on the left preserves environment entailment**

If $\Gamma \dashv \Delta$ and $\Gamma = \Gamma_1, \Gamma_2$, then $\Gamma_1 \dashv \Delta$.

*Proof.* By induction on the derivation of $\Gamma \dashv \Delta$ we find that $\Gamma_1 \dashv \Delta_1$ where $\Delta_1$ is either $\Delta$ or a prefix of $\Delta$ and Lemma 4.3.5 finishes the case. ■

**Theorem 4.3.7: Translation preserves substitution**

$|\sigma S| = |\sigma||S|$

*Proof.* By structural induction on $S$.

**Case** $S = X$

If $X \notin \mathsf{dom}(\sigma)$ this is trivial, otherwise $\sigma = [\dots, T/X, \dots]$ and $|\sigma| = [\dots, |T|/|X|, \dots]$ for some $T$. Hence, $|\sigma X| = |T| = |\sigma||T|$.

**Case** $S = C[\overline{T}]$

By definition,

$$|\sigma C[\overline{T}]| = |C[\overline{\sigma T}]| = \mathsf{ct}.C \wedge \bigwedge \overline{X = |\sigma T|}$$
$$|\sigma||C[\overline{T}]| = \mathsf{ct}.C \wedge \bigwedge \overline{X = |\sigma||T|}$$

By the IH, $\overline{|\sigma T| = |\sigma||T|}$ which lets us finish the case. ∎

**Lemma 4.3.8**

Given $\Gamma = (\overline{X_C <: N_C}, \mathsf{this} : C[\overline{X_C}])$, **class** $C[\overline{X_B}] \triangleleft B[\overline{U}]$, $\mathsf{tparams}(B) = \overline{X_B <: N_B}$ and $\Gamma \dashv\Vdash \Delta$, then

1. $\Delta \vdash \overline{|X_B| =:= |U|}$
2. $\Delta \vdash \overline{|X_B| <: |N_B|}$

*Proof.* We first prove part 1. then use that result to prove part 2.

By definition, $[\![C]\!] = \dots \wedge \mathsf{baseArgs}(C)$ and $\mathsf{baseArgs}(C) = \left(\bigwedge \overline{X_B = |U|}\right) \wedge \dots$. Hence,

$$\dfrac{\dfrac{\dfrac{}{\Delta_{[\mathsf{this}]} \vdash \mathsf{this} :_! [\![C]\!]}\ (\text{Sub, EE-This})}{\Delta_{[\mathsf{this}]} \vdash \mathsf{this} :_! \overline{(X_B = |U|)}}\ (\text{Sub, 2.4.5})}{\Delta \vdash \overline{|X_B| =:= |U|}}\ (\text{Sel1, Sel2})$$

Let $\Gamma_1 = \overline{X_C <: N_C}$. By inversion, $\vdash C\ \mathsf{ok}$ implies $\Gamma_1 \vdash B[\overline{U}]\ \mathsf{wf}$ which implies $\Gamma_1 \vdash \overline{U <: \sigma N_B}$ where $\sigma = [\overline{U/X_B}]$. Hence,

$$\dfrac{\dfrac{\dfrac{\dfrac{}{\Gamma \vdash \overline{U <: \sigma N_B}}\ (\text{Weaken})}{\Delta \vdash \overline{|U| <: |\sigma N_B|}}\ (4.3.11)}{\Delta \vdash \overline{|\sigma||X_B| <: |\sigma||N_B|}}\ (4.3.7) \qquad \Delta \vdash \overline{|X_B| =:= |U|}}{\Delta \vdash \overline{|X_B| <: |N_B|}}\ (\text{Trans, 2.4.6})$$

∎

**Theorem 4.3.9: Well-formedness preservation**

If $\Gamma \dashv\!\vdash \Delta$ and $\Gamma \vdash S$ wf then $\Delta \vdash |S|$ wf.

*Proof.* By induction on the derivation of $\Gamma \vdash S$ wf.

**Case** $\Gamma \vdash \mathsf{Object}$ wf (WF-Object)

> $|\mathsf{Object}| = \mathsf{ct.Object}$ is well-formed since $\mathsf{ct} \in \mathsf{dom}(\Delta)$ by EE-Empty and Lemma 4.3.6.

**Case** $$\frac{X \in \mathsf{dom}(\Gamma)}{\Gamma \vdash X \text{ wf}} \ (\text{WF-Var})$$

> By Lemma 4.3.6 and inversion of EE-Typs we must have $\Delta \vdash |X| <: |N|$ for some $N$ which implies $\Delta|X|$ wf since DOT subtyping is only defined on well-formed types.

**Case** $$\frac{\textbf{class}\ C[\overline{X_C <: N}] \lhd P \ldots \quad \sigma = [\overline{T/X_C}] \quad \Gamma \vdash \overline{T} \text{ wf} \quad \Gamma \vdash \overline{T <: \sigma N}}{\Gamma \vdash C[\overline{T}] \text{ wf}} \ (\text{WF-Class})$$

> By definition, $|C[\overline{T}]| = \mathsf{ct}.C \wedge \{_ \Rightarrow \overline{X_C = |T|}\}$. By the IH, $\Delta \vdash \overline{|T|}$ wf and $\Delta \vdash \mathsf{ct}.C$ wf since $\mathsf{ct} \in \mathsf{dom}(\Delta)$.

■

**Lemma 4.3.10**

Given $\mathsf{tparams}(C) = \overline{X <: N}$, $\Gamma \vdash C[\overline{T}]$ wf and $\Gamma \dashv\!\vdash \Delta$, then $\Delta \vdash |C[\overline{T}]| <: \{_ \Rightarrow |\sigma|[\![C]\!]\}$ where $\sigma = [\overline{T/X}]$.

*Proof.* We have $|C[\overline{T}]| = \mathsf{ct}.C \wedge \{_ \Rightarrow \overline{X = |T|}\}$.

$$\dfrac{\dfrac{\dfrac{\dfrac{}{\Delta_{[\mathsf{ct}]} \vdash \mathsf{ct} :_! [\![CT]\!]}\ (4.3.6,\text{EE-Empty})}{\Delta_{[\mathsf{ct}]} \vdash \mathsf{ct} :_! (C = \{\mathsf{this} \Rightarrow [\![C]\!], \overline{X : \bot \,..\, |N|}\})}\ (\text{Sub}, 2.4.5)}{\Delta \vdash \mathsf{ct}.C <: \{\mathsf{this} \Rightarrow [\![C]\!], \overline{X : \bot \,..\, |N|}\}}\ (\text{Sel1})}{\Delta \vdash \mathsf{ct}.C <: \{\mathsf{this} \Rightarrow [\![C]\!]\}}\ (\text{Trans}, \text{BindX}, 2.4.5)$$

Hence, by transitivity, width and depth subtyping we only need to show that

$$\Delta \vdash \{\mathsf{this} \Rightarrow [\![C]\!]\} \wedge \{_ \Rightarrow \overline{X = |T|}\} <: \{_ \Rightarrow |\sigma|[\![C]\!]\}$$

As in the example given in subsection 4.3.1, this requires using And-Bind, but in And-Bind the bound variable of the recursive types involved must all be equal. This is doable since we're working up to α-renaming, but we need to be careful: this might be bound in $\Delta$ and free in $|T|$, therefore we cannot rename _ to this. Instead, we rename this to a fresh variable $z$:

$$\{\mathsf{this} \Rightarrow [\![C]\!]\} = \{z \Rightarrow [z/\mathsf{this}][\![C]\!]\}$$

By inversion of $\vdash C$ ok via GT-Class, only $\overline{X}$ may be free in the types appearing in $CT(C)$, therefore this is equivalent to

$$\{z \Rightarrow \tau[\![C]\!]\} \text{ where } \tau = [\overline{z.X/|X|}]$$

Furthermore, we note that $|\sigma|[\![C]\!] = [\overline{|T|/|X|}][\![C]\!] = [\overline{|T|/z.X}](\tau[\![C]\!])$.

Let $\Delta_1 = \Delta, z : \tau[\![C]\!] \wedge \bigwedge \overline{X = |T|}$. Then

$$\dfrac{\dfrac{\dfrac{\dfrac{\dfrac{}{\Delta_1 \vdash z :_! \overline{(X = |T|)}}\ (\textsc{Sub}, \textsc{Var})}{\Delta_1 \vdash \overline{z.X =:= |T|}}\ (\textsc{Sel1}, \textsc{Sel2})}{\Delta_1 \vdash \tau[\![C]\!] <: [\overline{|T|/z.X}](\tau[\![C]\!])}\ (2.4.6)}{\Delta \vdash \{z \Rightarrow \tau[\![C]\!], \overline{X = |T|}\} <: \{_ \Rightarrow |\sigma|[\![C]\!]\}}\ (\textsc{BindX}, \textsc{And11})}{\Delta \vdash \{z \Rightarrow \tau[\![C]\!]\} \wedge \{_ \Rightarrow \overline{X = |T|}\} <: \{_ \Rightarrow |\sigma|[\![C]\!]\}}\ (\textsc{And-Bind})$$

■

**Theorem 4.3.11: Subtyping preservation**

If $\Gamma \dashv\vdash \Delta$, $\Gamma \vdash S$ wf and $\Gamma \vdash S <: T$ then $\Delta \vdash |S| <: |T|$.

*Proof.* By Theorem 4.3.9, $\Delta \vdash |S|$ wf. By Lemma 4.2.4, $\Gamma \vdash T$ wf so by Theorem 4.3.9 again $\Delta \vdash |T|$ wf. We proceed by induction on the derivation of $\Gamma \vdash S <: T$.

**Case** $\dfrac{\Gamma(Z) = Q}{\Gamma \vdash Z <: Q}$ (GS-Var)

> We must have $\Gamma = \Gamma_1, \overline{X <: N}, \Gamma_2$ where $Z = X_i$, $Q = N_i$. Then by Lemma 4.3.6, $\Gamma_1, \overline{X <: N} \vdash \Delta$ and EE-Typs finishes the case.

**Case** $\Gamma \vdash S <: S$ (GS-Refl)

> By Theorem 4.3.9, $\Delta \vdash |S|$ wf and Refl finishes the case.

**Case** $\dfrac{\textbf{class}\ C[\overline{X_C <: N}](\ldots) \triangleleft B[\overline{U}] \ldots \qquad \sigma = [\overline{T/X_C}]}{\Gamma \vdash C[\overline{T}] <: B[\overline{\sigma U}]}$ (GS-Class)

> By definition, $|B[\overline{\sigma U}]| = \text{ct}.B \wedge \{_ \Rightarrow \overline{X_B = |\sigma U|}\}$ so by And2 we only need to show that $|C[\overline{T}]|$ is a subtype of each operand of the intersection:
>
> $$\dfrac{\dfrac{\Delta_{[\text{ct}]} \vdash \text{ct} :_! [\![CT]\!]}{\Delta \vdash \text{ct}.C <: \text{ct}.B}\ (\textsc{Sel1}, \textsc{Sub}, 2.4.5)}{\Delta \vdash |C[\overline{T}]| <: \text{ct}.B}\ (\textsc{And11})$$

$$\dfrac{\dfrac{\dfrac{}{\Delta \vdash \{_ \Rightarrow |\sigma|[\![C]\!]\} <: \{_ \Rightarrow \overline{X_B = |\sigma||U|}\}}\ (\textsc{BindX}, 2.4.5)}{\Delta \vdash \{_ \Rightarrow |\sigma|[\![C]\!]\} <: \{_ \Rightarrow \overline{X_B = |\sigma U|}\}}\ (4.3.7)}{\Delta \vdash |C[\overline{T}]| <: \{_ \Rightarrow \overline{X_B = |\sigma U|}\}}\ (\textsc{Trans}, 4.3.10)$$

**Case** $\dfrac{\Gamma \vdash S <: U \quad \Gamma \vdash U <: T}{\Gamma \vdash S <: T}$ (GS-Trans)

$$\dfrac{\dfrac{}{\Delta \vdash |S| <: |U|}\ (\text{IH}) \quad \dfrac{\dfrac{}{\Gamma \vdash U \text{ wf}}\ (4.2.4)}{\Delta \vdash |U| <: |T|}\ (\text{IH})}{\Delta \vdash |S| <: |T|}\ (\textsc{Trans})$$

■

**Lemma 4.3.12: Class translation preserves value parameters**

If $\Gamma \dashv\!\!\mid \Delta$, $\Gamma \vdash N$ wf and $\text{vparams}(N) = \overline{f : U}$, then $\Delta \vdash \overline{|N| <: (f() : |U|)}$

*Proof.* By inversion of $\text{vparams}(N)$. Case G-Object is trivial.

**Case** $\dfrac{\textbf{class}\ C[\overline{X <: N}](\overline{f : U'}) \ldots \quad \sigma = [\overline{T/X}]}{\text{vparams}(C[\overline{T}]) \coloneqq \overline{f : \sigma U'}}$ (G-Class)

For all $i$ in bounds:

$$\dfrac{\dfrac{\dfrac{[\![C]\!] = \ldots \wedge [\![f_i : U_i]\!] \wedge \ldots}{\Delta \vdash |\sigma|[\![C]\!] <: |\sigma|[\![f_i : U_i]\!]}\ (\textsc{Bind1}, 2.4.5)}{\Delta \vdash |\sigma|[\![C]\!] <: (f_i() : |\sigma U_i|)}\ (4.3.7)}{\Delta \vdash |C[\overline{T}]| <: (f_i() : |\sigma U_i|)}\ (\textsc{Trans}, \textsc{Bind1}, 4.3.10)$$

■

**Lemma 4.3.13: Class translation preserves methods**

If $\Gamma \dashv\!\!\mid \Delta$, $\Gamma \vdash N$ wf and $\text{mtype}(m, N) = [\overline{Y <: P}] \to (\overline{y : U}) \to U_0$, then $\Delta \vdash |N| <: (m(\text{mtag} : \overline{|Y <: P|}, \overline{y : |U|}) : |U_0|)$.

*Proof.* By induction on the derivation of $\text{mtype}_\Gamma(m, N)$.

$$\textbf{Case}\quad \frac{\textsf{class}\ C[\overline{X <: N}]\ \dots\ \{\overline{M}\} \quad \sigma = [\overline{S/X}] \qquad (\textsf{def}\ m[\overline{Y <: P'}](\overline{x : U'}) : U'_0 = \dots) \in \overline{M}}{\textsf{mtype}(m,\ C[\overline{T}]) \coloneqq [\overline{Y <: \sigma P'}] \to (\overline{x : \sigma U'}) \to \sigma U'_0}\ (\text{GM-Class})$$

This case mirrors case G-Class of Lemma 4.3.12.

$$\cfrac{\cfrac{\cfrac{[\![C]\!] = \dots \wedge [\![m]\!]_C \wedge \dots}{\Delta \vdash |\sigma|[\![C]\!] <: |\sigma|[\![m]\!]_C}\ (\text{Bind1, 2.4.5})}{\Delta \vdash |\sigma|[\![C]\!] <: (m(\textsf{mtag} : |\overline{Y <: \sigma P'}|, \overline{y : |\sigma U|}) : |\sigma U_0|)}\ (4.3.7)}{\Delta \vdash |C[\overline{T}]| <: (m(\textsf{mtag} : |\overline{Y <: \sigma P'}|, \overline{y : |\sigma U|}) : |\sigma U_0|)}\ (\text{Trans, Bind1, 4.3.10})$$

$$\textbf{Case}\quad \frac{\textsf{class}\ C[\overline{X <: N}](\dots) \lhd P\ \{\overline{M}\} \quad \sigma = [\overline{T/X}] \qquad (\textsf{def}\ m\ \dots) \notin \overline{M}}{\textsf{mtype}(m,\ C[\overline{T}]) \coloneqq \textsf{mtype}(m,\ \sigma P)}\ (\text{GM-Super})$$

$$\cfrac{\cfrac{\cfrac{}{\Gamma \vdash C[\overline{T}] <: \sigma P}\ (\text{GS-Class})}{\Delta \vdash |C[\overline{T}]| <: |\sigma P|}\ (4.3.11) \qquad \cfrac{\cfrac{}{\Gamma \vdash \sigma P\ \textsf{wf}}\ (4.2.4)}{\Delta \vdash |\sigma P| <: (m(\textsf{mtag} : |\overline{Y <: P}|, \overline{y : |U|}) : |U_0|)}\ (\text{IH})}{\Delta \vdash |C[\overline{T}]| <: (m(\textsf{mtag} : |\overline{Y <: P}|, \overline{y : |U|}) : |U_0|)}\ (\text{Trans})$$

■

**Lemma 4.3.14: Method translation preserves overriding relationship**

Given $\textsf{class}\ C[\overline{X_C <: N_C}] \lhd B[\overline{U}]\{\overline{M}\}$, $\Gamma = \overline{X_C <: N_C}, \textsf{this} : C[\overline{X_C}]$ and $\Gamma \dashv\!\!| \ \Delta$, then $m \in \textsf{mnames}(B)$ implies $\Delta \vdash [\![m]\!]_C <: [\![m]\!]_B$.

*Proof.* Let

$$\begin{aligned} \textsf{tparams}(B) &= \overline{X_B <: N_B} \\ \textsf{mtype}(m,\ B[\overline{X_B}]) &= [\overline{Z <: Q}] \to (\overline{y : V}) \to V_0 \\ \textsf{mtype}(m,\ C[\overline{X_C}]) &= [\overline{Y <: P}] \to (\overline{x : U}) \to U_0 \end{aligned}$$

then $\textsf{mtype}(m, B[\overline{U}]) = |\overline{Z <: \sigma Q}] \to (\overline{y : \sigma V}) \to \sigma V_0$ by observation. We proceed by inversion on the derivation of $\textsf{mtype}(m,\ C[\overline{X_C}])$.

$$\textbf{Case}\quad \frac{(\textsf{def}\ m[\overline{Y <: P}](\overline{x : U}) : U_0 = e_0) \in \overline{M}}{\textsf{mtype}(m,\ C[\overline{X_C}]) \coloneqq [\overline{Y <: P}] \to (\overline{x : U}) \to U_0}\ (\text{GM-Class})$$

Let $\Gamma_m = \Gamma, \overline{Y <: P}$ and $\Delta_m = \Delta, \textsf{mtag} : |\overline{Y <: P}|$, then $\Gamma_m \dashv\!\!| \ \Delta_m$ by EE-Typs. By inversion, $\vdash C\ \textsf{ok}$ implies $\Gamma \vdash m\ \textsf{ok}$ implies $\textsf{override}_\Gamma(m, C[\overline{X_C}], B[\overline{U}])$ which implies that $\overline{Y = \sigma Z}, \overline{P = \sigma Q}$, $\overline{x = y}$, $\overline{U = \sigma V}$ and $\Gamma_m \vdash U_0 <: \sigma V_0$, hence

$$
\dfrac{
 \dfrac{
  \dfrac{
   \dfrac{
    \dfrac{}{\Delta_m \vdash \overline{|X_B| =:= U}}\,(4.3.8)
   }{\Delta_m \vdash \overline{|\sigma Q| <: |Q|}}\,(4.3.7,\ 2.4.6)
  }{\Delta_m \vdash (Y : \bot\,..\,|Q|) <: (Y : \bot\,..\,|P|)}\,(\textsc{Typ})
 }{\Delta \vdash |\overline{Y <: Q}| <: |\overline{Y <: P}|}\,(\textsc{BindX})
 \qquad
 \dfrac{
  \dfrac{
   \dfrac{
    \dfrac{\Gamma_m \vdash U_0 <: \sigma V_0}{\Delta_m \vdash |U_0| <: |\sigma V_0|}\,(4.3.11)
   }{\Delta_m \vdash |U_0| <: |\sigma||V_0|}\,(4.3.7)
  }{\Delta_m \vdash |U_0| <: |V_0|}\,(\textsc{Trans},\ 2.4.6)
 }{\Delta_m, \overline{y : |V|} \vdash |U_0| <: |V_0|}\,(\textsc{Weaken})
}{\Delta \vdash [\![m]\!]_C <: [\![m]\!]_B}\,(\textsc{Fun'},\ \textsc{Narrow})
$$

**Case** $\dfrac{(\textbf{def}\ m \ldots) \notin \overline{M}}{\text{mtype}(m, C[\overline{T}]) := \text{mtype}(m, \sigma P)}$ (GM-Super)

In this case, $[\![m]\!]_C = |\sigma|[\![m]\!]_B$ by inspection so we only need to show that $\Delta \vdash |\sigma|[\![m]\!]_B <: [\![m]\!]_B$ which follows by Theorem 4.3.7 and Lemma 4.3.8.

■

**Lemma 4.3.15**

Given **class** $C[\overline{X_C <: N_C}](\ldots) \lhd B[\overline{U}]$, $\text{tparams}(B) = \overline{X_B <: N_B}$, $\Gamma = \overline{X_C <: N_C}$, $\text{this} : C[\overline{X_C}]$ and $\Gamma \dashv \Delta$, then $\Delta \vdash [\![C]\!]^{|\overline{X_C}|} <: [\![B]\!]^{|\overline{X_B}|}$.

*Proof.* By definition, we want to show:

$$
\begin{aligned}
\Delta \vdash\, & [\![\text{vparams}(C[\overline{X_C}])]\!] \wedge [\![\text{mnames}(C)]\!]_C \wedge \text{baseArgs}(C) \wedge \bigwedge \overline{X_C = |X_C|} <: \\
& [\![\text{vparams}(B[\overline{X_B}])]\!] \wedge [\![\text{mnames}(B)]\!]_B \wedge \text{baseArgs}(B) \wedge \bigwedge \overline{X_B = |X_B|}
\end{aligned}
$$

After proving the following claims, we can finish the proof by width and depth subtyping.

**Claim 1:** $\Delta \vdash [\![\text{vparams}(C[\overline{X_C}])]\!] <: [\![\text{vparams}(B[\overline{X_B}])]\!]$

Let $\sigma_B = [\overline{U/X_B}]$ and note that $\Delta \vdash \overline{|X_B| =:= |U|}$ by Lemma 4.3.8. $\vdash C$ ok implies that $\text{vparams}(C[\overline{X_C}]) = ((\sigma_B\, \text{vparams}(B[\overline{X_B}])), \ldots)$ so by Theorem 4.3.7 and observation, $[\![\text{vparams}(C[\overline{X_C}])]\!] = |\sigma_B|[\![\text{vparams}(B[\overline{X_B}])]\!] \wedge T$ for some $T$. Finally, we find $\Delta \vdash |\sigma_B|[\![\text{vparams}(B[\overline{X_B}])]\!] \wedge T <: [\![\text{vparams}(B[\overline{X_B}])]\!]$ by width subtyping and Lemma 2.4.6.

**Claim 2:** $\Delta \vdash [\![\text{mnames}(C)]\!]_C <: [\![\text{mnames}(B)]\!]_B$

By definition, $\text{mnames}(C) = (\text{mnames}(B), \ldots)$ so $[\![\text{mnames}(C)]\!]_C = [\![\text{mnames}(B)]\!]_C \wedge T$ for some $T$ and we only need to prove that $\Delta \vdash [\![m]\!]_C <: [\![m]\!]_B$ for all $m \in \text{mnames}(B)$. Lemma 4.3.14 finishes the claim.

**Claim 3:** $\Delta \vdash \text{baseArgs}(C) <: \text{baseArgs}(B)$

By definition, $\text{baseArgs}(C) = \ldots \wedge \text{baseArgs}(B)$, so this follows by width subtyping.

**Claim 4:** $\Delta \vdash \mathsf{baseArgs}(C) <: \overline{X_B = |X_B|}$

We have $\mathsf{baseArgs}(C) = \left(\bigwedge \overline{X_B = |U|}\right) \wedge \ldots$. Hence,

$$\dfrac{\dfrac{\dfrac{}{\Delta \vdash \mathsf{this} :_! \overline{X_B = |U|}}\ (\textsc{Sub})}{\Delta \vdash \overline{|U| =:= |X_B|}}\ (\textsc{Sel1}, \textsc{Sel2})}{\Delta \vdash \mathsf{baseArgs}(C) <: \overline{X_B = |X_B|}}\ (2.4.5, \textsc{Typ})$$

∎

**Lemma 4.3.16**

Given $\mathsf{tparams}(C) = \overline{X_C <: N_C}$, then $|\varnothing| \vdash \{\mathsf{this} \Rightarrow [\![C]\!]^{\overline{|X_C|}}, \overline{X_C : \bot \,..\, |N_C|}\} <: \mathsf{ct}.C$.

*Proof.* Since $\vdash CT$ ok and $C \in \mathsf{dom}(CT)$, there exists a a sequence of classes $D$ such that $D_1 = C$, $D_n = \mathsf{Object}$ and **class** $D_i[\ldots] \lhd D_{i+1}[\ldots]$ for all $i$. We prove by induction on the length $n$ ($\geq 2$) of the sequence.

**Case** $(n = 2)$ **class** $C[\overline{X_C <: N_C}](\ldots) \lhd \mathsf{Object}\ldots$

$$\dfrac{\dfrac{\dfrac{\dfrac{}{|\varnothing| \vdash \mathsf{ct} :_! [\![CT]\!]}\ (\textsc{Var}) \qquad \dfrac{}{|\varnothing| \vdash [\![CT]\!] <: (C : (\{\mathsf{this} \Rightarrow [\![C]\!]\}) \,..\, \top)}\ (2.4.5, \textsc{Typ})}{|\varnothing| \vdash \mathsf{ct} :_! (C = \{\mathsf{this} \Rightarrow [\![C]\!], \overline{X_C : \bot \,..\, |N_C|}\})}\ (\textsc{Sub})}{|\varnothing| \vdash \{\mathsf{this} \Rightarrow [\![C]\!], \overline{X_C : \bot \,..\, |N_C|}\} <: \mathsf{ct}.C}\ (\textsc{Sel2})}{|\varnothing| \vdash \{\mathsf{this} \Rightarrow [\![C]\!]^{\overline{|X_C|}}, \overline{X_C : \bot \,..\, |N_C|}\} <: \mathsf{ct}.C}\ (\textsc{Trans}, \textsc{BindX})$$

**Case** $(n > 2)$ **class** $C[\overline{X_C <: N_C}](\ldots) \lhd B[\overline{U}]\ldots$

It is easy to see that $|\varnothing| \vdash \mathsf{ct}.B \wedge \{\mathsf{this} \Rightarrow [\![C]\!], \overline{X_C : \bot \,..\, |N_C|}\} <: \mathsf{ct}.C$ so by transitivity and And2 we only need to prove

$$|\varnothing| \vdash \{\mathsf{this} \Rightarrow [\![C]\!]^{\overline{|X_C|}}, \overline{X_C : \bot \,..\, |N_C|}\} <: \mathsf{ct}.B$$

Let $\mathsf{tparams}(B) = \overline{X_B <: N_B}$. By the IH, $|\varnothing| \vdash \{\mathsf{this} \Rightarrow [\![B]\!]^{\overline{|X_B|}}, \overline{X_B : \bot \,..\, |N_B|}\} <: \mathsf{ct}.B$, so by transitivity we only need to prove $|\varnothing| \vdash \{\mathsf{this} \Rightarrow [\![C]\!]^{\overline{|X_C|}}, \overline{X_C : \bot \,..\, |N_C|}\} <: \{\mathsf{this} \Rightarrow [\![B]\!]^{\overline{|X_B|}}, \overline{X_B : \bot \,..\, |N_B|}\}$. Let $\Delta = |\varnothing|, \mathsf{this} : [\![C]\!]^{\overline{|X_C|}} \wedge \overline{X_C : \bot \,..\, |N_C|}$. Then,

$$
\dfrac{
\dfrac{\dfrac{}{|\overline{X_C <: N_C}, \text{this} : C[\overline{X_C}]| \dashv\vdash \Delta}\;(\textsc{EE-This})}{\Delta \vdash \llbracket C \rrbracket^{|\overline{X_C}|} <: \llbracket B \rrbracket^{|\overline{X_B}|}}\;(4.3.15)
\quad
\dfrac{\dfrac{\dfrac{}{\Delta \vdash \overline{|U| <: |N_B|}}\;(4.3.8)}{\Delta \vdash \overline{X_B = |U|} <: \overline{X_B : \bot \mathinner{..} |N_B|}}\;(\textsc{Typ})}{\Delta \vdash \llbracket C \rrbracket <: \overline{X_B : \bot \mathinner{..} |N_B|}}\;(2.4.5)
}{
\dfrac{\Delta \vdash \llbracket C \rrbracket^{|\overline{X_C}|} <: \llbracket B \rrbracket^{|\overline{X_B}|} \wedge \overline{X_B : \bot \mathinner{..} |N_B|}}{|\varnothing| \vdash \{\text{this} \Rightarrow \llbracket C \rrbracket^{|\overline{X_C}|}, \overline{X_C : \bot \mathinner{..} |N_C|}\} <: \{\text{this} \Rightarrow \llbracket B \rrbracket^{|\overline{X_B}|}, \overline{X_B : \bot \mathinner{..} |N_B|}\}}\;(\textsc{BindX}, \textsc{And11})
}\;(2.4.5)
$$

■

**Lemma 4.3.17: this translation is type-preserving**

If $\Gamma = \overline{X <: N}, \text{this} : C[\overline{X}]$ and $\Gamma \dashv\vdash \Delta$, then $\Delta \vdash \text{this} : |C[\overline{X}]|$.

*Proof.* By inversion of $\Gamma \dashv\vdash \Delta$, we have $\Delta \vdash \text{this} : \llbracket C \rrbracket^{|\overline{X}|}$. Hence,

$$
\dfrac{
\dfrac{
\dfrac{
\Delta \vdash \text{this} : \llbracket C \rrbracket^{|\overline{X}|}
\quad
\dfrac{\dfrac{\dfrac{}{\Delta \vdash \overline{|X| <: |N|}}\;(4.3.11)}{\Delta \vdash \overline{(X = |X|)} <: \overline{(X : \bot \mathinner{..} |N|)}}\;(\textsc{Typ})}{\Delta \vdash \llbracket C \rrbracket^{|\overline{X}|} <: \llbracket C \rrbracket^{|\overline{X}|} \wedge \bigwedge \overline{X : \bot \mathinner{..} |N|}}\;(2.4.5)
}{\Delta \vdash \text{this} : \llbracket C \rrbracket^{|\overline{X}|} \wedge \bigwedge \overline{X : \bot \mathinner{..} |N|}}\;(\textsc{Sub})
}{\Delta \vdash \text{this} : \{\text{this} \Rightarrow \llbracket C \rrbracket^{|\overline{X}|}, \overline{X : \bot \mathinner{..} |N|}\}}\;(\textsc{VarPack})
}{\Delta \vdash \text{this} : |C[\overline{X}]|}\;(\textsc{Sub}, \textsc{Weaken}, 4.3.16)
$$

■

**Theorem 4.3.18: Typing translation is type-preserving**

If $\Gamma \dashv\vdash \Delta$ and $\Gamma \vdash e : T$, then $\Delta \vdash |e|_\Gamma : |T|$.

*Proof.* By induction on the derivation of $\Gamma \vdash e : T$.

**Case** $\dfrac{\Gamma(x) = T}{\Gamma \vdash x : T}\;(\textsc{GT-Var})$

We can distinguish two sub-cases:

- If $x = \text{this}$, then by EE-This we must have $T = N$ and Lemma 4.3.17 finishes the case.
- Otherwise, by EE-Var we must have $\Delta(x) = |T|$ and Var finishes the case.

**Case** $$\frac{\Gamma \vdash e_0 : T_0 \qquad \mathsf{vparams}(\mathsf{bound}_\Gamma(T_0)) = \overline{f : T}}{\Gamma \vdash e_0.f_i : T_i}\ (\text{GT-GETTER})$$

We have $|e_0.f_i|_\Gamma = |e_0|_\Gamma.f_i()$. Let $U = \mathsf{bound}_\Gamma(T_0)$. Then,

$$\dfrac{\dfrac{}{\Delta \vdash e_0 : |T_0|}\ (\text{IH}) \qquad \dfrac{\dfrac{\dfrac{}{\Gamma \vdash T_0 <: U}\ (4.2.1)}{\Delta \vdash |T_0| <: |U|}\ (4.3.11) \qquad \dfrac{}{\Delta \vdash |U| <: (f_i() : |T_i|)}\ (4.3.12)}{\Delta \vdash |T_0| <: (f_i() : |T_i|)}\ (\text{TRANS})}{\dfrac{\Delta \vdash e_0 : (f_i() : T_i)}{\Delta \vdash e_0.f_i() : T_i}\ (\text{TAPP'})}\ (\text{SUB})$$

**Case** $$\frac{\begin{array}{c}\Gamma \vdash e_0 : T_0 \qquad \mathsf{mtype}(m,\ \mathsf{bound}_\Gamma(T_0)) = [\overline{Y <: P}] \to (\overline{y : U}) \to U_0 \\ \sigma = [\overline{V/Y}] \qquad \Gamma \vdash \overline{V}\ \mathsf{wf},\ \overline{V <: \sigma P},\ \overline{e : S},\ \overline{S <: \sigma U}\end{array}}{\Gamma \vdash e_0.m[\overline{V}](\overline{e}) : \sigma U_0}\ (\text{GT-INVK})$$

We have $|e_0.m[\overline{V}](\overline{e})|_\Gamma = \textbf{let}\ x_{\mathsf{mtag}} = \{_ \Rightarrow \overline{Y = |V|}\}\ \textbf{in}\ |e_0|_\Gamma.m(x_{\mathsf{mtag}}, \overline{|e|_\Gamma})$. By Lemma 4.3.13 and following a similar reasoning than in the previous case we find

$$\Delta \vdash |e_0|_\Gamma : (m(\mathsf{mtag} : |\overline{Y <: P}|, \overline{y : |U|}) : |U_0|)$$

Let $\tau = [\overline{x_{\mathsf{mtag}}.Y/|Y|}]$ and $\Delta_m = \Delta, x_{\mathsf{mtag}} : \{_ \Rightarrow \overline{Y = |V|}\}$. Note that $|\sigma| = [\overline{|V|/x_{\mathsf{mtag}}.Y}]\tau$ and that we can always weaken $\Delta$ to $\Delta_m$. Then,

$$\dfrac{\dfrac{\dfrac{\dfrac{\dfrac{\dfrac{\dfrac{\dfrac{\dfrac{}{\Delta_m \vdash \overline{|V| <: |\sigma P|}}\ (4.3.11)}{\Delta_m \vdash \overline{|V| <: |\sigma||P|}}\ (4.3.7)}{\Delta_m \vdash \overline{|V| <: \tau|P|}}\ (2.4.6)}{\Delta_m \vdash x_{\mathsf{mtag}} : (Y : \bot\ ..\ \tau|P|)}\ (\text{SUB, TYP})}{\Delta_m \vdash x_{\mathsf{mtag}} : \{\mathsf{mtag} \Rightarrow \overline{Y : \bot\ ..\ |P|}\}}\ (\text{VARPACK})}{\Delta_m \vdash |e_0|_\Gamma.m(x_{\mathsf{mtag}}) : \tau((\overline{y : |U|}) \Rightarrow |U_0|)}\ (\text{TAPPVAR})}{\Delta_m \vdash |e_0|_\Gamma.m(x_{\mathsf{mtag}}) : |\sigma|((\overline{y : |U|}) \Rightarrow |U_0|)}\ (\text{SUB, 2.4.6})}{\Delta_m \vdash |e_0|_\Gamma.m(x_{\mathsf{mtag}}) : ((\overline{y : |\sigma U|}) \Rightarrow |\sigma U_0|)}\ (4.3.7) \qquad \dfrac{}{\Delta_m \vdash \overline{|e|_\Gamma : |\sigma U|}}\ (\text{SUB, IH})}{\Delta_m \vdash |e_0|_\Gamma.m(x_{\mathsf{mtag}}, \overline{|e|_\Gamma}) : |\sigma U_0|}\ (\text{TAPP'})$$

**Case** $\dfrac{\Gamma \vdash C[\overline{T}]\ \mathsf{wf} \quad \mathsf{vparams}(C[\overline{T}]) = \overline{f : U} \quad \Gamma \vdash \overline{e : S}, \overline{S <: U}}{\Gamma \vdash \mathbf{new}\, C[\overline{T}](\overline{e}) : C[\overline{T}]}$ (GT-New)

If $C$ = Object then this follows directly by TNew, Sub and Top. Otherwise, we have $|\mathbf{new}\, C[\overline{T}](\overline{e})|_\Gamma = (\mathbf{let}\ x_{\mathsf{ctag}} = \{_ \Rightarrow \overline{X = |T|}\}\ \mathbf{in}\ \mathsf{ct.new}_C(x_{\mathsf{ctag}}, \overline{|e|_\Gamma}))$. Let $\mathsf{tparams}(C) = \overline{X <: \ldots}$ and $\mathsf{vparams}(C[\overline{X}]) = \overline{f : U'}$. Then we must have $U = \sigma U'$ where $\sigma = [\overline{T/X}]$. It is easy to see that

$$\Delta \vdash \mathsf{ct} : (\mathsf{new}_C(\mathsf{ctag} : |\overline{X <: N}|, \overline{f_{\mathsf{param}} : \tau|U'|}) : \tau|C[\overline{X}]|) \quad \text{where } \tau = [\overline{\mathsf{ctag}.X/|X|}]$$

and the rest of the case proceeds much like the previous case with $\Delta_m = \Delta, x_{\mathsf{ctag}} : \{_ \Rightarrow \overline{X = |T|}\}$.

■

**Theorem 4.3.19: Class entailment implies parent entailment**

Given $\mathbf{class}\, C[\overline{X_C <: N_C}](\ldots) \lhd B[\overline{U}]$, $\mathsf{tparams}(B) = \overline{X_B <: N_B}$, $\Gamma_C = \overline{X_C <: N_C}$, $\mathsf{this} : C[\overline{X_C}]$ and $\Gamma_B = \overline{X_B <: N_B}$, $\mathsf{this} : B[\overline{X_B}]$, then $\Gamma_C \dashv \Delta$ implies $\Gamma_B \dashv \Delta$.

*Proof.* By inversion of $\Gamma_C \dashv \Delta$ via EE-This we have $\Delta_{[\mathsf{this}]} \vdash \mathsf{this} :_{(!)} [\![C]\!]^{|\overline{X}|}$.

$$\dfrac{\dfrac{\overline{\dfrac{}{\Delta \vdash |X_B| <: |N_B|}\ (4.3.8)}}{\overline{X_B <: N_B} \dashv \Delta}\ (\text{EE-Typs}) \qquad \dfrac{\Delta \vdash \mathsf{this} :_{(!)} [\![C]\!]^{|\overline{X_C}|} \quad \dfrac{}{\Delta \vdash [\![C]\!]^{|\overline{X_C}|} <: [\![B]\!]^{|\overline{X_B}|}}\ (4.3.15)}{\Delta \vdash \mathsf{this} :_{(!)} [\![B]\!]^{|\overline{X_B}|}}\ (\text{Sub})}{\Gamma_B \dashv \Delta}\ (\text{EE-This})$$

■

**Lemma 4.3.20: Method translation is well-typed**

Given $\mathsf{tparams}(C) = \overline{X <: N}$, $\Gamma = \overline{X <: N}$, $\mathsf{this} : C[\overline{X}]$, $\Gamma \dashv \Delta$, $\mathsf{mtype}(m, C[\overline{X}]) = [\overline{Y <: P}] \to (\overline{x : U}) \to U_0$ and $\mathsf{mbody}(m, C[\overline{X}]) = e_0$, then $\Delta \vdash (\!| m |\!)_C : [\![m]\!]_C$.

*Proof.* By induction on the derivations of $\mathsf{mtype}(m, C[\overline{X}])$ and $\mathsf{mbody}(m, C[\overline{X}])$.

**Case** $\dfrac{\mathbf{class}\, C[\overline{X <: N}] \ldots \{\overline{M}\} \quad (\mathbf{def}\ m[\overline{Y <: P}](\overline{x : T}) : T_0 = e_0) \in \overline{M}}{\mathsf{mtype}(m, C[\overline{X}]) \coloneqq [\overline{Y <: P}] \to (\overline{x : T}) \to T_0 \quad \mathsf{mbody}(m, C[\overline{X}]) \coloneqq \sigma e_0}$ (GM-Class)

Let $\Gamma_m = \Gamma, \overline{Y <: P}, \overline{x : T}$ and $\Delta_m = \Delta, \mathsf{mtag} : |\overline{Y <: P}|, \overline{x : |T|}$, then $\Gamma_m \dashv \Delta_m$ by EE-Typs. By

inversion, $\vdash C$ ok implies $\Gamma \vdash m$ ok implies $\Gamma_m \vdash e_0 : E_0$, $E_0 <: U_0$ and

$$\dfrac{\dfrac{}{\Delta_m \vdash |e_0|_{\Gamma_m} : |E_0|}\,(4.3.18) \qquad \dfrac{}{\Delta_m \vdash |E_0| <: |U_0|}\,(4.3.11)}{\dfrac{\Delta_m \vdash |e_0|_{\Gamma_m} : |U_0|}{\Delta \vdash (\!| m |\!)_C : [\![ m ]\!]_C}\,(\text{DFun'})}\,(\text{Sub})$$

**Case**
$$\dfrac{\begin{array}{c}\textbf{class}\ C[\overline{X_C <: N_C}](\ldots) \lhd B[\overline{U}]\ \{\overline{M}\} \\ (\textbf{def}\ m \ldots) \notin \overline{M}\end{array}}{\begin{array}{c}\mathsf{mtype}(m, C[\overline{X_C}]) \coloneqq \mathsf{mtype}(m, P) \\ \mathsf{mbody}(m, C[\overline{X_C}]) \coloneqq \mathsf{mbody}(m, P)\end{array}}\,(\text{GM-Super})$$

Let $\mathsf{tparams}(B) = [\overline{X_B <: N_B}]$, $\Gamma_B = \overline{X_B <: N_B}$, $\mathsf{this} : B[\overline{X_B}]$ and $\sigma = [\overline{|U|/|X_B|}]$. By observation and 4.3.7 we must have

$$(\!| m |\!)_C = |\sigma|(\!| m |\!)_B$$
$$[\![ m ]\!]_C = |\sigma|[\![ m ]\!]_B$$

Then,

$$\dfrac{\dfrac{\dfrac{}{\Gamma_B \dashv\vdash \Delta}\,(4.3.19)}{\Delta \vdash (\!| m |\!)_B : [\![ m ]\!]_B}\,(\text{IH}) \qquad \dfrac{}{\Delta \vdash \overline{|X_B| =:= |U|}}\,(4.3.8)}{\Delta \vdash |\sigma|(\!| m |\!)_B : |\sigma|[\![ m ]\!]_B}\,(2.4.7)$$

∎

**Lemma 4.3.21: Class translation is well-typed**

Suppose $\mathsf{tparams}(C) = \overline{X <: N}$ and $\Gamma = (\overline{X <: N}$, $\mathsf{this} : C[\overline{X}])$ and let $|\Gamma| = \Delta$, $\mathsf{this} : [\![ C ]\!]^{\tau|X|}$ where $\tau = [\overline{\mathsf{ctag}.X/|X|}]$. Then, $\Delta \vdash \{\mathsf{this} \Rightarrow (\!| C |\!)^{\tau|X|}\} : \{\mathsf{this} \Rightarrow [\![ C ]\!]^{\tau|X|}\}$.

*Proof.* By TNew, this is true if the following claims are all true.

**Claim 1:** $|\Gamma| \vdash (\!| f : U |\!) : [\![ f : U ]\!] \quad \forall (f : U) \in \mathsf{vparams}(C[\overline{X}])$

By Var, we have $|\Gamma| \vdash \overline{f_{\mathsf{param}} : \tau|U|}$. By Lemma 2.4.6, $|\Gamma| \vdash \overline{f_{\mathsf{param}} : |U|}$ and DFun finishes the claim.

**Claim 2:** $|\Gamma| \vdash (\!| m |\!)_C : [\![ m ]\!]_C \quad \forall m \in \mathsf{mnames}(C)$

By Theorem 4.3.4, $\Gamma \dashv\vdash |\Gamma|$ and Lemma 4.3.20 finishes the claim.

**Claim 3:** $(X_i = \mathsf{ctag}.X_i) : (X_i = \mathsf{ctag}.X_i) \quad \forall X_i \in \overline{X}$

> By DTyp

∎

**Lemma 4.3.22: Class table translation is well-typed**

$\varnothing \vdash_{\text{DOT}} \{\mathsf{ct} \Rightarrow \llparenthesis CT \rrparenthesis\} : \{\mathsf{ct} \Rightarrow \llbracket CT \rrbracket\}$.

*Proof.* After proving the following claims for each **class** $C[\overline{X <: N}](\overline{f : U})$ in $CT$, we can finish the proof by TNew.

**Claim 1:**

$$|\varnothing| \vdash (C = \{\mathsf{this} \Rightarrow \llbracket C \rrbracket, \overline{X : \bot \,..\, |N|}\}) : (C = \{\mathsf{this} \Rightarrow \llbracket C \rrbracket, \overline{X : \bot \,..\, |N|}\})$$

> By DTyp.

**Claim 2:**

$$|\varnothing| \vdash (\mathsf{new}_C(\mathsf{ctag}, \overline{f_{\mathsf{param}}}) = \{\mathsf{this} \Rightarrow \llparenthesis C[\overline{X}] \rrparenthesis\}) : (\mathsf{new}_C(\mathsf{ctag} : \{\mathsf{this} \Rightarrow \overline{X : \bot \,..\, |N|}\}, \overline{f_{\mathsf{param}} : \tau|U|}) : \tau|C[\overline{X}]|)$$

where $\tau = [\overline{\mathsf{ctag}.X/|X|}]$.

> Let $\Delta = |\varnothing|, \mathsf{ctag} : \{\mathsf{this} \Rightarrow \overline{X : \bot \,..\, |N|}\}, \overline{f_{\mathsf{param}} : \tau|U|}$. Then,
>
> $$\dfrac{\dfrac{}{\Delta \vdash \{\mathsf{this} \Rightarrow \llparenthesis C \rrparenthesis^{\tau|\overline{X}|}\} : \{\mathsf{this} \Rightarrow \llbracket C \rrbracket^{\tau|\overline{X}|}\}}\,(4.3.21) \qquad \dfrac{\dfrac{\dfrac{\dfrac{\overline{\dfrac{}{|\overline{X <: N}, \mathsf{this} : C[\overline{X}]| \vdash \mathsf{ctag}.X <: |N|}}\,(\textsc{Sel1}, \textsc{Var})}{\overline{|\overline{X <: N}, \mathsf{this} : C[\overline{X}]| \vdash (X = \tau|X|) <: (X : \bot \,..\, |N|)}}\,(\textsc{Typ})}{\Delta \vdash \{\mathsf{this} \Rightarrow \llbracket C \rrbracket^{\tau|\overline{X}|}\} <: \{\mathsf{this} \Rightarrow \llbracket C \rrbracket, \overline{X : \bot \,..\, |N|}\}}\,(\textsc{BindX})}{\Delta \vdash \{\mathsf{this} \Rightarrow \llbracket C \rrbracket^{\tau|\overline{X}|}\} <: \mathsf{ct}.C}\,(\textsc{Trans})}{\Delta \vdash \{\mathsf{this} \Rightarrow \llbracket C \rrbracket^{\tau|\overline{X}|}\} <: \tau|C[\overline{X}]|}\,(\textsc{And2}, \textsc{Bind1})}{\Delta \vdash \{\mathsf{this} \Rightarrow \llparenthesis C \rrparenthesis^{\tau|\overline{X}|}\} : \tau|C[\overline{X}]|}\,(\textsc{Sub})$$
>
> And DFun' finishes the case.

∎

**Theorem 4.3.23: Program translation is type-preserving**

If $\varnothing \vdash_{\text{FGJ}} T$ wf and $\varnothing \vdash_{\text{FGJ}} e : T$ then $\varnothing \vdash_{\text{DOT}}$ **let** $\mathsf{ct} = \{\mathsf{ct} \Rightarrow \llparenthesis CT \rrparenthesis\}$ **in** $|e|_{\varnothing} : |T|$.

*Proof.*

$$\dfrac{\dfrac{}{\varnothing \vdash \{\mathsf{ct} \Rightarrow (\!|\, CT \,|\!)\} : \{\mathsf{ct} \Rightarrow [\![ CT ]\!]\}}\ (4.3.22) \qquad \dfrac{\dfrac{\varnothing \vdash e : T}{\mathsf{ct} : [\![ CT ]\!] \vdash |e|_{\varnothing} : |T|}\ (4.3.18)}{\mathsf{ct} : \{\mathsf{ct} \Rightarrow [\![ CT ]\!]\} \vdash |e|_{\varnothing} : |T|}\ (\textsc{EnvPackTp})}{\varnothing \vdash \textbf{let}\ \mathsf{ct} = \{\mathsf{ct} \Rightarrow (\!|\, CT \,|\!)\}\ \textbf{in}\ |e|_{\varnothing} : |T|}\ (\textsc{Let})$$

■

# 5 Pathless Scala

In this chapter, we present Pathless Scala (PS).[1] PS extends cast-less FGJ with multiple inheritance via traits and with intersection types in the style of DOT. As its name indicate, PS lacks path-dependent types and can thus be seen as a stepping stone on the way to Dependent Scala in Chapter 7. To develop a type-preserving translation scheme from PS to DOT we once again need to extend DOT, this time with a new typing rule And-I'. In the process of proving the extended DOT sound, we end up having to generalize the definition of type soundness used in [Rompf and Amin 2016, Theorem 1] which did not imply the usual property of *preservation*.

## 5.1 Syntax

**Figure 5.1: PS: Syntax**

| | |
|---|---|
| $x, y, z$ | Variable |
| $B, C, D, E$ | Class name |
| $f, g$ | Class parameter |
| $m$ | Method name |
| $X_C$ | Class variable |
| $X_m$ | Method variable |
| $X, Y, Z ::= X_C \mid X_m$ | Type variable |
| $N, P, Q ::= C[\overline{T}]$ | Non-variable |
| $S, T, U, V ::=$ | Type |
| $\quad X \mid N \mid S \,\&\, T$ | |
| $\Gamma ::=$ | Context |
| $\quad \varnothing \mid \Gamma, x : T \mid \Gamma, \overline{X <: N}$ | |
| $L ::=$ | Class declaration |
| $\quad$ **class** $C[\overline{X_C <: N}](\overline{f : T}) \lhd P(\overline{f}), \overline{Q}\ \{\overline{M}\}$ | |
| $\quad$ **trait** $C[\overline{X_C <: N}] \lhd \overline{Q}\ \{\overline{H}; \overline{M}\}$ | |
| $H ::=$ | Abstract method |
| $\quad$ **def** $m[\overline{X_m <: N}](\overline{x : T}) : T_0$ | |
| $M ::=$ | Concrete method |
| $\quad H = e_0$ | |
| $e ::=$ | Expression |
| $\quad x$ | variable |
| $\quad e.f$ | parameter access |
| $\quad e_0.m[\overline{T}](\overline{e})$ | method call |
| $\quad$ **new** $C[\overline{T}](\overline{e})$ | object |
| $\sigma, \tau ::= [\overline{T/X}]$ | Type substitution |

We call $S \,\&\, T$ the *intersection* of $S$ and $T$.

[1] Part of this chapter is revised and extended from [Martres 2021].

A PS class is either a *proper* class (declared using the keyword "**class**") or a trait (declared using the keyword "**trait**"). Proper classes must extend exactly one other proper class as before, but both proper classes and traits can extend zero, one or many traits. Traits cannot extend proper classes syntactically but are semantically considered subtypes of Object.[2] Compared to proper classes, traits do not have constructor parameters[3] and cannot be constructed using **new**, but they can have methods declared without a body which we call *abstract* and which must be implemented in sub-classes of the traits. For convenience, we define in Figure 5.2 lookup functions returning the parents and the method declarations of either classes or traits as well as functions used to determine whether a given class name $C$ corresponds to a proper class or trait.

**Figure 5.2: PS: Lookup functions (part 1)**

**Parent classes** $\boxed{\mathsf{parents}(N) = \overline{P}}$

$$\mathsf{parents}(\mathsf{Object}) \coloneqq \varnothing$$

$$\frac{\textbf{class}\ C \lhd P(\ldots), \overline{Q}\ \{\overline{M}\} \quad \sigma = [\overline{T/X}]}{\mathsf{parents}(C[\overline{T}]) \coloneqq \sigma P, \sigma\overline{Q}}$$

$$\frac{\textbf{trait}\ C[\overline{X <: N}] \lhd \overline{P}\ \{\overline{H}; \overline{M}\} \quad \sigma = [\overline{T/X}]}{\mathsf{parents}(C[\overline{T}]) \coloneqq \mathsf{Object}, \sigma\overline{P}}$$

**$C$ is a proper class** $\boxed{\mathsf{isProperClass}(C)}$

$$\frac{\textbf{class}\ C \ldots}{\mathsf{isProperClass}(C)}$$

**Method declarations** $\boxed{\mathsf{mdecls}(N) = \overline{M}}$

$$\mathsf{mdecls}(\mathsf{Object}) \coloneqq \varnothing$$

$$\frac{\textbf{class}\ C \lhd P(\ldots), \overline{Q}\ \{\overline{M}\} \quad \sigma = [\overline{T/X}]}{\mathsf{mdecls}(C[\overline{T}]) \coloneqq \sigma\overline{M}}$$

$$\frac{\textbf{trait}\ C[\overline{X <: N}] \lhd \overline{P}\ \{\overline{H}; \overline{M}\} \quad \sigma = [\overline{T/X}]}{\mathsf{mdecls}(C[\overline{T}]) \coloneqq \sigma\overline{H}, \sigma\overline{M}}$$

**$C$ is a trait** $\boxed{\mathsf{isTrait}(C)}$

$$\frac{\textbf{trait}\ C \ldots}{\mathsf{isTrait}(C)}$$

## 5.2 Subtyping and well-formedness

The subtyping rules for intersections in Figure 5.3 mirror the DOT rules And11, And12 and And2 such that the subtyping relationship defined by these rules induces a partial order in which $T_1$ **&** $T_2$ is the *greatest lower bound* of $T_1$ and $T_2$. The introduction of intersection types means that syntactically distinct types can now be mutual subtypes like $T$ and $T$ **&** $T$. This motivates an additional rule PS-Inv which lets us relate $C[T]$ with $C[T\ \&\ T]$.

Without surprise, WFP-And (in Figure 5.4) considers an intersection type to be well-formed if both of its operands are well-formed.

[2] This is a restriction from real Scala where a trait may explicitly extend a class.
[3] Scala used to have the same restriction until Scala 3: [Odersky et al. 2022].

**Figure 5.3: PS: Subtyping**

$\boxed{\Gamma \vdash S <: T}$

GS-Refl and GS-Trans are carried over from Figure 4.2.

$$\frac{P \in \mathsf{parents}(C[\overline{T}])}{\Gamma \vdash C[\overline{T}] <: [\overline{T/X}]P} \quad \text{(PS-Class)}$$

$$\frac{\Gamma \vdash \overline{S <: T}, \overline{T <: S}}{\Gamma \vdash C[\overline{S}] <: C[\overline{T}]} \quad \text{(PS-Inv)}$$

$$\frac{\Gamma \vdash S_1 <: T}{\Gamma \vdash S_1 \;\&\; S_2 <: T} \quad \text{(PS-And11)}$$

$$\frac{\Gamma \vdash S_2 <: T}{\Gamma \vdash S_1 \;\&\; S_2 <: T} \quad \text{(PS-And12)}$$

$$\frac{\Gamma \vdash S <: T_1, S <: T_2}{\Gamma \vdash S <: T_1 \;\&\; T_2} \quad \text{(PS-And2)}$$

**Figure 5.4: PS: Well-formedness**

**Well-formed type** $\boxed{\Gamma \vdash T \ \mathsf{wf}}$

We extend Figure 4.3 with:

$$\frac{\Gamma \vdash T_1, T_2 \ \mathsf{wf}}{\Gamma \vdash T_1 \;\&\; T_2 \ \mathsf{wf}} \quad \text{(WFP-And)}$$

## 5.3 Typing

### 5.3.1 Expression typing

The expression typing rules from FGJ (Figure 4.6) can be carried over as-is, only the helper functions need to be generalized (in Figure 5.5) to handle intersection types.

#### Generalizing bound

$\mathsf{bound}_\Gamma(T)$ is still defined to return a non-variable upper-bound of $T$, but now this upper-bound is allowed to be an intersection of applied class types. This requires generalizing both vparams and mtype.

#### Generalizing vparams

G-Object and G-Class can be carried over without changes.

**Figure 5.5: PS: Lookup functions (part 2)**

The definitions from Figure 4.5 are carried over.

**Non-variable upper bound of type** $\boxed{\mathsf{bound}_\Gamma(T) \coloneqq \& \overline{N}}$

$$\mathsf{bound}_\Gamma(S \,\&\, T) \coloneqq \mathsf{bound}_\Gamma(S) \,\&\, \mathsf{bound}_\Gamma(T) \qquad \text{(B-AND)}$$

**Type parameters lookup** $\boxed{\mathsf{tparams}(C) \coloneqq \overline{X <: N}}$

$$\frac{\textbf{trait}\ C[\overline{X <: N}]\ \ldots}{\mathsf{tparams}(C) \coloneqq \overline{X <: N}}$$

**Value parameters lookup** $\boxed{\mathsf{vparams}(T) \coloneqq \overline{f : T}}$

$$\frac{\mathsf{isTrait}(N)}{\mathsf{vparams}(N) \coloneqq \varnothing} \qquad \text{(PG-TRAIT)}$$

$$\frac{\mathsf{vparams}(T_2) \subseteq \mathsf{vparams}(T_1)}{\mathsf{vparams}(T_1 \,\&\, T_2) \coloneqq \mathsf{vparams}(T_1)} \ \text{(PG-ANDL)} \qquad \frac{\mathsf{vparams}(T_1) \subseteq \mathsf{vparams}(T_2)}{\mathsf{vparams}(T_1 \,\&\, T_2) \coloneqq \mathsf{vparams}(T_2)} \ \text{(PG-ANDR)}$$

**Method type lookup** $\boxed{\mathsf{mtype}(m, T) \coloneqq [\overline{Y <: P}] \to (\overline{x : T}) \to T_0}$

$$\frac{(\textbf{def}\ m[\overline{Y <: P}](\overline{x : U}) : U_0 = e_0) \in \mathsf{mdecls}(C[\overline{T}])}{\mathsf{mtype}(m, C[\overline{T}]) \coloneqq [\overline{Y <: P}] \to (\overline{x : U}) \to U_0} \qquad \text{(PM-IMPL)}$$

$$\frac{\mathsf{parents}(N) = \overline{P} \qquad (\textbf{def}\ m\ \ldots) \notin \mathsf{mdecls}(N)}{\mathsf{mtype}(m, N) \coloneqq \mathsf{mtype}(m, \& \overline{P})} \qquad \text{(PM-SUPER)}$$

$$\frac{\begin{array}{c}\mathsf{mtype}(m, T_1) = [\overline{Y <: P}] \Rightarrow (\overline{x : S}) \Rightarrow V_1 \\ \mathsf{mtype}(m, T_2) = [\overline{Y <: P}] \Rightarrow (\overline{x : S}) \Rightarrow V_2\end{array}}{\mathsf{mtype}(m, T_1 \,\&\, T_2) \coloneqq [\overline{Y <: P}] \Rightarrow (\overline{x : S}) \Rightarrow V_1 \,\&\, V_2} \qquad \text{(PM-ANDLR)}$$

$$\frac{\begin{array}{c}\mathsf{mtype}(m, T_1)\ \text{defined} \\ \mathsf{mtype}(m, T_2)\ \text{undefined}\end{array}}{\mathsf{mtype}(m, T_1 \,\&\, T_2) \coloneqq \mathsf{mtype}(m, T_1)} \ \text{(PM-ANDL)} \qquad \frac{\begin{array}{c}\mathsf{mtype}(m, T_1)\ \text{undefined} \\ \mathsf{mtype}(m, T_2)\ \text{defined}\end{array}}{\mathsf{mtype}(m, T_1 \,\&\, T_2) \coloneqq \mathsf{mtype}(m, T_2)} \ \text{(PM-ANDR)}$$

PG-Trait reflects the fact that traits cannot have value parameters.

PG-AndL and PG-AndR assume that in an intersection, the value parameters of one of the two operands will be a subset of the value parameters of the other. This makes sense since traits cannot have value parameters and PS does not allow inheriting from multiple unrelated classes. While it is possible to construct an intersection type where the operands are unrelated classes, no value of such a type exists, so leaving vparams undefined in that case is not an issue.

#### Generalizing mtype

Given x : L & R and the class table:

```
trait L { def foo(): A }
trait R { def foo(): B }
```

What is the type of x.foo()? In Java this would be an error, even though it is possible to construct a class that override both of these methods via covariant overriding. The problem is that there is no Java type representing the greatest lower bound of A and B, whereas as we've seen above in Scala this is simply A & B. This motivates the definiton of PM-AndLR. It is completed by PM-AndL and PM-AndR which handle the easy cases where the method is only defined on one side of the intersection.

GM-Super is replaced by PM-Super which handles multiple parents, and GM-Class is replaced by PM-Impl which handles both proper classes and traits.

### 5.3.2 Declaration typing

#### Abstract methods in proper classes

Methods in a proper class can either be declared in the class or inherited. The syntax of proper classes forces declared methods to be concrete, but methods inherited from a trait may be abstract. One might assume that a method is considered abstract in a class if there are only abstract declarations of this method among its base types. However, both Java and Scala 3 allow "re-abstracting" a method. For example in,

```
trait Base { def foo(): Object = ... }
trait Sub ◁ Base { def foo(): Object }
class A ◁ Object, Sub {}
class B ◁ Object, Base, Sub {}
```

A and B have the same linearization so we'd expect them to be equivalent, but in fact an inherited method is considered abstract in a class if it is abstract among all the direct parents of this class, so A is not well-formed since it only inherits an abstract foo from Sub.

To model this, we define the mutually recursive $\mathsf{mnames}_{con}(N)$ and $\mathsf{mnames}_{abs}(N)$ in Figure 5.6 to be the sets of names of respectively concrete and abstract members of $N$. PT-Class in Figure 5.7 then takes care of checking that $\mathsf{mnames}_{abs}$ is empty for proper classes.

**Figure 5.6: PS: Lookup functions (part 3)**

**Concrete and abstract method names lookup** $\boxed{\mathsf{mnames}(C) := \overline{m}}$

$$\frac{\overline{P} = \mathsf{parents}(N) \qquad \mathsf{mdecls}(N) = \overline{\mathbf{def}\ m_{abs}\ ...};\ \overline{\mathbf{def}\ m_{con}\ ... = ...}}{\begin{array}{l}\mathsf{mnames}_{con}(N) := \overline{m_{con}} \cup (\overline{\mathsf{mnames}_{con}(P)} \setminus \overline{m_{abs}}) \\ \mathsf{mnames}_{abs}(N) := \overline{m_{abs}} \cup (\overline{\mathsf{mnames}_{abs}(P)} \setminus \mathsf{mnames}_{con}(P))\end{array}}$$

**Method names lookup** $\boxed{\mathsf{mnames}(C) := \overline{m}}$

$$\mathsf{mnames}(N) := \mathsf{mnames}_{abs}(N) \cup \mathsf{mnames}_{con}(N)$$

**Figure 5.7: PS: Typing rules**

The expression typing rules from Figure 4.6 are carried over.

**Method typing** $\boxed{\Gamma \vdash m \text{ ok}}$

$$\frac{\begin{array}{c}\Gamma = \overline{X <: N}, \mathsf{this} : C[\overline{X}] \\ \mathsf{mtype}(m, C[\overline{X}]) = [\overline{Y <: P}] \to (\overline{x : U}) \to U_0 \\ \Gamma, \overline{Y <: P} \vdash \overline{U}, U_0, \overline{P} \text{ wf} \\ \mathsf{mbody}(m, C[\overline{X}]) = e_0 \text{ implies } \Gamma, \overline{Y <: P}, \overline{x : U} \vdash e_0 : E_0, E_0 <: U_0 \\ Q \in \mathsf{parents}(C[\overline{X}]) \text{ implies } \mathsf{override}_\Gamma(m, C[\overline{X}], Q)\end{array}}{\Gamma \vdash m \text{ ok}} \quad \text{(PT-Method)}$$

**Class typing** $\boxed{\vdash C \text{ ok}}$

$$\frac{\begin{array}{c}\mathbf{class}\ C[\overline{X <: N}](\overline{g : U}, \overline{f : T}) \lhd P(\overline{g}), \overline{Q}\ \{\overline{\mathbf{def}\ m\ ...}\} \\ \mathcal{L}(C[\overline{X}]) \text{ defined} \quad \mathsf{isProperClass}(P) \quad \overline{\mathsf{isTrait}(Q)} \\ \Gamma = \overline{X <: N}, \mathsf{this} : C[\overline{X}] \\ \Gamma \vdash \overline{N}, \overline{U}, \overline{T}, P, \overline{Q} \text{ wf} \quad \Gamma \vdash \overline{m} \text{ ok} \quad \mathsf{vparams}(P) = \overline{g : U} \\ \mathsf{mnames}_{abs}(C) = \varnothing \quad m' \in \mathsf{mnames}(C) \text{ implies } \mathsf{isValid}_\Gamma(m')\end{array}}{\vdash C \text{ ok}} \quad \text{(PT-Class)}$$

$$\frac{\begin{array}{c}\mathbf{trait}\ C[\overline{X <: N}] \lhd \overline{Q}\ \{\overline{\mathbf{def}\ m\ ...}\} \\ \mathcal{L}(C[\overline{X}]) \text{ defined} \quad \overline{\mathsf{isTrait}(Q)} \\ \Gamma = \overline{X <: N}, \mathsf{this} : C[\overline{X}] \\ \Gamma \vdash \overline{N}, \overline{Q} \text{ wf} \quad \Gamma \vdash \overline{m} \text{ ok} \\ m' \in \mathsf{mnames}(C) \text{ implies } \mathsf{isValid}_\Gamma(m')\end{array}}{\vdash C \text{ ok}} \quad \text{(PT-Trait)}$$

### Linearization and method implementer

The *base types* of a class are determined by the reflexive transitive closure of the parents function. With Scala traits, unlike Java interfaces, the *order* in which they are inherited matters. Since the same trait may be indirectly inherited multiple times, Scala defines a canonical order of the base types of a class called its *linearization*.

[Odersky and Zenger 2005] defines linearization for class names $C$, but we find it more convenient to generalize it to applied class types $N$:

$$\frac{N_1,\ \ldots,\ N_n = \mathsf{parents}(N)}{\mathcal{L}(N) \coloneqq N,\ \mathcal{L}(N_n) \vec{+} \ldots \vec{+} \mathcal{L}(N_1)}$$

Where $\vec{+}$ denotes concatenation with elements on the right replacing identical elements of the left operand. It is illegal to inherit the same class twice if it is applied to different type arguments[4] and so we leave $\vec{+}$ undefined in that case:

$$\varnothing \vec{+} \overline{N} \coloneqq \overline{N}$$

$$\frac{N_0 \in \overline{N_r}}{(N_0, \overline{N_l}) \vec{+} \overline{N_r} \coloneqq \overline{N_l} \vec{+} \overline{N_r}}$$

$$\frac{N_0 = C_0[\ldots] \quad \overline{N_r} = \overline{C_r[\ldots]} \quad C_0 \notin \overline{C_r}}{(N_0, \overline{N_l}) \vec{+} \overline{N_r} \coloneqq N_0, (\overline{N_l} \vec{+} \overline{N_r})}$$

PT-Class and PT-Trait ensure that $\mathcal{L}$ is defined on all well-formed types.

We will use linearization to determine which base type of $N$ contains the implementation of $m$ that will be called at runtime which we dub the *implementer* of $m$ in $N$ written $\mathsf{mimpl}(m, N)$ which we use to redefine mbody (contrast with Figure 4.5):

$$\frac{(\mathbf{def}\ m[\overline{Y <: P}](\overline{x : U}) : U_0 = e_0) \in \mathsf{mdecls}(\mathsf{mimpl}(m, N))}{\mathsf{mbody}(m, N) \coloneqq e_0} \qquad \text{(PMB-All)}$$

We motivate the definition of mimpl with an example. Consider the following class table:

```
class One {}; class Two {}
trait Base { def foo(): Object }
trait Sub1 ◁ Base { def foo(): Object = new One }
trait Sub2 ◁ Base { def foo(): Object = new Two }
class A ◁ Object, Sub1, Sub2
```

[4] In real Scala this is in fact possible with variant type parameters. Even with invariant type parameters, we could allow $C[T]$ as well as $C[T \,\&\, T]$ but this would require taking the environment as input in the definition of $\vec{+}$ to do subtyping checks. This would complicate our presentation for little benefits.

The equivalent class table in Java (using **`interface`** instead of **`trait`**) would be illegal: both `Sub1` and `Sub2` contain a concrete implementation of `foo` and neither trait overrides the other. But this is legal Scala[5] and `(`**`new`**` A).foo()` will evaluate to **`new`**` Two()` because `Sub2` precedes `Sub1` in the linearization of A.

In general, concrete methods override abstract methods in both Java and Scala, but if we compare a concrete method $M$ defined in $C$ with another concrete method $M'$ defined in $D$ then:

- In Java, $M$ overrides $M'$ if $D$ is a base type of $C$.
- In Scala, $M$ overrides $M'$ **in** $\boldsymbol{N}$ if $C$ precedes $D$ **in** $\mathcal{L}(\boldsymbol{N})$. Since a type $P$ will always appear before its parent in any linearization involving $P$, this generalizes the Java rule.

Based on this specification, we can define mimpl as:

$$\mathsf{mimpl}(m,\ N) := \mathsf{mimpl}'(m,\ \mathcal{L}(N))$$

$$\mathsf{mimpl}'(m,\ (N, \overline{P})) := \begin{cases} N & \text{if } (\mathbf{def}\ m \ldots = \ldots) \in \mathsf{mdecls}(m,\ N_1) \\ \mathsf{mimpl}'(m,\ \overline{P}) & \text{otherwise.} \end{cases}$$

In the example above we have $\mathcal{L}$(`A`) = `A`, `Sub2`, `Sub1`, `Object` and so we find mimpl(`foo`, `A`) = `Sub2` as expected.

### Valid overrides

For a class $C$ to be well-typed, it is not enough for `mimpl` to be defined for all its members, we must also check that the implementations chosen are *valid* overrides. As in FGJ, a valid override must *match* the type of all the methods with the same name in its base types, meaning the type and term parameters must be equal (up to $\alpha$-renaming) and the result type is allowed to vary covariantly. But on top of that, the override must not be *accidental*, a concept specific to Scala illustrated in the following example.

This class table is not well-typed in Scala:

```
class One {}; class Two {}
trait Base { def foo(): Object }
trait Sub1 ◁ Base { def foo(): Object = ... }
trait Unrelated { def foo(): Object }
trait Sub2 ◁ Unrelated { def foo(): Object = ... }
class A ◁ Object, Sub1, Sub2
```

Although we have mimpl(`foo`, `A`) = `Sub2` and $\mathsf{override}_\Gamma(m,$ `Sub2`, `Sub1`) defined, the compiler

[5] To be precise, `foo` in `Sub2` needs to be declared with the **`override`** keyword for `A` to be well-typed, but we do not model this in our calculus: when translating code from PS into real Scala, **`override`** should be added everywhere it is legal to do so as determined by the Scala Language Specification [Odersky et al. 2021a, § 5.2.3].

complains[6]:

```
method foo in trait Sub2 cannot override a concrete member without
a third member that's overridden by both (this rule is designed to
prevent “accidental overrides”)
```

In other words, when $N$ overrides a concrete member $m$ defined in $P$, we must ensure that $N$ and $P$ have a common base type which also declares $m$ as specified by noAccidentalOverride in Figure 5.8.

**Figure 5.8: PS: Overriding**

$\mathsf{override}_\Gamma$ is carried over from Figure 4.4.

**$m$ is valid in $\Gamma$** $\quad$ $\boxed{\mathsf{isValid}_\Gamma(m)}$

$$\frac{\begin{array}{l}\Gamma = \overline{X <: N}, \mathsf{this} : C[\overline{X}] \\ P = \mathsf{mimpl}(m, C[\overline{X}]) \\ Q \in \mathcal{L}(C[\overline{X}]) \text{ implies:} \\ \quad \bullet\ \mathsf{override}_\Gamma(m, P, Q) \\ \quad \bullet\ \mathsf{noAccidentalOverride}(m, P, Q)\end{array}}{\mathsf{isValid}_\Gamma(m)}$$

$$\frac{\begin{array}{l}\Gamma = \overline{X <: N}, \mathsf{this} : C[\overline{X}] \\ \mathsf{mimpl}(m, C[\overline{X}]) \text{ undefined} \\ Q \in \mathcal{L}(C[\overline{X}]) \text{ implies:} \\ \quad \bullet\ \mathsf{override}_\Gamma(m, C[\overline{X}], Q)\end{array}}{\mathsf{isValid}_\Gamma(m)}$$

**$m$ in N does not accidentally override $m$ in P** $\quad$ $\boxed{\mathsf{isValid}_\Gamma(m)}$

$$\frac{\begin{array}{c}\mathsf{mimpl}(m, P) \text{ defined} \\ m \in \mathsf{mnames}(Q) \text{ for some } Q \in \mathcal{L}(N) \cap \mathcal{L}(P)\end{array}}{\mathsf{noAccidentalOverride}(m, N, P)}$$

$$\frac{\mathsf{mimpl}(m, Q) \text{ undefined}}{\mathsf{noAccidentalOverride}(m, N, Q)}$$

## 5.4 Meta-theory

Lemmas 4.2.2 to 4.2.4 easily carry over to Pathless Scala. Lemma 4.2.1 also carries over with a slightly different statement to account for the different result type of bound:

**Lemma 5.4.1: Correctness of bound**

If $\mathsf{bound}_\Gamma(S) = T$, then $\Gamma \vdash S <: T$.

*Proof.* By induction on the derivation of $\mathsf{bound}_\Gamma(S)$. We only show the additional case compared to Lemma 4.2.1.

[6] after adding **`override`** to the definition of `foo` in `Sub2`

**Case** $\mathsf{bound}_\Gamma(S_1 \mathbin{\&} S_2) \coloneqq \mathsf{bound}_\Gamma(S_1) \mathbin{\&} \mathsf{bound}_\Gamma(S_2)$ (B-And)

> We have $\mathsf{bound}_\Gamma(S_1 \mathbin{\&} S_2) = T_1 \mathbin{\&} T_2$. By the IH, $\Gamma \vdash S_1 <: T_1$ and $\Gamma \vdash S_2 <: T_2$. Lemma 2.4.5 finishes the case.

■

## 5.5 Translation

We extend the translation scheme from Section 4.3 to support intersection types, traits, and multiple inheritance in Figure 5.9.

Type translation is easy: PS intersections map directly onto DOT intersections and the existing rule for applied class type TR-Class does not need to be changed to handle traits. Expression translation does not require any change to the existing rules from Figure 4.7.

Unlike with proper classes, we do not define a declaration translation $(\!| C |\!)$ for traits: this isn't necessary since traits do not have constructors and the translation already takes care of copying over inherited method bodies. Instead, we manually define $[\![C]\!]$ for traits which requires a corresponding definition of $[\![m]\!]_C$.

To represent multiple inheritance, we generalize the class table translation to keep track of all parents $\overline{B[...]}$ of a class $C$ in its type tag via an intersection: $\mathsf{ct}.C = (\bigwedge \mathsf{ct}.B \wedge ...)$. We similarly generalize $\mathsf{baseArgs}(N)$ to handle multiple parents.

### 5.5.1 Required addition to DOT

Recall our example from subsection 5.3.1:

```
trait L { def foo(): A }
trait R { def foo(): B }
```

We defined mtype such that if $\Gamma = x : \mathsf{L} \mathbin{\&} \mathsf{R}$, then $\Gamma \vdash x.\mathtt{foo}() : \mathsf{A} \mathbin{\&} \mathsf{B}$. If typing preservation holds, we should thus be able to derive $|\Gamma| \vdash x.\mathtt{foo}() : |\mathsf{A}| \wedge |\mathsf{B}|$. Using the same approach as in Lemma 4.3.13, we can see that,

$$|\Gamma| \vdash |\mathsf{L}| <: (\mathtt{foo}() : |\mathsf{A}|)$$
$$|\Gamma| \vdash |\mathsf{R}| <: (\mathtt{foo}() : |\mathsf{B}|)$$

Intuitively, we would then like to conclude that $|\Gamma| \vdash |\mathsf{L}| \wedge |\mathsf{R}| <: (\mathtt{foo}() : |\mathsf{A}| \wedge |\mathsf{B}|)$ but DOT lacks a subtyping rule that would let us distribute the intersection type inside the method type and we have not been able to extend the existing DOT mechanization with such a rule. We conjecture that DOT can be extended with such a rule since it is standard in type systems with intersection types [Barendregt, Coppo, and Dezani-Ciancaglini 1983]. We will discuss missing subtyping rules in DOT in more details in subsection 8.1.2.

Thankfully, all hope is not a lost: we can take inspiration from wfDOT and try to compensate

**Figure 5.9: Translating PS types, expressions and definitions to DOT**

All definitions from Figure 4.7 are carried over.
Getter, method, class and environment translation from Figure 4.8 are carried over.

**Type Translation** $\boxed{|T| \coloneqq T_{\text{DOT}}}$

$$|T_1 \mathbin{\&} T_2| \coloneqq |T_1| \wedge |T_2|$$

**Trait Method Translation** $\boxed{[\![m]\!]_C \coloneqq T}$

$$\frac{\textbf{trait}\ C[\overline{X <: N}] \ldots \qquad \mathsf{mtype}(m, C[\overline{X}]) = [\overline{Y <: P}] \to (\overline{x : U}) \to U_0}{[\![m]\!]_C \coloneqq m(\mathsf{mtag} : |\overline{Y <: P}|, \overline{x : |U|}) : |U_0|}$$

**Trait Translation** $\boxed{[\![C]\!] \coloneqq T}$

$$\frac{\textbf{trait}\ C[\overline{X <: N}] \ldots}{\begin{array}{l}[\![C]\!] \ \coloneqq [\![\mathsf{mnames}(C)]\!]_C \wedge \mathsf{baseArgs}(C) \\ [\![C]\!]^{\overline{T}} \coloneqq [\![C]\!] \wedge \{_ \Rightarrow \overline{X = T}\}\end{array}}$$

**Class Table Translation** $\boxed{(\!|\,CT\,|\!) \coloneqq \overline{d_{\text{DOT}}}}$

$$(\!|\,\varnothing\,|\!) \coloneqq (\mathsf{Object} = \top)$$

$$\frac{L_C = \textbf{class}\ C[\overline{X_C <: N}](\overline{f : U}) \lhd B[\ldots], \overline{D[\ldots]} \qquad \tau = [\overline{\mathsf{ctag}.X_C / |X_C|}]}{\begin{array}{l}(\!|\,\overline{L}, L_C\,|\!) \coloneqq (\!|\,\overline{L}\,|\!), C = \mathsf{ct}.B \wedge \bigwedge \overline{\mathsf{ct}.D} \wedge \{\mathsf{this} \Rightarrow [\![C]\!], \overline{X_C : \bot \,..\, |N|}\}, \\ \mathsf{new}_C(\mathsf{ctag} : |\overline{X_C <: N}|, \overline{f_{\mathsf{param}} : \tau|U|}) : \tau|C[\overline{X_C}]| = \{\mathsf{this} \Rightarrow (\!|\,C\,|\!)^{\tau|\overline{X_C}|}\}\end{array}}$$

$$\frac{L_C = \textbf{trait}\ C[\overline{X <: N}] \lhd \overline{B[\ldots]}}{(\!|\,\overline{L}, L_C\,|\!) \coloneqq (\!|\,\overline{L}\,|\!), C = \bigwedge \overline{\mathsf{ct}.B} \wedge \{\mathsf{this} \Rightarrow [\![C]\!], \overline{X : \bot \,..\, |N|}\}}$$

**Arguments of Base Types** $\boxed{\mathsf{baseArgs}(C) \coloneqq T_{\text{DOT}}}$

$$\frac{\mathsf{parents}(N) = \overline{B[\overline{S}]} \qquad \overline{\mathsf{tparams}(B) = \overline{X <: \ldots}}}{\mathsf{baseArgs}(N) \coloneqq \bigwedge \overline{\overline{X = |S|},\ \mathsf{baseArgs}(B[\overline{X}])}}$$

weak subtyping rules by stronger typing rules: it is easy to show that $|\Gamma| \vdash x.\texttt{foo}() : |\mathsf{A}|$ and $|\Gamma| \vdash x.\texttt{foo}() : |\mathsf{B}|$, so we should be able to deduce $|\Gamma| \vdash x.\texttt{foo}() : |\mathsf{A}| \wedge |\mathsf{B}|$.

Recall that wfDOT, unlike oopslaDOT, defines the following rule:

$$\frac{\Gamma \vdash x : T \quad \Gamma \vdash x : U}{\Gamma \vdash x : T \wedge U} \tag{And-I}$$

This isn't quite what we want: this rule only applies to variable $x$ so it won't help us give a more precise type to $x.\texttt{foo}()$, but it's a step in the right direction and it turns out to be relatively easy to add to the mechanization.[7] What we really need is[8]

$$\frac{\Gamma \vdash t : T \quad \Gamma \vdash t : U}{\Gamma \vdash t : T \wedge U} \tag{And-I'}$$

Perhaps surprisingly, adding this rule to the existing mechanization is much more challenging. To understand why, we must first briefly describe the operational semantics in [Rompf and Amin 2016, Figure 2].

The syntax of the calculus is extended with concrete variables $y$ and stores $\rho = \overline{y : d}$ mapping concrete variables to declarations. The store typing relation $\rho\ \Gamma \vdash t : T$ extends the regular typing relation $\Gamma \vdash t : T$ with an extra rule to ascribe a type to $y$ based on the value $\rho(y)$. The small-step reduction relation $\rho_1\ t_1 \longrightarrow t_2\ \rho_2$ take a store $\rho_1$ and a term $t_1$ as input and non-deterministically outputs a new term $t_2$ in an extended store $\rho_2$.

**Definition 5.5.1: DOT: Reduction relation**

**Reduction** $\boxed{\rho_1\ t_1 \longrightarrow t_2\ \rho_2}$

As in Section 2.2, the superscript in $t^x$ emphasizes that $x$ may appear free in $t$.

$$\begin{array}{lllll}
\rho & \{z \Rightarrow \overline{d^z}\} & \longrightarrow & v & \rho,\ (v : \overline{d^v}) \quad \text{with } v \text{ fresh} \\
\rho & v_1.m(v_2) & \longrightarrow & t^{v_2} & \rho,\ (v : \overline{d^v}) \quad \text{if } \rho(v1) \ni (m(x) = t^x) \\
\rho_1 & e[t_1] & \longrightarrow & e[t_2] & \rho_2 \quad \text{if } \rho 1\ t1 \longrightarrow t2\ \rho 2 \\
 & \multicolumn{4}{l}{\text{where } e ::= [\,] \mid [\,].m(t) \mid v.m([\,])}
\end{array}$$

With these definitions in mind, we can state the main type safety theorem:

**Theorem 5.5.2: DOT: Type Safety (original version)**

$$\begin{aligned}
&\forall \rho, t, T.\ \text{if } (\rho\ \varnothing \vdash t : T),\ \text{ then:} \\
&\quad \textit{either } \exists y.\ (t = y \text{ and } y \in \mathsf{dom}(\rho)) \\
&\quad \textit{or } \exists \rho_1, t_1.\ ((\rho\ t \longrightarrow t_1\ \rho_1) \text{ and } (\rho_1\ \varnothing \vdash t_1 : T))\,.
\end{aligned}$$

[7] See https://github.com/smarter/minidot/commit/a832f266757ee7af154de5f12be972637549080b

[8] This was first noted by [Hu 2019]. This rule is also present in [Barendregt, Coppo, and Dezani-Ciancaglini 1983].

In other words, given an empty context and a store $\rho$, if $t$ has type $T$ then *either* $t$ is a concrete value $y$ in the store $\rho$, *or* $t$ can be reduced to some term $t_1$ in a store $\rho_1$ such that $t_1$ preserves the type $T$.

This definition of type safety is peculiar: it combines together *progress* and *preservation* [Wright and Felleisen 1994] but it is weaker than the usual definition of preservation which normally applies to all possible reductions. This is explicitly called out in [Rompf and Amin 2016, Section 6]:

> "Note that Definition 1 assumes deterministic execution. Otherwise the statement would need to be modified to consider all possible following configurations."

This weaker statement naturally leads to a weaker induction hypothesis and this is where our attempt at adding And-I' runs into troubles.

**Theorem 5.5.3**

oopslaDOT extended with And-I' is sound.

*Proof sketch.* The original proof of Theorem 5.5.2 goes by induction on the derivation of $\rho\ \varnothing \vdash t : T$. Since store typing extends the regular typing judgment, we now have an extra case to handle.

**Case** $$\frac{\rho\ \varnothing \vdash t : T \quad \rho\ \varnothing \vdash t : U}{\rho\ \varnothing \vdash t : T \wedge U}\ (\text{And-I'})$$

Suppose $t = y$, then by inversion we must have $y \in \mathsf{dom}(\rho)$ which finishes the case. Otherwise, by the IH we have $\rho_1$, $t_1$, $\rho_2$, $t_2$ such that

$$(\rho\ t \longrightarrow t_1\ \rho_1) \text{ and } (\rho_1\ \varnothing \vdash t_1 : T)$$
$$(\rho\ t \longrightarrow t_2\ \rho_2) \text{ and } (\rho_2\ \varnothing \vdash t_2 : U)$$

To complete the case, we need to find some $\rho'$, $t'$ such that

$$(\rho\ t \longrightarrow t'\ \rho') \text{ and } (\rho'\ \varnothing \vdash t' : T \wedge U)$$

But since the definition of the reduction relation does not specify an evaluation order, we cannot prove that $(\rho\ t \longrightarrow t_1\ \rho_1)$ and $(\rho\ t \longrightarrow t_2\ \rho_2)$ imply $t_1 = t_2$ and $\rho_1 = \rho_2$, so we are stuck. ◊

To remedy this, we must generalize the type safety statement to subsume the usual preservation property:

**Theorem 5.5.4: DOT: Type Safety (generalized version)**

$$\forall \rho, t, T. \text{ if } (\rho\ \varnothing \vdash t : T), \text{ then we have } \textbf{both}:$$

1. *either* $\exists y. (t = y \text{ and } y \in \mathsf{dom}(\rho))$
   *or* $\exists \rho_1, t_1.(\rho\ t \longrightarrow t_1\ \rho_1)$,
2. **and** $\forall \rho_2, t_2. ((\rho\ t \longrightarrow t_2\ \rho_2) \text{ implies } (\rho_2\ \varnothing \vdash t_2 : T))$ .

*Proof.* The updated definition of `type_safety` is part of
https://github.com/smarter/minidot/commit/cee565e9452095ae3788f92cd912fd1733b8d54b. ■

Finally, we can complete our proof:

**Theorem 5.5.5**

oopslaDOT with the type safety definition from Theorem 5.5.4 can be soundly extended with AND-I'.

*Proof.* By induction on the derivation of $\rho\ \varnothing \vdash t : T$ as before.

**Case** $\dfrac{\rho\ \varnothing \vdash t : T \quad \rho\ \varnothing \vdash t : U}{\rho\ \varnothing \vdash t : T \wedge U}$ (AND-I')

We prove each part of the theorem separately. Part 1. follows directly by the IH. For part 2., by the IH we find that

$$\forall \rho_2, t_2. ((\rho\ t \longrightarrow t_2\ \rho_2) \text{ implies } (\rho_2\ \varnothing \vdash t_2 : T) \textbf{ and } (\rho_2\ \varnothing \vdash t_2 : U))$$

And so AND-I' finishes the case.
The mechanized version of this proof is also part of
https://github.com/smarter/minidot/commit/cee565e9452095ae3788f92cd912fd1733b8d54b.
■

For the record, we note that adding AND-I' to oopslaDOT is not enough to recover all possible uses of AND-I in wfDOT, because AND-I' does not cover the strict typing judgment $\Gamma \vdash x :_! T$. While this did not end up being needed in our proofs, we did mechanize this generalization in https://github.com/smarter/minidot/commit/0f146a40c24d7b34a2100fe6d56dca1e6400d968.

### 5.5.2 Meta-theory

This is where the work we did in previous chapters starts to pay off: most of the proof of type-preserving translation detailed in subsection 4.3.2 can be easily adapted to PS. We explicitly detail a few lemmas and theorems.

**Theorem 5.5.6: Subtyping preservation**

If $\Gamma \dashv\vdash \Delta$, $\Gamma \vdash S$ wf and $\Gamma \vdash S <: T$ then $\Delta \vdash |S| <: |T|$.

*Proof.* By induction on the derivation of $\Gamma \vdash S <: T$, cases GS-REFL, GS-VAR and GS-TRANS proceed like the corresponding case in Theorem 4.3.11. Case PS-CLASS proceeds like GS-CLASS. Cases PS-AND11, PS-AND12 and PS-AND2 proceed by the IH on each premise followed respectively by AND11, AND12 and AND2.

**Case** $\dfrac{\Gamma \vdash \overline{S <: T}, \overline{T <: S}}{\Gamma \vdash C[\overline{S}] <: C[\overline{T}]}$ (PS-INV)

> Let $\mathsf{tparams}(C) = \overline{X <: \ldots}$, then
>
> $$\dfrac{\dfrac{\dfrac{}{\Delta \vdash \overline{S <: T}, \overline{T <: S}}\ (\text{IH})}{\Delta \vdash (\overline{X = S}) <: (\overline{X = T})}\ (\text{TYP})}{\Delta \vdash |C[\overline{S}]| <: |C[\overline{T}]|}\ (2.4.5,\ \text{BINDX})$$

■

**Lemma 5.5.7: Class translation preserves value parameters**

If $\Gamma \dashv\!\!\mid \Delta$, $\Gamma \vdash T$ wf and $\mathsf{vparams}(T) = \overline{f : U}$, then $\Delta \vdash \overline{|T| <: (f() : |U|)}$

*Proof.* By induction on the derivation of $\mathsf{vparams}(T)$. We only show the additional cases compared to Lemma 4.3.12. Case PG-TRAIT is trivial. Case PG-ANDR is symmetrical to PG-ANDL.

**Case** $\dfrac{\mathsf{vparams}(T_2) \subseteq \mathsf{vparams}(T_1)}{\mathsf{vparams}(T_1\ \&\ T_2) \coloneqq \mathsf{vparams}(T_1)}$ (PG-ANDL)

> By the IH, $\Delta \vdash \overline{|T_1| <: (f() : |U|)}$ and by AND11, $\Delta \vdash |T_1\ \&\ T_2| <: |T_1|$. GS-TRANS finishes the case.

■

As we discussed in subsection 5.5.1, Lemma 4.3.13 cannot be directly carried over given the subtyping rules of DOT, but it can be replaced by a lemma on typing derivations that makes use of AND-I'.

**Lemma 5.5.8: Class translation preserves methods**

If $\Gamma \dashv\!\!\mid \Delta$, $\Gamma \vdash T_0$ wf and and $\Delta \vdash t_0 : |T_0|$, $\overline{t : |\sigma U|}$, $|V| <: |\sigma P|$ where $\sigma = [\overline{V/Y}]$ then $\mathsf{mtype}(m, T_0) = [\overline{Y <: P}] \to (\overline{x : U}) \to U_0$ implies $\Delta, x_{\mathsf{mtag}} : \{_ \Rightarrow \overline{Y = |V|}\} \vdash t_0.m(x_{\mathsf{mtag}}, \overline{t}) : |\sigma U_0|$.

*Proof.* By induction on the derivation of $\mathsf{mtype}(m, T_0)$.

**Case** $$\frac{(\textbf{def}\ m[\overline{Y <: P}](\overline{x : U}) : U_0 = \underset{\sim\sim}{e_0}) \in \mathsf{mdecls}(C[\overline{T}])}{\mathsf{mtype}(m, C[\overline{T}]) \coloneqq [\overline{Y <: P}] \to (\overline{x : U}) \to U_0}\ (\text{PM-Impl})$$

> By the same reasoning used in case GM-Class of Lemma 4.3.13, we find $|\Gamma| \vdash |C[\overline{T}]| <: (m(\mathsf{mtag} : |\overline{Y <: P}|, \overline{x : |U|}) : |U_0|)$. So by subsumption, $|\Gamma| \vdash t_0 : (m(\mathsf{mtag} : |\overline{Y <: P}|, \overline{y : |U|}) : |U_0|)$ and the rest of the case proceeds like case GT-Invk of Theorem 4.3.18.

**Case** $$\frac{\mathsf{parents}(N) = \overline{P} \quad (\textbf{def}\ m \ldots) \notin \mathsf{mdecls}(N)}{\mathsf{mtype}(m, N) \coloneqq \mathsf{mtype}(m, \ \&\overline{P})}\ (\text{PM-Super})$$

> By PS-Class and PS-And2, $\Gamma \vdash N <: \&\overline{P}$, so by Theorem 5.5.6 and subsumption, $|\Gamma| \vdash t_0 : |\&\overline{P}|$. The IH finishes the case.

**Case** $$\frac{\mathsf{mtype}_\Gamma(m, T_1) = [\overline{Y <: P}] \Rightarrow (\overline{x : S}) \Rightarrow V_L \qquad \mathsf{mtype}_\Gamma(m, T_2) = [\overline{Y <: P}] \Rightarrow (\overline{x : S}) \Rightarrow V_R}{\mathsf{mtype}_\Gamma(m, T_1 \,\&\, T_2) \coloneqq [\overline{Y <: P}] \Rightarrow (\overline{x : S}) \Rightarrow V_L \,\&\, V_R}\ (\text{PM-AndLR})$$

> We have $|\Gamma| \vdash t_0 : |T_1| \wedge |T_2|$ so by subsumption, $|\Gamma| \vdash t_0 : |T_1|, t_0 : |T_2|$ and by the IH,
>
> $$|\Gamma|, x_{\mathsf{mtag}} : \{_ \Rightarrow \overline{Y = |V|}\} \vdash t_0.m(x_{\mathsf{mtag}}, t) : |\sigma V_L|$$
> $$|\Gamma|, x_{\mathsf{mtag}} : \{_ \Rightarrow \overline{Y = |V|}\} \vdash t_0.m(x_{\mathsf{mtag}}, t) : |\sigma V_R|$$
>
> Therefore by And-I',
>
> $$|\Gamma|, x_{\mathsf{mtag}} : \{_ \Rightarrow \overline{Y = |V|}\} \vdash t_0.m(x_{\mathsf{mtag}}, t) : |\sigma V_L| \wedge |\sigma V_R|$$
>
> By definition, $|\sigma V_L| \wedge |\sigma V_R| = |\sigma V_L \,\&\, \sigma V_R)| = |\sigma(V_L \,\&\, V_R)|$ which finishes the case.

■

**Theorem 5.5.9: Typing translation is type-preserving**

If $\Gamma \dashv\vdash \Delta$ and $\Gamma \vdash e : T$, then $\Delta \vdash |e|_\Gamma : |T|$.

*Proof.* By induction on the derivation of $\Gamma \vdash e : T$. All cases but GT-Invk proceed as in Theorem 4.3.18.

$$\textbf{Case}\quad \frac{\Gamma \vdash e_0 : T_0 \quad \mathsf{mtype}(m,\ \mathsf{bound}_\Gamma(T_0)) = [\overline{Y <: P}] \to (\overline{y : U}) \to U_0 \\ \sigma = [\overline{V/Y}] \quad \Gamma \vdash \overline{V}\ \mathsf{wf}, \overline{V <: \sigma P}, \overline{e : S}, \overline{S <: \sigma U}}{\Gamma \vdash e_0.m[\overline{V}](\overline{e}) : \sigma U_0}\ (\text{GT-Invk})$$

We have $|e_0.m[\overline{V}](\overline{e})|_\Gamma = \textbf{let}\ x_{\mathsf{mtag}} = \{_ \Rightarrow \overline{Y = |V|}\}\ \textbf{in}\ |e_0|_\Gamma.m(x_{\mathsf{mtag}}, \overline{|e|_\Gamma})$. By the IH, $\Delta \vdash |e_0|_\Gamma : |T_0|, \overline{e : |S|}$. Let $T_0' = \mathsf{bound}_\Gamma(T_0)$, then by subsumption, Lemma 5.4.1 and Theorem 5.5.6 we have $\Delta \vdash |e_0|_\Gamma : |T_0'|$ and Lemma 5.5.8 finishes the case.

■

**Lemma 5.5.10: Class table translation is well-typed**

$\varnothing \vdash_{\text{DOT}} \{\mathsf{ct} \Rightarrow (\!| CT |\!)\} : \{\mathsf{ct} \Rightarrow [\![ CT ]\!]\}$.

*Proof.* Generalizing the proof of Lemma 4.3.22 to handle traits is easy since traits are translated like classes but have no constructors. ■

**Theorem 5.5.11: Program translation is type-preserving**

If $\varnothing \vdash_{\text{PS}} T$ wf and $\varnothing \vdash_{\text{PS}} e : T$ then $\varnothing \vdash_{\text{DOT}} \textbf{let}\ \mathsf{ct} = \{\mathsf{ct} \Rightarrow (\!| CT |\!)\}\ \textbf{in}\ |e|_\varnothing : |T|$.

*Proof.* Like Theorem 4.3.23 but using Theorem 5.5.9 and Lemma 5.5.10. ■

# 6 Pathless Lattice Scala

**Figure 6.1: PLS: Syntax**

$$
\begin{array}{ll}
x, y, z & \text{Variable} \\
B, C, D, E & \text{Class name} \\
f, g & \text{Class parameter} \\
m & \text{Method name} \\
X_C & \text{Class variable} \\
X_m & \text{Method variable} \\
X, Y, Z ::= X_C \mid X_m & \text{Type variable} \\
N, P, Q ::= C[\overline{T}] & \text{Non-variable} \\
S, T, U, V ::= & \text{Type} \\
\quad X \mid N \mid S \,\&\, T \mid S \,|\, T & \\
 & \\
\Gamma ::= & \text{Context} \\
\quad \varnothing \mid \Gamma, x : T \mid \Gamma, \overline{X <: N} &
\end{array}
$$

$$
\begin{array}{lll}
L ::= & & \text{Class declaration} \\
\quad \textbf{class}\, C[\overline{X_C <: N}](\overline{f : T}) \lhd P(\overline{f}), \overline{Q}\, \{\overline{M}\} & & \\
\quad \textbf{trait}\, C[\overline{X_C <: N}] \lhd \overline{Q}\, \{\overline{H}; \overline{M}\} & & \\
H ::= & & \text{Abstract method} \\
\quad \textbf{def}\ m[\overline{X_m <: N}](\overline{x : T}) : T_0 & & \\
M ::= & & \text{Concrete method} \\
\quad H = e_0 & & \\
b ::= & & \text{Boolean literal} \\
\quad \textbf{true} \mid \textbf{false} & & \\
e ::= & & \text{Expression} \\
\quad x & & \text{variable} \\
\quad e.f & & \text{parameter access} \\
\quad e_0.m[\overline{T}](\overline{e}) & & \text{method call} \\
\quad \textbf{new}\, C[\overline{T}](\overline{e}) & & \text{object} \\
\quad b & & \text{boolean} \\
\quad \textbf{if}\ e_0\ \textbf{then}\ e_1\ \textbf{else}\ e_2 & & \text{conditional} \\
 & & \\
\sigma, \tau ::= [\overline{T/X}] & & \text{Type substitution}
\end{array}
$$

In this chapter, we present Pathless Lattice Scala (PLS), an extension of the Pathless Scala calculus which completes the subtyping lattice by adding union types and a bottom type Nothing. To motivate the need for union types, we simultaneously introduce the standard conditional form **if** $e_0$ **then** $e_1$ **else** $e_2$ and a Boolean type.

The additional subtyping rules for union types end up invalidating some of the meta-theory of PS. We compensate for this by introducing a new *partial well-formedness* judgment which we make use of in the type-preserving translation proof. The proof from PS is otherwise readily adapted. The more complex member selection rules for union types motivate the introduction of an algorithmic subtyping relation to keep our typing judgment implementable.

## 6.1 Syntax

We call $S \mid T$ the *union* of $S$ and $T$. In a conditional expression **if** $e_0$ **then** $e_1$ **else** $e_2$, $e_0$ must be a Boolean,

Like Object, Boolean and Nothing are valid class names while not being defined in the class table $CT$. For subtyping and linearization to work with Boolean we extend the definition of parents from Figure 5.2 with,[1]

$$\mathsf{parents}(\mathsf{Boolean}) \coloneqq \mathsf{Object}$$

Nothing is not a valid input to most lookup functions including parents since member selection on Nothing is never well-typed in Scala, instead it is special-cased in the subtyping judgment in Figure 6.3 with rule LS-Nothing.

## 6.2 Declarative subtyping and well-formedness

In FGJ (and by extension PS), the well-formedness judgment makes use of the subtyping judgment: an applied class type $C[\overline{T}]$ is only well-formed if its type arguments $\overline{T}$ conform to the substituted upper-bounds of the corresponding type parameters. By contrast, in DOT it's the subtyping judgment which (implicitly) makes use of the well-formedness judgment: only well-formed types may appear in a DOT subtyping judgment. This impedance mismatch required us to make use of Lemma 4.2.4 to prove subtyping preservation. But while this lemma can be carried over to PS, it no longer holds in PLS due to the additional subtyping rules LS-Or21 and LS-Or22 defined in Figure 6.3.

In both of these rules, the conclusion involves a type which does not appear in any premise and for which we therefore cannot infer well-formedness. We could try to handle this by explicitly requiring the types that appear "out of thin air" to be well-formed:

$$\frac{\Gamma \vdash S <: T_1 \quad \boxed{\Gamma \vdash T_2 \ \mathsf{wf}}}{\Gamma \vdash S <: T_1 \mid T_2}\ (\text{LS-Or21-Alt}) \qquad \frac{\Gamma \vdash S <: T_2 \quad \boxed{\Gamma \vdash T_1 \ \mathsf{wf}}}{\Gamma \vdash S <: T_1 \mid T_2}\ (\text{LS-Or22-Alt})$$

But that would make well-formedness and subtyping mutually recursive which would complicate our proofs. To break the cycle, we define a notion of *partial well-formedness* in Figure 6.2 which more closely matches DOT well-formedness: $T$ is partially well-formed in $\Gamma$ if all free variables in $T$ are defined in $\Gamma$. We also reuse the well-formedness convention from the presentation of DOT in subsection 2.2.1: all subtyping and typing rules implicitly require the types involved to be partially well-formed. It is easy to show that a partially well-formed PLS type translates to a well-formed DOT type (Theorem 6.5.1).

[1] In real Scala, primitive classes such as Boolean are subtypes of AnyVal, not Object, and the true top type is Any. We do not model this additional complexity here. Note that this hierarchy might change in the future as the JVM might retrofit primitives to extend Object [Dan Smith 2022].

**Figure 6.2: PLS: Partial Well-formedness**

**Free variables** $\boxed{\mathsf{fv}(T) := \{\overline{X}\}}$

$$\mathsf{fv}(X) := \{X\}$$

$$\mathsf{fv}(C[\overline{T}]) := \bigcup \overline{\mathsf{fv}(T)}$$

$$\mathsf{fv}(T_1 \,\&\, T_2) := \mathsf{fv}(T_1) \cup \mathsf{fv}(T_2)$$

$$\mathsf{fv}(T_1 \mid T_2) := \mathsf{fv}(T_1) \cup \mathsf{fv}(T_2)$$

**Partially Well-formed Type** $\boxed{\Gamma \vdash T \ \mathsf{pwf}}$

$$\frac{\mathsf{fv}(T) \subseteq \mathsf{dom}(\Gamma)}{\Gamma \vdash T \ \mathsf{pwf}}$$

**Partially Well-formed Environment** $\boxed{\Gamma \ \mathsf{pwf}}$

$$\varnothing \ \mathsf{pwf} \qquad \frac{\Gamma, \overline{X <: N} \vdash \overline{N} \ \mathsf{pwf}}{\Gamma, \overline{X <: N} \ \mathsf{pwf}} \qquad \frac{\Gamma \vdash T \ \mathsf{pwf}}{\Gamma, x : T \ \mathsf{pwf}}$$

As expected, well-formedness implies partial well-formedness (Lemma 6.4.1). While having an extra judgment might seem inelegant, this split closely matches the behavior of the Scala compiler where most bound-checks are deferred to a compiler phase after typechecking to avoid cycles that could lead to compiler crashes.

### 6.2.1 Algorithmic subtyping

Until now, every subtyping judgment we've defined has been declarative and not algorithmic, in particular they all included a transitivity rule. Declarative judgments are convenient when working on the meta-theory, but to really model the behavior of the language as it is implemented, we should ensure that subtyping can actually be implemented by defining an

**Figure 6.3: PLS: Declarative Subtyping**

All rules from Figure 5.3 are carried over.

$\boxed{\Gamma \vdash S <: T}$

$$\Gamma \vdash \mathsf{Nothing} <: T \qquad \text{(LS-Nothing)}$$

$$\frac{\Gamma \vdash S_1 <: T, S_2 <: T}{\Gamma \vdash S_1 \mid S_2 <: T} \qquad \text{(LS-Or1)}$$

$$\frac{\Gamma \vdash S <: T_1}{\Gamma \vdash S <: T_1 \mid T_2} \ \text{(LS-Or21)} \qquad \frac{\Gamma \vdash S <: T_2}{\Gamma \vdash S <: T_1 \mid T_2} \ \text{(LS-Or22)}$$

**Figure 6.4: PLS: Well-formedness**

All rules from from Figure 5.4 are carried over.

**Well-formed type** $\boxed{\Gamma \vdash T \text{ wf}}$

$$\Gamma \vdash \mathsf{Nothing}\ \text{wf}\ (\text{WFL-Nothing}) \qquad \Gamma \vdash \mathsf{Boolean}\ \text{wf}\ (\text{WFL-Boolean})$$

$$\frac{\Gamma \vdash T_1, T_2 \text{ wf}}{\Gamma \vdash T_1 \mid T_2 \text{ wf}} \quad (\text{WFL-Or})$$

algorithmic judgment. Such a judgment will also come in handy in the next section, where we will make use of algorithmic subtyping in the definition of the function baseTypes to ensure that it is algorithmic itself.

Typically, algorithmic subtyping judgments are designed to be *syntax-driven*, where the conclusion of separate rules do not overlap. But if we only want to demonstrate that an algorithmic implementation is possible without regards for its complexity, this is not necessary: if multiple rules are applicable, an implementation can simply try them all in order until one succeeds. We only need to ensure that all rules are *mode-correct* as defined in [Dunfield and Krishnaswami 2021, § 3.1]:

> "A rule is mode-correct if there is a strategy for recursively deriving the premises such that two conditions hold:
>
> 1. The premises are mode-correct: for each premise, every input meta-variable is known (from the inputs to the rule's conclusion and the outputs of earlier premises).
>
> 2. The conclusion is mode-correct: if all premises have been derived, the outputs of the conclusion are known."

The only rule in our system which does not satisfy these conditions is GS-Trans. To eliminate it without losing expressiveness, we replace the rules GS-Var and PS-Class (which both reveal the upper-bound of a type) by rules AS-Var and AS-Class in Figure 6.5. The key difference is that the new rules additionally recurse on the revealed upper-bound. Other rules are left unchanged except for the use of $\Vdash$ over $\vdash$.

We prove that algorithmic subtyping is sound with respect to declarative subtyping in Theorem 6.4.4 and we conjecture that it is complete in Conjecture 6.4.6.

## 6.3 Typing

Declaration typing is unchanged from PS. The expression typing rules for booleans and conditionals in Figure 6.6 are unsurprising. The definition of bound needs to be extended to handle unions, and here it is helpful to carefully study the behavior of Scala once again.

**Figure 6.5: PLS: Algorithmic Subtyping**

$$\boxed{\Gamma \Vdash S <: T}$$

When multiple rules are applicable, the algorithm picks the first one.

$$\Gamma \Vdash S <: S \qquad \text{(AS-Refl)}$$

$$\Gamma \Vdash \mathsf{Nothing} <: T \qquad \text{(AS-Nothing)}$$

$$\frac{\Gamma(X) = N \quad \Gamma \Vdash N <: T}{\Gamma \Vdash X <: T} \qquad \text{(AS-Var)}$$

$$\frac{\Gamma \Vdash \overline{S =:= T}}{\Gamma \Vdash C[\overline{S}] <: C[\overline{T}]} \qquad \text{(AS-Inv)}$$

$$\frac{P \in \mathsf{parents}(C[\overline{S}]) \quad \Gamma \Vdash P <: B[\overline{T}]}{\Gamma \Vdash C[\overline{S}] <: B[\overline{T}]} \qquad \text{(AS-Class)}$$

$$\frac{\Gamma \Vdash S <: T_1, S <: T_2}{\Gamma \Vdash S <: T_1 \;\&\; T_2} \text{(AS-And2)} \qquad \frac{\Gamma \Vdash S_1 <: T, S_2 <: T}{\Gamma \Vdash S_1 \mid S_2 <: T} \text{(AS-Or1)}$$

$$\frac{\Gamma \Vdash S_1 <: T}{\Gamma \Vdash S_1 \;\&\; S_2 <: T} \text{(AS-And11)} \qquad \frac{\Gamma \Vdash S <: T_1}{\Gamma \Vdash S <: T_1 \mid T_2} \text{(AS-Or21)}$$

$$\frac{\Gamma \Vdash S_2 <: T}{\Gamma \Vdash S_1 \;\&\; S_2 <: T} \text{(AS-And12)} \qquad \frac{\Gamma \Vdash S <: T_2}{\Gamma \Vdash S <: T_1 \mid T_2} \text{(AS-Or22)}$$

Given `x : L | R` and the class table,

```
trait L { def foo(): A }
trait R { def foo(): B }
```

Can we attribute a type to `x.foo()`? Early on during its development, the Scala 3 compiler answered this positively and typed `x.foo()` as `A | B`. But this was later changed to emit an error because `foo()` is not defined in a common base class of `L` and `R`.[2] One argument in favor of this restriction is that in a typical *Design by Contract* approach [Meyer 1992], the behavior of a method in a class or trait is determined not just by its type but by the *contract* that every implementation of the method must conform too. A contract is usually specified informally as documentation comments and may include required pre-conditions and guaranteed post-conditions.[3] Methods with the same name defined in unrelated traits need not adhere to any common contract, so figuring out the behavior of `x.foo()` would force users to manually

[2] See https://github.com/lampepfl/dotty/pull/1550#pullrequestreview-2438518 for the historical discussion of this change.

[3] A good example of design by contract in the wild is `java.lang.Comparable`.

**Figure 6.6: PLS: Typing rules**

The definitions from Figure 5.7 are carried over.

**Expression typing** $\boxed{\Gamma \vdash e : T}$

$$\Gamma \vdash b : \mathsf{Boolean} \quad \text{(LT-Bool)}$$

$$\frac{\Gamma \vdash e_0 : \mathsf{Boolean} \qquad \Gamma \vdash e_1 : T_1 \quad \Gamma \vdash e_2 : T_2}{\Gamma \vdash \textbf{if } e_0 \textbf{ then } e_1 \textbf{ else } e_2 : T_1 \mid T_2} \quad \text{(LT-Cond)}$$

determine the union of the contracts of `foo()` in `L` and in `R`. In fact, these contracts might even be mutually exclusive, making all calls to `x.foo()` illegal.

We can replicate the behavior of Scala 3 by defining $\mathsf{bound}_\Gamma(S \mid T)$ to be the intersection of the common *base types* of $S$ and $T$. Our definition makes use of a baseTypes helper function which generalizes linearization to arbitrary types.[4]

Note that this definition of bound is not quite as expressive as we'd like, given `x: Foo[A]` | `Foo[B]` and the class table,

```
trait Foo[X] {
  def foo(): X
}
```

We'd like `x.foo()` to have type `A` | `B`, but this expression doesn't typecheck because the only common parent class of the union is Object. In actual Scala this isn't a problem because the compiler can take advantage of use-site variance [Igarashi and Viroli 2006; Odersky et al. 2021b] to approximate `Foo[A]` | `Foo[B]` as `Foo[? >: A & B <: A | B]`. Extending our calculus to support use-site variance remains future work.

## 6.4 Meta-theory

**Lemma 6.4.1: Well-formedness implies partial well-formedness**

$\Gamma \vdash T$ wf implies $\Gamma \vdash T$ pwf

*Proof.* Straightforward induction on the derivation of $\Gamma \vdash T$ wf. ■

[4] Note that unlike in the definition of linearization in Subsection 5.3.2, we use list union $\cup$ in place of the stricter $\vec{+}$ since we do not want to prevent selections on prefixes of type $C[S] \,\&\, C[T]$ even if we cannot prove in the current context that $S$ and $T$ are equal.

**Figure 6.7: PLS:** bound **and** baseTypes

The definitions of bound and baseTypes from Figure 5.5 are carried over.

$$\boxed{\mathsf{bound}_\Gamma(T) \mathrel{:=} \&\,\overline{N}}$$

$$\mathsf{bound}_\Gamma(S \mid T) \mathrel{:=} \&\,\mathsf{baseTypes}_\Gamma(S \mid T) \qquad \text{(B-Or)}$$

$$\boxed{\mathsf{baseTypes}_\Gamma(T) \mathrel{:=} \overline{N}}$$

$$\mathsf{baseTypes}_\Gamma(X) \mathrel{:=} \mathsf{baseTypes}_\Gamma(\Gamma(X)) \qquad \text{(BT-Var)}$$

$$\mathsf{baseTypes}_\Gamma(N) \mathrel{:=} \mathcal{L}(N) \qquad \text{(BT-Class)}$$

$$\mathsf{baseTypes}_\Gamma(S \,\&\, T) \mathrel{:=} \mathsf{baseTypes}_\Gamma(S) \cup \mathsf{baseTypes}_\Gamma(T) \qquad \text{(BT-And)}$$

$$\frac{\Gamma \mapsto T <: S}{\mathsf{baseTypes}_\Gamma(S \mid T) \mathrel{:=} \mathsf{baseTypes}_\Gamma(S)} \qquad \text{(BT-Or1)}$$

$$\frac{\Gamma \mapsto S <: T}{\mathsf{baseTypes}_\Gamma(S \mid T) \mathrel{:=} \mathsf{baseTypes}_\Gamma(T)} \qquad \text{(BT-Or2)}$$

$$\frac{\mathsf{baseTypes}_\Gamma(S) = \overline{P} \qquad \mathsf{baseTypes}_\Gamma(T) = \overline{P'}}{\mathsf{baseTypes}_\Gamma(S \mid T) \mathrel{:=} \left[Q \in \overline{P} \,\middle|\, \exists Q' \in \overline{P'}.\,\Gamma \mapsto Q <: Q',\, Q' <: Q\right]} \qquad \text{(BT-Or)}$$

**Lemma 6.4.2: Correctness of** baseTypes

If $N \in \mathsf{baseTypes}_\Gamma(T)$, then $\Gamma \vdash T <: N$.

*Proof.* By induction on $\mathsf{baseTypes}_\Gamma(T)$. Case BT-Or2 mirrors case BT-Or1.

**Case** $\mathsf{baseTypes}_\Gamma(X) \mathrel{:=} \mathsf{baseTypes}_\Gamma(\Gamma(X))$ (BT-Var)

> By GS-Var, $\Gamma \vdash X <: \Gamma(X)$ and by the IH, $\Gamma \vdash \Gamma(X) <: N$. GS-Trans finishes the case.

**Case** $\mathsf{baseTypes}_\Gamma(P) \mathrel{:=} \mathcal{L}(P)$ (BT-Class)

> By definition $\mathsf{parents}(P) \subseteq \mathcal{L}(P)$ and PS-Class finishes the case.

**Case** $\mathsf{baseTypes}_\Gamma(T_1 \,\&\, T_2) \mathrel{:=} \mathsf{baseTypes}_\Gamma(T_1) \cup \mathsf{baseTypes}_\Gamma(T_2)$ (BT-And)

> Either $N \in \mathsf{baseTypes}_\Gamma(T_1)$ in which case $\Gamma \vdash T_1 <: N$ and PS-And11 finishes the case or $N \in \mathsf{baseTypes}_\Gamma(T_2)$ in which case $\Gamma \vdash T_2 <: N$ and PS-And12 finishes the case.

**Case** $\dfrac{\Gamma \vdash T <: S}{\mathsf{baseTypes}_\Gamma(S \mid T) := \mathsf{baseTypes}_\Gamma(S)}$ (BT-Or1)

$$\dfrac{\dfrac{\Gamma \vdash T <: S}{\Gamma \vdash S \mid T <: S}\text{(LS-Or1)} \quad \dfrac{}{\Gamma \vdash S <: N}\text{(IH)}}{\Gamma \vdash S \mid T <: N}\text{(GS-Trans)}$$

**Case** $\dfrac{\mathsf{baseTypes}_\Gamma(S) = \overline{P} \quad \mathsf{baseTypes}_\Gamma(T) = \overline{P'}}{\mathsf{baseTypes}_\Gamma(S \mid T) := \left[Q \in \overline{P} \;\middle|\; \exists Q' \in \overline{P'}. \Gamma \vdash Q <: Q', Q' <: Q\right]}$ (BT-Or)

By definition, there exists $N' \in \mathsf{baseTypes}_\Gamma(T_2)$ such that $\Gamma \vdash N <: N', N' <: N$.

$$\dfrac{\dfrac{N \in \mathsf{baseTypes}_\Gamma(S)}{\Gamma \vdash S <: N}\text{(IH)} \quad \dfrac{\dfrac{N' \in \mathsf{baseTypes}_\Gamma(T)}{\Gamma \vdash T <: N'}\text{(IH)} \quad \Gamma \vdash N' <: N}{\Gamma \vdash T <: N}\text{(GS-Trans)}}{\Gamma \vdash S \mid T <: N}\text{(LS-Or1)}$$

■

**Lemma 6.4.3: Correctness of bound**

If $\mathsf{bound}_\Gamma(S) = T$, then $\Gamma \vdash S <: T$.

*Proof.* By induction on the derivation of $\mathsf{bound}_\Gamma(S)$. We only show the additional case compared to Lemma 5.4.1.

**Case** $\mathsf{bound}_\Gamma(S \mid T) := \& \, \mathsf{baseTypes}_\Gamma(S \mid T)$ (B-Or)

Let $\overline{N} = \mathsf{baseTypes}_\Gamma(S \mid T)$. By Lemma 6.4.2, $\Gamma \vdash \overline{S \mid T <: N}$ and repeated uses of PS-And2 finish the case.

■

**Theorem 6.4.4: Soundness of algorithmic subtyping**

If $\Gamma \mapsto S <: T$ then $\Gamma \vdash S <: T$.

*Proof.* By straightforward induction on the derivation of $\Gamma \mapsto S <: T$. ■

**Conjecture 6.4.5: Transitivity of algorithmic subtype relation**

If $\Gamma \mapsto S <: T$ and $\Gamma \mapsto T <: U$ then $\Gamma \mapsto S <: U$.

*Proof sketch.* [Kennedy and Pierce 2007, Appendix B] proves transitivity of algorithmic subtyping for a calculus similar to FGJ but with definition-site variance, we believe this argument could be adapted to our calculus.

Suppose the derivation of $\Gamma \Vdash S <: T$ has size $m$ and the derivation of $\Gamma \Vdash T <: U$ has size $n$. We proceed by induction on $m+n$, with a case analysis on the final rules of both derivations.

In the original proof, the difficult case involves the equivalent of AS-Inv on the left and AS-Class on the right:

$$\frac{\Gamma \Vdash \overline{S' =:= T'}}{\Gamma \Vdash C[\overline{S'}] <: C[\overline{T'}]} \ (\text{AS-Inv}) \qquad \frac{P' \in \mathsf{parents}(C[\overline{T'}]) \quad \Gamma \Vdash P' <: B[\overline{V}]}{\Gamma \Vdash C[\overline{T'}] <: B[\overline{U'}]} \ (\text{AS-Class})$$

By definition, $P' = [\overline{T'/X}]P$ where $P \in \mathsf{parents}(C[\overline{X}])$. If we can show that $\Gamma \Vdash [\overline{S'/X}]P <: B[\overline{V}]$ then we can finish the case by AS-Class. In the original proof, this is done by showing that for all $V, V'$, if there is a derivation of $\Gamma \Vdash [\overline{S'/X}]V <: V'$ that has size $< n$, then $\Gamma \Vdash [\overline{T'/X}]V <: V'$ is derivable. This requires a nested induction on the derivation of $\Gamma \Vdash [\overline{S'/X}]V <: V'$ that makes judicious use of the outer IH (hence the size requirement on the derivation of $\Gamma \Vdash [\overline{S'/X}]V <: V'$).

Given the sheer number of (sub-)cases involved and since we will anyway abandon transitivity when we add type members in Chapter 7 to match the behavior of Scala, we did not attempt to complete this proof. ◊

**Conjecture 6.4.6: Algorithmic subtype is complete**

If $\Gamma \vdash S <: T$, then $\Gamma \Vdash S <: T$.

*Proof sketch.* By induction on the derivation of $\Gamma \vdash S <: T$. Case GS-Trans relies on Conjecture 6.4.5. ◊

## 6.5 Translation

Our encoding of Boolean is similar to the one presented in [Amin, Grütter, et al. 2016, § 5] except we do not try to hide the implementation details of the type since this is not required for type-preservation.

### 6.5.1 Meta-theory

We only show the most interesting changes compared to subsection 5.5.2.

**Theorem 6.5.1: Partial well-formedness preservation**

If $\Gamma \dashv\!\!\!\mid \Delta$ and $\Gamma \vdash S$ pwf then $\Delta \vdash |S|$ wf.

*Proof.* We have $\mathsf{fv}(S) \subseteq \mathsf{dom}(\Gamma)$ and we need to prove $\mathsf{fv}(|S|) \subseteq \mathsf{dom}(\Delta)$. We proceed by induction on the derivation of $\mathsf{fv}(S)$. We only show the base case as all others follow directly by the IH.

**Figure 6.8: Translating PLS types, expressions and definitions to DOT**

All definitions from Figure 5.9 are carried over.

**Type Translation** $|T| := T_{\text{DOT}}$

$$|T_1 \mid T_2| := |T_1| \vee |T_2|$$

$$|\text{Boolean}| := \text{ct.Boolean}$$

$$|\text{Nothing}| := \bot$$

**Expression Translation** $|e|_\Gamma := t_{\text{DOT}}$

$$|\textbf{true}|_\Gamma := \text{ct.true}()$$

$$|\textbf{false}|_\Gamma := \text{ct.false}()$$

$$\frac{x_{\text{mtag}} \text{ is fresh} \quad \Gamma \vdash \textbf{if } e_0 \textbf{ then } e_1 \textbf{ else } e_2 : T}{|\textbf{if } e_0 \textbf{ then } e_1 \textbf{ else } e_2|_\Gamma := \textbf{let } x_{\text{mtag}} = \{_ \Rightarrow \overline{\text{A} = |T|}\} \textbf{ in } |e_0|_\Gamma.\text{if}(x_{\text{mtag}}, |e_1|_\Gamma, |e_2|_\Gamma)}$$

**Class Table Translation** $(\!| CT |\!) := \overline{d_{\text{DOT}}}$

$(\!| \varnothing |\!) :=$
$\quad \text{Object} = \top,$
$\quad \text{Boolean} = (\text{if}(\text{mtag} : \{_ \Rightarrow \text{A} : \bot .. \top\}, \text{t} : \text{mtag.A}, \text{f} : \text{mtag.A}) : \text{mtag.A}),$
$\quad \text{true}() : \text{ct.Boolean} = \{_ \Rightarrow \text{if}(\text{mtag}, \text{t}, \text{f}) = \text{t}\},$
$\quad \text{false}() : \text{ct.Boolean} = \{_ \Rightarrow \text{if}(\text{mtag}, \text{t}, \text{f}) = \text{f}\}$

**Case** $\text{fv}(X) := \{X\}$

> Since $X \in \text{dom}(\Gamma)$, we have $\Gamma \vdash X <: N$ for some $N$ by GS-Var. By Lemma 4.3.6 and inversion of EE-Typs, we must have $\Delta \vdash |Z| <: |N|$ and therefore $\Delta \vdash |X|$ wf since DOT subtyping rules only apply to well-formed types.

■

**Theorem 6.5.2: Subtyping preservation**

Suppose $\text{ct} \in \text{dom}(\Delta)$, $\Gamma$ pwf and for all $X \in \text{dom}(\Gamma)$, $\Gamma(X) = N$ implies $\Delta \vdash |X| <: |N|$. Then $\Gamma \Vdash S <: T$ implies $\Delta \vdash |S| <: |T|$.

*Proof.* Because PLS subtyping is only defined on partially well-formed types, we must have $\Gamma \vdash S, T$ pwf, so by Theorem 6.5.1, $\Delta \vdash |S|, |T|$ wf. We proceed by induction on the derivation of $\Gamma \vdash S <: T$ like in Theorem 5.5.6. The additional cases easily follow by the IH.

■

**Theorem 6.5.3: Typing translation is type-preserving**

If $\Gamma \dashv\vdash \Delta$ and $\Gamma \vdash e : T$, then $\Delta \vdash |e|_\Gamma : |T|$.

*Proof.* By induction on the derivation of $\Gamma \vdash e : T$. We only show the additional cases compared to Theorem 5.5.9.

**Case** $\Gamma \vdash b : \mathsf{Boolean}$ (LT-Bool)

If $b = \textbf{true}$ then $|b|_\Gamma = \mathsf{ct.true()}$ and

$$\dfrac{\dfrac{\dfrac{}{\Delta \vdash \mathsf{ct} : [\![CT]\!]}\ (\textsc{Var})}{\Delta \vdash \mathsf{ct} : \{\mathsf{true()} : |\mathsf{Boolean}|\}}\ (\textsc{Sub})}{\Delta \vdash \mathsf{ct.true()} : |\mathsf{Boolean}|}\ (\textsc{TApp'})$$

Otherwise $b = \textbf{false}$ and the derivation proceeds similarly.

**Case** $\dfrac{\Gamma \vdash e_0 : \mathsf{Boolean} \quad \Gamma \vdash e_1 : T_1 \quad \Gamma \vdash e_2 : T_2}{\Gamma \vdash \textbf{if}\ e_0\ \textbf{then}\ e_1\ \textbf{else}\ e_2 : T_1 \mid T_2}$ (LT-Cond)

We have $|\textbf{if}\ e_0\ \textbf{then}\ e_1\ \textbf{else}\ e_2|_\Gamma = \textbf{let}\ x_{\mathsf{mtag}} = \{_ \Rightarrow \mathsf{A} = |T_1 \mid T_2|\}\ \textbf{in}\ |e_0|_\Gamma.\mathsf{if}(x_{\mathsf{mtag}}, |e_1|_\Gamma, |e_2|_\Gamma)$. By the IH,

$$\Delta \vdash |e_0|_\Gamma : |\mathsf{Boolean}|,\ |e_1|_\Gamma : |T_1|,\ |e_2|_\Gamma : |T_2|$$

By Theorem 6.5.2 and Sub,

$$\Delta \vdash |e_1|_\Gamma : |T_1 \mid T_2|,\ |e_2|_\Gamma : |T_1 \mid T_2|$$

Let $\Delta_1 = \Delta, x_{\mathsf{mtag}} : \{_ \Rightarrow \mathsf{A} = |T_1 \mid T_2|\}$, then by TApp' and Sub,

$$\Delta_1 \vdash |e_0|_\Gamma.\mathsf{if}(x_{\mathsf{mtag}}, |e_1|_\Gamma, |e_2|_\Gamma) : |T_1 \mid T_2|$$

And Let finishes the case. ■

**Lemma 6.5.4: Class table translation is well-typed**

$\varnothing \vdash_{\text{DOT}} \{\mathsf{ct} \Rightarrow (\!|CT|\!)\} : \{\mathsf{ct} \Rightarrow [\![CT]\!]\}$.

*Proof.* To generalize Lemma 5.5.10, we only need to show that our additions to the class table typecheck: we can type Boolean by DTyp and "true" as well as "false" by TNew and DFun'. ■

**Theorem 6.5.5: Program translation is type-preserving**

If $\varnothing \vdash_{\text{PLS}} T$ wf and $\varnothing \vdash_{\text{PLS}} e : T$ then $\varnothing \vdash_{\text{DOT}} \textbf{let}\ \text{ct} = \{\text{ct} \Rightarrow (\!| CT |\!)\}\ \textbf{in}\ |e|_{\varnothing} : |T|$.

*Proof.* Like Theorem 4.3.23 but using Theorem 6.5.3 and Lemma 6.5.4. ■

# 7 Dependent Scala

In this chapter, we present Dependent Scala (DS), an extension of the Pathless Lattice Scala calculus with type members. While our type-preserving translation forces us to define rather complex declarative subtyping rules, we are able to define sound and simple algorithmic subtyping rules that match the behavior of the Scala compiler.

**Figure 7.1: DS: Syntax**

| | |
|---|---|
| $x, y, z$ | Variable |
| $B, C, D, E$ | Class name |
| $L$ | Type label |
| $f, g$ | Class parameter |
| $m$ | Method name |
| $X_C$ | Class variable |
| $X_m$ | Method variable |
| $X, Y, Z ::= X_C \mid X_m$ | Type variable |
| $N, P, Q ::= C[\overline{T}]$ | Class type |
| $S, T, U, V ::=$ | Type |
| $X \mid N \mid S \,\&\, T \mid S \,\mathbf{\mid}\, T \mid x.L$ | |
| $\Gamma ::=$ | Context |
| $\varnothing \mid \Gamma, x : T \mid \Gamma, \overline{X <: N}$ | |
| $CD ::=$ | Class declaration |
| $\textbf{class}\ C[\overline{X_C <: N}](\overline{f : T}) \triangleleft P(\overline{f}), \overline{Q}\ \{\overline{TD};\ \overline{M}\}$ | |
| $\textbf{trait}\ C[\overline{X_C <: N}] \triangleleft \overline{Q}\ \{\overline{TD};\ \overline{H};\ \overline{M}\}$ | |
| $TD ::=$ | Type declaration |
| $\textbf{type}\ L >: S <: T$ | |
| $H ::=$ | Abstract method |
| $\textbf{def}\ m[\overline{X_m <: N}](\overline{x : T}) : T_0$ | |
| $M ::=$ | Concrete method |
| $H = e_0$ | |
| $b ::=$ | Boolean literal |
| $\textbf{true} \mid \textbf{false}$ | |
| $e ::=$ | Expression |
| $x$ | variable |
| $e.f$ | parameter access |
| $x_0.m[\overline{T}](\overline{x})$ | method call |
| $\textbf{new}\ C[\overline{T}](\overline{e})$ | object |
| $b$ | boolean |
| $\textbf{if}\ e_0\ \textbf{then}\ e_1\ \textbf{else}\ e_2$ | conditional |
| $\{\textbf{val}\ x = e_1;\ e_2\}$ | local block |
| $\sigma, \tau ::= [\overline{T/X}]$ | Type substitution |
| $\theta ::= [\overline{y/x}]$ | Variable substitution |

## 7.1 Syntax

To simplify this presentation, we impose a syntactical restriction that was not present in our previous calculi: method calls may only involve variables as receiver and variables as arguments (just like applications in wfDOT). We compensate for this loss of expressiveness by introducing local block expressions $\{\textbf{val}\, x = e_1;\, e_2\}$ which we can use to desugar regular method calls:

**Definition 7.1.1: Method call desugaring**

$$\frac{x_0 \text{ is fresh} \quad \overline{x \text{ is fresh}}}{e_0.m[\overline{T}](\overline{e}) \rightsquigarrow \{\textbf{val}\, x_0 = e_0;\, \overline{\textbf{val}\, x = e};\, x_0.m[\overline{T}](\overline{x})\}}$$

This will not affect the semantics of our programs, but it means that the receiver of a method must always be evaluated before its arguments because of the translation strategy we will use for local blocks (Figure 7.12) and the way the reduction relation of DOT is defined (Definition 5.5.1). Alternatively, we could have kept arbitrary method calls by generalizing DT-Invk to introduce fresh variables if necessary and run avoidance on them like DT-Block does. This would be closer to the actual compiler implementation but would make our typing judgment and proofs related to it more complex for no obvious benefits.

## 7.2 Declarative subtyping and well-formedness

We extend the free variable judgment to account for free term variables (the definition of pwf itself stays as-is).

In Scala, unlike DOT, a type selection $x.L$ is only well-formed if $x$ actually has a type member named $L$.

**Figure 7.2: PLS: (Partial) Well-formedness**

All definitions from Figures 6.2 and 6.4 are carried over.

**Free variables** $\boxed{fv(T) \coloneqq \{\overline{X}, \overline{x}\}}$

$$\mathsf{fv}(x.L) \coloneqq \{x\}$$

**Well-formed type** $\boxed{\Gamma \vdash T \;\mathsf{wf}}$

$$\frac{\Gamma \vdash x : T \quad \mathsf{ttype}(x.L,\ \mathsf{bound}_\Gamma(T)) \text{ defined}}{\Gamma \vdash x.L \;\mathsf{wf}} \quad \text{(WFD-TSel)}$$

In the previous chapters, we were able to augment our subtyping relationship to handle intersection and union by simply adopting the corresponding DOT rules, but this simple recipe will not work here. Recall the DOT type selection rule Sel1:

$$\frac{\Gamma_{[x]} \vdash x :_! (L : \bot .. T)}{\Gamma \vdash x.L <: T} \quad (\textsc{Sel1})$$

What rule could we define in our source calculus that would correspond to Sel1? Let's try to deconstruct the premise of this rule:

1. $x$ is typed in a truncated context eliminating all bindings to the right of $x$. We can mirror this in our source calculus but this means we'll need to be careful about the interplay of context truncation and the environment entailment relation we use in proofs (Lemma 7.6.5).

2. $x$ is typed using the "strict typing" judgment which prevents uses of VarPack. Typing in our source calculus is even stricter: there is no subsumption rules, and the type of a variable is simply its type in the context (GS-Var). So we should be able to translate $\Gamma_{[x]} \vdash_{\text{DS}} x : U$, $U <: V$ into $\Delta_{[x]} \vdash_{\text{DOT}} x :_! |V|$ by relying on subtyping preservation to show that $\Delta_{[x]} \vdash_{\text{DOT}} |U| <: |V|$.

3. $x$ is typed as a type member declaration $(L : \bot .. T)$. Declarations are not types in our source calculus, but we can look up such declarations in a class given its name. We define tdecls in Figure 7.4 for this purpose.

Based on these considerations, we can come up with the following rule:

$$\frac{\Gamma_{[x]} \vdash x : T \quad \Gamma_{[x]} \vdash T <: C[\overline{U}] \qquad (\mathsf{type}\, L >: S_1 <: S_2) \in \mathsf{tdecls}(C) \quad \sigma = [\overline{U/X}] \quad \theta = [x/\mathsf{this}]}{\Gamma \vdash x.L <: \sigma(\theta S_2)} \quad (\textsc{DS-Sel1-Unproven})$$

Note that as in previous calculi, we need to substitute type variables by type parameters when looking up a member in some prefix $x$, but since the bounds of a type member may refer to another type member, "this" may appear free in the bounds and must be substituted by $x$.

Unfortunately, we have not been able to extend our subtyping preservation proof to work with DS-Sel1-Unproven. The issue is that given $\Gamma(\mathsf{this}) = D[\overline{X}]$ and $x = \mathsf{this}$, then $\Delta \dashv\vdash \Gamma$ only implies $\Delta \vdash \mathsf{this} :_! [\![D]\!]^{|\overline{X}|}$ (via EE-This) and not $\Delta \vdash \mathsf{this} :_! |D[\overline{X}]|$, and strict typing prevents us from using VarPack to recover the more precise type here. So the reasoning we used in point 2 above to recover DOT subsumption from DS subtyping breaks down.

To work around this technical issue, we define separate subtyping rules DS-SelThis1 and DS-SelThis2 for type selections on this in Figure 7.3. These rules rely on the type member lookup function ttype from Figure 7.4 which we now turn our attention to.

In Scala, the way we determine the bounds of a type member is analogous to the way we determine the parameter types and result type of a method, and so our definition of ttype naturally mirrors mtype, but unlike in past calculi, ttype also takes the prefix $x$ as input to perform the substitution we mentioned above.

**Figure 7.3: DS: Subtyping**

$$\boxed{\Gamma \vdash S <: T}$$

$$\frac{\Gamma \vdash \mathsf{this} : C[\overline{X}] \qquad \mathsf{ttype}(\mathsf{this}.L,\, C[\overline{X}]) = S_1 \mathrel{..} S_2}{\Gamma \vdash \mathsf{this}.L <: S_2} \quad \text{(DS-SelThis1)}$$

$$\frac{\Gamma \vdash \mathsf{this} : C[\overline{X}] \qquad \mathsf{ttype}(\mathsf{this}.L,\, C[\overline{X}]) = S_1 \mathrel{..} S_2}{\Gamma \vdash S_1 <: \mathsf{this}.L} \quad \text{(DS-SelThis2)}$$

$$\frac{x \neq \mathsf{this} \quad \Gamma \vdash x : T \quad \Gamma_{[x]} \vdash T <: C[\overline{U}] \qquad (\mathsf{type}\, L >: S_1 <: S_2) \in \mathsf{tdecls}(C) \quad \sigma = [\overline{U/X}] \quad \theta = [x/\mathsf{this}]}{\Gamma \vdash x.L <: \sigma(\theta S_2)} \quad \text{(DS-SelOther1)}$$

$$\frac{x \neq \mathsf{this} \quad \Gamma \vdash x : T \quad \Gamma_{[x]} \vdash T <: C[\overline{U}] \qquad (\mathsf{type}\, L >: S_1 <: S_2) \in \mathsf{tdecls}(C) \quad \sigma = [\overline{U/X}] \quad \theta = [x/\mathsf{this}]}{\Gamma \vdash \sigma(\theta S_1) <: x.L} \quad \text{(DS-SelOther2)}$$

When looking up the bounds of a type member defined in both operands of an intersection in TT-AndLR, the returned bounds must "fit" within the bounds of each operand. This is accomplished by taking the union of the lower bounds and the intersection of the upper bounds. But note that nothing prevents the resulting bounds from being absurd, like Object .. Nothing. Combined with subtyping transitivity this gives rise to the infamous "bad bounds" problem [Rompf and Amin 2016, § 4.3]. This is where our choice of DOT as a compilation target really starts to shine since it shields us from having to worry about this in our own proofs.

The lack of symmetry between DS-SelThis1 and DS-SelOther1 is unsatisfying, it would be nicer if we could use ttype everywhere, but here again we run into technical difficulties as we would need to simultaneously prove results about ttype and subtyping preservation. Thankfully, none of the issues we've encountered in this section apply to the algorithmic subtyping judgment we study next.

## 7.3 Algorithmic subtyping

We only need two rules for algorithmic subtyping of type selections: AS-Sel1 and AS-Sel2 defined in Figure 7.5. These rules use ttype to determine the bounds of a type selection. Since ttype is only defined on non-variable types it cannot be directly called on the selection prefix. And since the rules need to be mode-correct, we cannot simply materialize an upper-bound "out of thin air" using subtyping. Instead, we rely on the lookup function bound to produce a valid input for ttype, just like we did for mtype in previous calculi.

The most striking feature of our new rules is that they do not involve any context truncation,

**Figure 7.4: DS: Type lookup functions**

**Type declarations lookup** $\boxed{\mathsf{tdecls}(C) = \overline{TD}}$

$$\frac{\left\{\begin{array}{c}\textbf{class}\\\textbf{trait}\end{array}\right\} C[...]\,\{\,\overline{TD},\ ...\,\}}{\mathsf{tdecls}(C) \mathrel{\mathbf{:=}} \overline{TD}}$$

**Type member names lookup** $\boxed{\mathsf{tnames}(C) \mathrel{\mathbf{:=}} \overline{A}}$

$$\frac{\mathsf{tdecls}(N) = \overline{\mathsf{type}\,L...} \qquad \overline{P} = \mathsf{parents}(N)}{\mathsf{tnames}(N) \mathrel{\mathbf{:=}} \overline{\mathsf{tnames}(P)} \cup \overline{L}}$$

**Type member lookup** $\boxed{\mathsf{ttype}(x.L,\, T) = S_1 \mathinner{..} S_2}$

$$\frac{\theta = [x/\mathsf{this}] \quad \sigma = [\overline{T/X}] \qquad (\mathsf{type}\,L >: S_1 <: S_2) \in \mathsf{tdecls}(C)}{\mathsf{ttype}(x.L,\, C[\overline{T}]) \mathrel{\mathbf{:=}} \sigma(\theta S_1) \mathinner{..} \sigma(\theta S_2)} \quad (\text{TT-Member})$$

$$\frac{\mathsf{parents}(N) = \overline{P} \quad (\mathsf{type}\,L\,...) \notin \mathsf{tdecls}(N)}{\mathsf{ttype}(x.L,\, C[\overline{T}]) \mathrel{\mathbf{:=}} \mathsf{ttype}(x.L,\, \&\,\overline{P})} \quad (\text{TT-Super})$$

$$\frac{\mathsf{ttype}(x.L,\, T_1) = S_1 \mathinner{..} S_2 \qquad \mathsf{ttype}(x.L,\, T_2) = S_1' \mathinner{..} S_2'}{\mathsf{ttype}(x.L,\, T_1 \,\&\, T_2) \mathrel{\mathbf{:=}} (S_1 \mid S_1') \mathinner{..} (S_2 \,\&\, S_2')} \quad (\text{TT-AndLR})$$

$$\frac{\mathsf{ttype}(x.L,\, T_1) = S_1 \mathinner{..} S_2 \qquad \mathsf{ttype}(x.L,\, T_2)\ \text{undefined}}{\mathsf{ttype}(x.L,\, T_1 \,\&\, T_2) \mathrel{\mathbf{:=}} S_1 \mathinner{..} S_2} \quad (\text{TT-AndL})$$

$$\frac{\mathsf{ttype}(x.L,\, T_1)\ \text{undefined} \qquad \mathsf{ttype}(x.L,\, T_2) = S_1 \mathinner{..} S_2}{\mathsf{ttype}(x.L,\, T_1 \,\&\, T_2) \mathrel{\mathbf{:=}} S_1 \mathinner{..} S_2} \quad (\text{TT-AndR})$$

and yet we are able to prove them sound with respect to the declarative subtyping rules in Theorem 7.5.6! The key to this trick lies in the expressiveness difference between the declarative and algorithmic rules.

In the previous chapter, we conjectured that the algorithmic subtyping relation was transitive and therefore complete (Conjecture 6.4.6). This is no longer true in Dependent Scala as illustrated by the following example,

**Figure 7.5: DS: Algorithmic Subtyping**

All rules from from Figure 6.5 are carried over.

$$\boxed{\Gamma \mapsto S <: T}$$

$$\frac{\Gamma \vdash x : U \quad \mathsf{ttype}(x.L,\ \mathsf{bound}_\Gamma(U)) = S_1 \mathinner{..} S_2 \quad \Gamma \mapsto S_2 <: T}{\Gamma \mapsto x.L <: T} \quad \text{(AS-SEL1)}$$

$$\frac{\Gamma \vdash x : U \quad \mathsf{ttype}(x.L,\ \mathsf{bound}_\Gamma(U)) = T_1 \mathinner{..} T_2 \quad \Gamma \mapsto S <: T_1}{\Gamma \mapsto S <: x.L} \quad \text{(AS-SEL2)}$$

```
trait A[S <: Object, T <: Object] {
  type M >: S <: T

  def id(x: S): T = x
}
```

Let $\Gamma = (\mathsf{S} <: \mathsf{Object}, \mathsf{T} <: \mathsf{Object}, \mathsf{this} : \mathsf{A[S, T]}, \mathsf{x} : \mathsf{S})$, then to ensure that the body of id is well-typed we show,

$$\frac{\dfrac{}{\Gamma \vdash \mathsf{S} <: \mathsf{this.M}}\ \text{(DS-SELTHIS2)} \qquad \dfrac{}{\Gamma \vdash \mathsf{this.M} <: \mathsf{T}}\ \text{(DS-SELTHIS1)}}{\Gamma \vdash \mathsf{S} <: \mathsf{T}}\ \text{(GS-TRANS)}$$

But this code isn't valid Scala. Indeed, it would not be practical for the compiler to consider the bound of every type member in scope for every subtype check.[1] Note that this loss of transitivity is not a fundamental loss of expressiveness. It is always possible to manually tell the compiler to consider a specific intermediate type:

```
def conv(x: S): this.M = x
def id(x: S): T = conv(x)
```

Thanks to this restriction, we can establish that context truncation preserves algorithmic subtyping (Lemma 7.5.3) which is key to the proof of soundness of algorithmic subtyping (Theorem 7.5.6).

The additional cases for bound and baseTypes in Figure 7.6 are straightforward, but BT-SEL

[1] On the other hand, if a subtyping check involves type selections, the compiler will consider each bound of each type member involved. [Nieto 2017] shows that this can lead to type-checking taking an amount of time exponential in the number of declared type members.

implies that baseTypes is now defined in terms of bound, and because of B-Or, bound was already defined in terms of baseTypes, making them mutually recursive. Furthermore, because of AS-Sel1 and AS-Sel2, algorithmic subtyping is now defined in terms of bound, and since BT-Or already relied on algorithmic subtyping, all three judgments are now mutually recursive. This isn't a problem per se but it means that some lemmas such as Lemma 7.5.3 will need to be proved by simultaneous induction on all three judgments at once, *c'est la vie*!

**Figure 7.6: DS:** bound **and** baseTypes

The definitions of bound and baseTypes from Figure 6.7 are carried over.

**bound of type** $\boxed{\mathsf{bound}_\Gamma(T) \coloneqq \&\overline{N}}$

$$\frac{\Gamma \vdash x : T \quad \mathsf{ttype}(x.L,\ \mathsf{bound}_\Gamma(T)) = S_1 \mathrel{..} S_2}{\mathsf{bound}_\Gamma(x.L) \coloneqq \mathsf{bound}_\Gamma(S_2)} \quad \text{(B-Sel)}$$

$\boxed{\mathsf{baseTypes}_\Gamma(T) \coloneqq \overline{N}}$

$$\frac{\Gamma \vdash x : T \quad \mathsf{ttype}(x.L,\ \mathsf{bound}_\Gamma(T)) = S_1 \mathrel{..} S_2}{\mathsf{baseTypes}_\Gamma(x.L) \coloneqq \mathsf{baseTypes}_\Gamma(S_2)} \quad \text{(BT-Sel)}$$

## 7.4 Typing

### 7.4.1 Expression Typing

**Figure 7.7: DS: Expression Typing rules**

$\boxed{\Gamma \vdash e : T}$

$$\frac{\begin{array}{c}\Gamma \vdash x_0 : T_0 \quad \mathsf{mtype}(x_0.m,\ \mathsf{bound}_\Gamma(T_0)) = [\overline{Y <: P}] \longrightarrow (\overline{x : U}) \longrightarrow U_0 \\ \sigma = [\overline{V}/\overline{Y}] \quad \Gamma \vdash \overline{V}\ \mathsf{wf},\ \overline{V <: \sigma P},\ \overline{x} : \overline{S},\ \overline{S <: \sigma U}\end{array}}{\Gamma \vdash x_0.m[\overline{V}](\overline{x}) : T_0} \quad \text{(DT-Invk)}$$

$$\frac{\Gamma \vdash e_1 : S \quad \Gamma, x : S \vdash e_2 : T \quad \Gamma, x : S \vdash T \Uparrow^x T'}{\Gamma \vdash \{\mathbf{val}\, x = e_1;\, e_2\} : T'} \quad \text{(DT-Block)}$$

#### Method calls

In previous chapters, we used the lookup function mtype$(m, T)$ to determine how to type a method selection $x.m$ when the type of $x$ is upper-bounded by $T$. mtype looks up the declared method type and takes care of substituting the class type variables based on $T$ to produce a valid type. But in Dependent Scala, this is not enough, a method type might refer to a local type member:

**Figure 7.8: DS: Redefinition of mtype**

**Method type lookup** $\boxed{\mathsf{mtype}(x.m,\ T) \coloneqq [\overline{Y <: P}] \to (\overline{x : U}) \to U_0}$

$$\frac{\theta = [x/\mathsf{this}] \quad \sigma = [\overline{T/X}] \qquad (\mathbf{def}\ m[\overline{Y <: P}](\overline{x : U}) : U_0 = e_0) \in \mathsf{mdecls}(C[\overline{X}])}{\mathsf{mtype}(x.m,\ C[\overline{T}]) \coloneqq [\overline{Y <: \sigma(\theta P)}] \to (\overline{y : \sigma(\theta U)}) \to \sigma(\theta U_0)} \quad \text{(DM-Impl)}$$

$$\frac{\mathsf{parents}(N) = \overline{P} \quad (\mathbf{def}\ m \ldots) \notin \mathsf{mdecls}(N)}{\mathsf{mtype}(x.m,\ N) \coloneqq \mathsf{mtype}(x.m,\ \&\overline{P})} \quad \text{(DM-Super)}$$

$$\frac{\mathsf{mtype}(x.m,\ T_1) = [\overline{Y <: P}] \Rightarrow (\overline{x : S}) \Rightarrow V_1 \qquad \mathsf{mtype}(x.m,\ T_2) = [\overline{Y <: P}] \Rightarrow (\overline{x : S}) \Rightarrow V_2}{\mathsf{mtype}(x.m,\ T_1 \,\&\, T_2) \coloneqq [\overline{Y <: P}] \Rightarrow (\overline{x : S}) \Rightarrow V_1 \,\&\, V_2} \quad \text{(DM-AndLR)}$$

$$\frac{\mathsf{mtype}(x.m,\ T_1)\ \text{defined} \qquad \mathsf{mtype}(x.m,\ T_2)\ \text{undefined}}{\mathsf{mtype}(x.m,\ T_1 \,\&\, T_2) \coloneqq \mathsf{mtype}(x.m,\ T_1)} \quad \text{(DM-AndL)}$$

$$\frac{\mathsf{mtype}(x.m,\ T_1)\ \text{undefined} \qquad \mathsf{mtype}(x.m,\ T_2)\ \text{defined}}{\mathsf{mtype}(x.m,\ T_1 \,\&\, T_2) \coloneqq \mathsf{mtype}(x.m,\ T_2)} \quad \text{(DM-AndR)}$$

```
trait Zero {
  type Elem >: Nothing <: Object
  def zero(): this.Elem
}
```

If x : Zero, then the type of x.zero() should be x.Elem and not this.Elem, so we need to substitute the prefix in the method type. Since mtype already does type substitution, it makes sense to extend it to also perform term substitution by keeping track of the prefix, this also mirrors how we defined ttype earlier. In the redefinition of mtype in Figure 7.8, only DM-Impl uses the prefix, the other rules simply pass it along in recursive calls and are otherwise identical to the rules in Figure 5.5.

Rule DT-Invk in Figure 7.7 looks deceptively similar to GT-Invk but is in fact much more powerful since it supports dependent method types. To avoid writing down explicit variable substitutions, we rely on the identification of terms up to $\alpha$-renaming to force the parameter names returned by mtype and the names of the variables passed as arguments to coincide. As an example, the following class table is well-typed:

```
class X ◁ Object{}
trait HasA { type A >: Nothing <: Object }
class HasX ◁ HasA { type A >: X <: X }
class Foo ◁ Object {
  def foo(hasA: HasA, a: hasA.A): hasA.A = a
  def bar(hasX: HasX, x: X): X = foo(hasX, x)
}
```

### Local block

The type of a local block $\{\mathbf{val}\, x = e_1;\, e_2\}$ must be a super-type of the type of $e_2$, but it cannot mention $x$ since it is not part of the enclosing context. This motivates the introduction in Figure 7.9 of algorithmic judgments for *variable avoidance* [Pierce and Turner 2000, § 5.3; Nieto 2017, § 4.3].

In the judgment $\Gamma \vdash S \Updownarrow^x T_1 \mathrel{..} T_2$, the inputs are $\Gamma$, $S$ and $x$ and the outputs are $T_1$ and $T_2$. The rules ensure that $x$ does not appear in either $T_1$ or $T_2$ and that $\Gamma \vdash T_1 <: S, S <: T_2$ as shown in Theorem 7.5.7. All rules but A-Absent implicitly assume that $x \in S$. This is not enough to make avoidance syntax-directed since A-Dealias and A-Super have the same inputs, but the output of the judgment is still deterministic because these rules have non-overlapping premises (we write $\Gamma \nVdash \overline{S <: S'}$ to mean "$\Gamma \vdash \overline{S <: S'}$ does not hold").

For convenience, we also define $\Gamma \vdash S \Downarrow^x T_1$ and $\Gamma \vdash S \Uparrow^x T_2$ which return respectively the lower-bound and upper-bound produced by avoidance.

Ultimately, we only use the upper-bound in DT-Invk, but defining both is necessary for rule A-Dealias which we motivate with the following example:

```
class C[T] ◁ Object {
  def c(): Object = new Object
}
class A ◁ Object {
  type M >: X <: X
}
class B ◁ Object {
  def foo(): C[X] =
    {val x = new A; new C[x.M]}.c()
}
```

Given $\Gamma \vdash$ `this` : `B`, `x` : `A`, we find $\Gamma \vdash$ **new** `C[x.M]` : `C[x.M]` by GT-New. But since `x.M` is both lower- and upper-bounded by `X`, we also have $\Gamma \vdash$ `C[x.M]` <: `C[X]` by PS-Inv. A-Dealias takes advantage of this to give a more precise type to the local block than just Object.

**Figure 7.9: DS: Variable avoidance in types**

**Promotion** $\boxed{\Gamma \vdash S \Uparrow^x T}$

$$\frac{\Gamma \vdash S \Updownarrow^x T_1 \mathrel{..} T_2}{\Gamma \vdash S \Uparrow^x T_2}$$

**Demotion** $\boxed{\Gamma \vdash S \Downarrow^x T}$

$$\frac{\Gamma \vdash S \Updownarrow^x T_1 \mathrel{..} T_2}{\Gamma \vdash S \Downarrow^x T_1}$$

**Avoidance** $\boxed{\Gamma \vdash S \Updownarrow^x T_1 \mathrel{..} T_2}$

$$\frac{x \notin \mathsf{fv}(S)}{\Gamma \vdash S \Updownarrow^x S \mathrel{..} S} \quad \text{(A-Absent)}$$

$$\frac{\Gamma \vdash S_1 \Updownarrow^x T_1 \mathrel{..} T_1' \quad \Gamma \vdash S_2 \Updownarrow^x T_2 \mathrel{..} T_2'}{\Gamma \vdash (S_1 \mathbin{\&} S_2) \Updownarrow^x (T_1 \mathbin{\&} T_2) \mathrel{..} (T_1' \mathbin{\&} T_2')} \quad \text{(A-And)}$$

$$\frac{\Gamma \vdash S_1 \Updownarrow^x T_1 \mathrel{..} T_1' \quad \Gamma \vdash S_2 \Updownarrow^x T_2 \mathrel{..} T_2'}{\Gamma \vdash (S_1 \mid S_2) \Updownarrow^x (T_1 \mid T_2) \mathrel{..} (T_1' \mid T_2')} \quad \text{(A-Or)}$$

$$\frac{\begin{array}{c}\Gamma \vdash x : T \quad \mathsf{ttype}(x.L,\ \mathsf{bound}_\Gamma(T)) = S_1 \mathrel{..} S_2 \\ \Gamma \vdash S_1 \Downarrow^x S_1' \quad \Gamma \vdash S_2 \Uparrow^x S_2'\end{array}}{\Gamma \vdash x.L \Updownarrow^x S_1' \mathrel{..} S_2'} \quad \text{(A-Sel)}$$

$$\frac{\Gamma \vdash \overline{S \Updownarrow^x S' \mathrel{..} S''} \quad \Gamma \vdash\!\!\blacktriangleright \overline{S <: S'}}{\Gamma \vdash C[\overline{S}] \Updownarrow^x C[\overline{S'}] \mathrel{..} C[\overline{S'}]} \quad \text{(A-Dealias)}$$

$$\frac{\begin{array}{c}\Gamma \vdash \overline{S \Updownarrow^x S' \mathrel{..} S''} \quad \Gamma \not\vdash\!\!\blacktriangleright \overline{S <: S'} \\ \textbf{class}\, C[\overline{X <: N}] \lhd B[\overline{U}] \\ \sigma = [\overline{S/X}] \quad \Gamma \vdash B[\overline{\sigma U}] \Uparrow^x T\end{array}}{\Gamma \vdash C[\overline{S}] \Updownarrow^x \mathsf{Nothing} \mathrel{..} T} \quad \text{(A-Super)}$$

Ideally, we would like avoidance to give us the "best" approximations possible for any given type. In particular, for promotion we might conjecture that,

> "If $\Gamma \vdash S \Uparrow^x T$ then $\Gamma \vdash\!\!\blacktriangleright S <: U$ implies $\Gamma \vdash\!\!\blacktriangleright T <: U$."

But this statement is false, as demonstrated by the following counter-example:

```
class Inv[X] ◁ Object; trait X; trait Y
trait HasA { type A >: X | Y <: X & Y }
trait HasB { type B >: X <: Y }
class HasBImpl(a: HasA) ◁ Object, HasB {
  type B = a.A
}
class Test ◁ Object {
  def foo(a: HasA): Inv[a.A] = {
    val b: HasB = new HasBImpl(a);
    new Inv[b.B]
  }
}
```

Given $\Gamma$ = this : `Test`, `a` : HasA, `b` : HasB, we find $\Gamma \vdash$ `Inv[b.B]` $\Uparrow^{\mathsf{b}}$ Object even though we can derive $\Gamma \Vdash$ `Inv[b.B]` <: `Inv[a.A]`. Once again, having wildcards would be helpful here since we could enhance avoidance such that $\Gamma \vdash$ `Inv[b.B]` $\Uparrow^{\mathsf{b}}$ `Inv[? >: X <: Y]`. This would be good enough since by wildcard capture we should be able to derive $\Gamma \Vdash$ `Inv[? >: X <: Y]` =:= `Inv[a.A]`.

### 7.4.2 Declaration Typing

Method typing is generalized in DT-Method to support dependent methods like the ones we saw in the previous subsection. In both proper classes and traits, we ensure (via DT-Class and DT-Trait) that the bounds of type declarations are well-formed and that all type members are valid overrides. Override checking for type members (in Figure 7.10) proceeds much like override checking for methods (in Figure 5.8), but there is no abstract/concrete distinction.

**Figure 7.10: DS: Overriding**

$$\frac{\begin{array}{c}\Gamma = \overline{X <: N}, \mathsf{this} : C[\overline{X}] \\ P \in \mathcal{L}(C[\overline{X}]) \text{ implies } \mathsf{override}_\Gamma(L, N, P) \\ \mathsf{isProperClass}(C) \text{ defined and } \mathsf{ttype}(\mathsf{this}.L, C[\overline{X}]) = S_1 \,..\, S_2 \text{ implies } \Gamma \Vdash S_1 <: S_2\end{array}}{\mathsf{isValid}_\Gamma(L)}$$

$$\frac{\begin{array}{l}\mathsf{ttype}(\mathsf{this}.L, P) \text{ defined implies:} \\ \bullet\ \mathsf{ttype}(\mathsf{this}.L, N) = S_1 \,..\, S_2 \\ \bullet\ \mathsf{ttype}(\mathsf{this}.L, P) = T_1 \,..\, T_2 \\ \bullet\ \Gamma \vdash T_1 <: S_1,\ S_2 <: T_2\end{array}}{\mathsf{override}_\Gamma(L, N, P)}$$

A type member overrides another if it has equal or more precise bounds. In proper classes only, isValid additionally checks that the lower bound of each type member is a subtype of its upper-bound (using algorithmic subtyping since this should be determined without relying on the bounds of the type member itself). This is critical for our translation: we need to ensure that there exists a valid instantiation of each type member, otherwise we won't be able to typecheck the translated constructor since type members of DOT objects are not allowed to be abstract.

**Figure 7.11: DS: Declaration Typing rules**

**Method typing** $\boxed{\Gamma \vdash m \text{ ok}}$

$$\frac{\begin{array}{c}\Gamma = \overline{X <: N}, \mathsf{this} : C[\overline{X}] \\ \mathsf{mtype}(\mathsf{this}.m, C[\overline{X}]) = [\overline{Y <: P}] \to (\overline{x : U}) \to U_0 \\ \Delta_0 = \Gamma, \overline{Y <: P} \quad \Delta_{i+1} = \Delta_i, x_i : U_i \\ \Delta_0 \vdash \overline{P} \text{ wf} \quad \Delta_i \vdash U_{i+1} \text{ wf} \quad \Delta_n \vdash U_0 \text{ wf} \\ \mathsf{mbody}(\mathsf{this}.m, C[\overline{X}]) = e_0 \text{ implies } \Delta_n \vdash e_0 : E_0, E_0 <: U_0 \\ Q \in \mathsf{parents}(C[\overline{X}]) \text{ implies } \mathsf{override}_\Gamma(m, C[\overline{X}], Q)\end{array}}{\Gamma \vdash m \text{ ok}} \quad \text{(DT-Method)}$$

**Class typing** $\boxed{\vdash C \text{ ok}}$

$$\frac{\begin{array}{c}\textbf{class}\, C[\overline{X <: N}](\overline{g : U}, \overline{f : T}) \lhd P(\overline{g}), \overline{Q}\, \{\overline{\textbf{type}\, A >: S_1 <: S_2;}\ \overline{\textbf{def}\ m \ldots}\} \\ \mathcal{L}(C[\overline{X}]) \text{ defined} \quad \mathsf{isProperClass}(P) \quad \overline{\mathsf{isTrait}(Q)} \\ \Gamma = \overline{X <: N}, \mathsf{this} : C[\overline{X}] \quad \Gamma \vdash \overline{S_1}, \overline{S_2} \text{ wf} \\ \overline{X <: N} \vdash \overline{N}, \overline{U}, \overline{T}, P, \overline{Q} \text{ wf} \quad \Gamma \vdash \overline{m} \text{ ok} \quad \mathsf{vparams}(P) = \overline{g : U} \\ \mathsf{mnames}_{abs}(C) = \varnothing \quad m' \in \mathsf{mnames}(C) \text{ implies } \mathsf{isValid}_\Gamma(m') \\ A' \in \mathsf{tnames}(C) \text{ implies } \mathsf{isValid}(A')\end{array}}{\vdash C \text{ ok}} \quad \text{(DT-Class)}$$

$$\frac{\begin{array}{c}\textbf{trait}\, C[\overline{X <: N}] \lhd \overline{Q}\, \{\overline{\textbf{type}\, A >: S_1 <: S_2;}\ \overline{\textbf{def}\ m \ldots}\} \\ \mathcal{L}(C[\overline{X}]) \text{ defined} \quad \overline{\mathsf{isTrait}(Q)} \\ \Gamma = \overline{X <: N}, \mathsf{this} : C[\overline{X}] \\ \overline{X <: N} \vdash \overline{N}, \overline{Q} \\ \Gamma \vdash \overline{S_1}, \overline{S_2} \text{ wf} \quad \Gamma \vdash \overline{m} \text{ ok} \\ m' \in \mathsf{mnames}(C) \text{ implies } \mathsf{isValid}_\Gamma(m') \\ A' \in \mathsf{tnames}(C) \text{ implies } \mathsf{isValid}_\Gamma(A')\end{array}}{\vdash C \text{ ok}} \quad \text{(DT-Trait)}$$

## 7.5 Meta-theory

**Lemma 7.5.1**

If $\overline{X <: N}, \mathsf{this} : C[\overline{X}] \vdash T \text{ pwf}$, $x \in \mathsf{dom}(\Gamma)$ and $\Gamma \vdash S \text{ pwf}$, then $\Gamma \vdash [\overline{S/X}]([x/\mathsf{this}]T) \text{ pwf}$.

*Proof.* We must have $\mathsf{fv}(T) \subseteq \{\overline{X}, \mathsf{this}\}$, therefore $\mathsf{fv}([\overline{S/X}]([x/\mathsf{this}]T)) \subseteq (\mathsf{fv}(\overline{S}) \cup \{x\}) \subseteq$

$\mathsf{dom}(\Gamma)$. ■

**Lemma 7.5.2: Partial Well-formedness of type member lookup**

If $\mathsf{ttype}(x.L, U) = S_1 \mathrel{..} S_2$, $x \in \mathsf{dom}(\Gamma)$ and $\Gamma \vdash U$ wf then $\Gamma \vdash S_1, S_2$ pwf.

*Proof.* By induction on the definition of $\mathsf{ttype}(x.L, U)$. We only show the base case as the others follow directly by the IH.

**Case** $$\frac{\theta = [x/\mathsf{this}] \quad \sigma = [\overline{T/X}] \qquad (\mathsf{type}\, L >: S_1 <: S_2) \in \mathsf{tdecls}(C)}{\mathsf{ttype}(x.L, C[\overline{T}]) \coloneqq \sigma(\theta S_1) \mathrel{..} \sigma(\theta S_2)} \text{(TT-Member)}$$

By inversion of $\vdash C$ ok via either DT-Class or DT-Trait,

$$\overline{X <: N}, \mathsf{this} : C[\overline{X}] \vdash S_1, S_2 \text{ pwf}$$

By inversion of $\Gamma \vdash C[\overline{T}]$ wf, we must have $\Gamma \vdash \overline{T}$ wf. Therefore by Lemma 7.5.1, $\Gamma \vdash \sigma(\theta S_1), \sigma(\theta S_2)$ pwf.

■

**Lemma 7.5.3: Context truncation preserves algorithmic subtyping**

Let $\Gamma = \Gamma_1, \Gamma_2$. If $\Gamma$ pwf and $\Gamma_1 \vdash S, T$ pwf, then

1. $\Gamma \mapsto S <: T$ implies $\Gamma_1 \mapsto S <: T$
2. $\mathsf{baseTypes}_\Gamma(S)$ defined implies $\mathsf{baseTypes}_\Gamma(S) = \mathsf{baseTypes}_{\Gamma_1}(S)$ and $\overline{\mathsf{baseTypes}_{\Gamma_1}(S)}$ pwf
3. $\mathsf{bound}_\Gamma(S)$ defined implies $\mathsf{bound}_\Gamma(S) = \mathsf{bound}_{\Gamma_1}(S)$ and $\mathsf{bound}_{\Gamma_1}(S)$ pwf

*Proof.* By simultaneous induction on the size of the derivations of $\Gamma \mapsto S <: T$, $\mathsf{baseTypes}_\Gamma(S)$ and $\mathsf{bound}_\Gamma(S)$. We only show a few cases.

**Case** $$\frac{\Gamma(X) = N \quad \Gamma \mapsto N <: T}{\Gamma \mapsto X <: T} \text{(AS-Var)}$$

By inversion of $\Gamma_1 \vdash X$ wf, $X \in \mathsf{dom}(\Gamma_1)$ and so by definition, $\Gamma_1(X) = \Gamma(X)$. By part 1. of the IH, $\Gamma_1 \mapsto N <: T$ and AS-Var finishes the case.

**Case** $\mathsf{baseTypes}_\Gamma(X) \coloneqq \mathsf{baseTypes}_\Gamma(\Gamma(X))$ (BT-Var)

By the same reasoning as in the previous case, $\Gamma_1(X) = \Gamma(X)$. Part 2. of the IH finishes the case.

**Case** $\mathsf{bound}_\Gamma(X) \coloneqq \Gamma(X)$ (B-Var)

Immediate from $\Gamma_1(X) = \Gamma(X)$.

$$\textbf{Case}\quad \frac{\Gamma \vdash x : U \quad \mathsf{ttype}(x.L,\ \mathsf{bound}_\Gamma(U)) = S_1 \mathrel{..} S_2 \qquad \Gamma \mapsto\ x.L <: S_1}{\Gamma \mapsto\ x.L <: T}\ (\text{AS-Sel1})$$

By inversion of $\Gamma_1 \vdash x.L$ pwf, $x \in \mathsf{dom}(\Gamma_1)$. By inversion of $\Gamma \vdash x : U$ via GT-Var, $\Gamma(x) = U$. Therefore by GT-Var again, $\Gamma_1 \vdash x : U$ and by inversion of $\Gamma_1$ pwf, $\Gamma_1 \vdash U$ pwf. Let $U' = \mathsf{bound}_\Gamma(U)$. By part 3. of the IH, $U' = \mathsf{bound}_{\Gamma_1}(U)$ and $\Gamma_1 \vdash U'$ pwf.
By Lemma 7.5.2, $\Gamma \vdash S_1$ wf so by part 1. of the IH, $\Gamma_1 \mapsto\ x.L <: S_1$ and AS-Sel1 finishes the case.

$$\textbf{Case}\quad \frac{\Gamma \vdash x : U \quad \mathsf{ttype}(x.L,\ \mathsf{bound}_\Gamma(U)) = S_1 \mathrel{..} S_2}{\mathsf{baseTypes}_\Gamma(x.L) \coloneqq \mathsf{baseTypes}_\Gamma(S_2)}\ (\text{BT-Sel})$$

By the same reasoning as in the previous case, $\Gamma_1 \vdash x : U, \Gamma_1 \vdash U$ pwf, $\mathsf{bound}_\Gamma(U) = \mathsf{bound}_{\Gamma_1}(U)$. By part 2. of the IH, $\mathsf{baseTypes}_\Gamma(S_2) = \mathsf{baseTypes}_{\Gamma_1}(S_2)$ and $\Gamma_1 \vdash \mathsf{baseTypes}_{\Gamma_1}(S_2)$ pwf. BT-Sel finishes the case.

$$\textbf{Case}\quad \frac{\Gamma \vdash y : U \quad \mathsf{ttype}(y.L,\ \mathsf{bound}_\Gamma(U)) = S_1 \mathrel{..} S_2}{\mathsf{bound}_\Gamma(y.L) \coloneqq \mathsf{bound}_\Gamma(S_2)}\ (\text{B-Sel})$$

Similar to the previous case but using part 3. of the IH and B-Sel.

■

**Lemma 7.5.4**

If $x \neq \mathsf{this}$, $\Gamma \vdash x : U$, $\Gamma_{[x]} \vdash U <: U'$, $\mathsf{ttype}(x.L, U') = S_1 \mathrel{..} S_2$ and $\Gamma_{[x]} \vdash S_1, S_2$ wf, then

1. $\Gamma \vdash x.L <: S_2$
2. $\Gamma \vdash S_1 <: x.L$

*Proof.* We only show part 1. since part 2. mirrors it. We proceed by induction on the derivation of $\mathsf{ttype}(x.L, U')$. Cases TT-Super, TT-AndL, TT-AndR follow by the IH and transitivity.

$$\textbf{Case}\quad \frac{\theta = [x/\mathsf{this}] \quad \sigma = [\overline{T/X}] \qquad (\mathsf{type}\, L >: S_1' <: S_2') \in \mathsf{tdecls}(C)}{\mathsf{ttype}(x.L, C[\overline{T}]) \coloneqq \sigma(\theta S_1') \mathrel{..} \sigma(\theta S_2')}\ (\text{TT-Member})$$

By DS-SelOther1.

**Case**
$$\frac{\mathsf{ttype}(x.L,\, T_1) = S_1 \mathrel{..} S_2 \qquad \mathsf{ttype}(x.L,\, T_2) = S_1' \mathrel{..} S_2'}{\mathsf{ttype}(x.L,\, T_1 \mathbin{\&} T_2) \coloneqq (S_1 \mathbin{|} S_1') \mathrel{..} (S_2 \mathbin{\&} S_2')}\ (\text{TT-AndLR})$$

$$\dfrac{\dfrac{\dfrac{}{\Gamma_{[x]} \vdash U <: T_1}\ (\text{Trans, PS-And11})}{\Gamma_{[x]} \vdash x.L <: S_2}\ (\text{IH 1.}) \qquad \dfrac{\dfrac{}{\Gamma_{[x]} \vdash U <: T_2}\ (\text{Trans, PS-And12})}{\Gamma_{[x]} \vdash x.L <: S_2'}\ (\text{IH 1.})}{\Gamma_{[x]} \vdash x.L <: S_2 \mathbin{\&} S_2'}\ (\text{PS-And2})$$

■

Because bound and baseTypes are now mutually recursive, Lemmas 6.4.2 and 6.4.3 must be proved simultaneously.

**Lemma 7.5.5**

1. If $N \in \mathsf{baseTypes}_\Gamma(S)$, then $\Gamma \vdash S <: N$.
2. If $\mathsf{bound}_\Gamma(S) = T$, then $\Gamma \vdash S <: T$.

*Proof.* By simultaneous induction on the size of the derivations of $\mathsf{baseTypes}_\Gamma(S)$ and $\mathsf{bound}_\Gamma(S)$. We only show a few cases.

**Case**
$$\frac{\Gamma \vdash x : U \qquad \mathsf{ttype}(x.L,\ \mathsf{bound}_\Gamma(U)) = S_1 \mathrel{..} S_2}{\mathsf{baseTypes}_\Gamma(x.L) \coloneqq \mathsf{baseTypes}_\Gamma(S_2)}\ (\text{BT-Sel})$$

By part 1 of the IH, $\Gamma \vdash S_2 <: N$. If $x = \mathsf{this}$, then $\mathsf{bound}_\Gamma(U) = U = C[\overline{X}]$ and by DS-SelThis1, $\Gamma \vdash x.L <: S_2$. Otherwise, by part 2. of the IH, $\Gamma \vdash U <: \mathsf{bound}_\Gamma(U)$ and by Lemma 7.5.4, $\Gamma \vdash x.L <: S_2$ too. Therefore in either case, GS-Trans finishes the case.

**Case**
$$\frac{\Gamma \vdash x : U \qquad \mathsf{ttype}(x.L,\ \mathsf{bound}_\Gamma(U)) = S_1 \mathrel{..} S_2}{\mathsf{bound}_\Gamma(x.L) \coloneqq \mathsf{bound}_\Gamma(S_2)}\ (\text{B-Sel})$$

By part 2 of the IH, $\Gamma \vdash S_2 <: T$ and the rest of the case proceeds like in the previous case.

■

**Theorem 7.5.6: Soundness of algorithmic subtyping**

If $\Gamma$ pwf, $\Gamma \mapsto S <: T$ then $\Gamma \vdash S <: T$.

*Proof.* By induction on the derivation of $\Gamma \mapsto S <: T$ as in Theorem 6.4.4. We only show the additional case AS-Sel1 since AS-Sel2 proceeds similarly.

**Case**
$$\frac{\Gamma \vdash x : U \quad \mathsf{ttype}(x.L,\ \mathsf{bound}_\Gamma(U)) = S_1 \mathrel{..} S_2 \quad \Gamma \Vdash S_2 <: T}{\Gamma \Vdash x.L <: T}\ (\text{AS-SEL1})$$

By inversion of $\Gamma \vdash x : U$ via GT-VAR, we have $\Gamma(x) = U$, therefore by inversion of $\Gamma$ pwf, $\Gamma_{[x]} \vdash U$ wf. Let $U' = \mathsf{bound}_\Gamma(U)$.

**Subcase** $x = \mathsf{this}$

In this case, $U' = U = C[\overline{X}]$ and

$$\frac{\dfrac{}{\Gamma \vdash \mathsf{this}.L <: S_2}\ (\text{DS-SELTHIS1}) \qquad \dfrac{}{\Gamma \vdash S_2 <: T}\ (\text{IH})}{\Gamma \vdash \mathsf{this}.L <: T}\ (\text{GS-TRANS})$$

**Subcase** $x \neq \mathsf{this}$

By Lemma 7.5.3, $U' = \mathsf{bound}_{\Gamma_{[x]}}(U)$ and $\Gamma_{[x]} \vdash U'$ pwf.

$$\frac{\dfrac{\dfrac{}{\Gamma_{[x]} \vdash U <: \mathsf{bound}_{\Gamma_{[x]}}(U)}\ (7.5.5) \qquad \dfrac{}{\Gamma_{[x]} \vdash S_2 \text{ pwf}}\ (7.5.2)}{\Gamma \vdash x.L <: S_2}\ (7.5.4) \qquad \dfrac{}{\Gamma \vdash S_2 <: T}\ (\text{IH})}{\Gamma \vdash x.L <: T}\ (\text{GS-TRANS})$$

∎

**Theorem 7.5.7: Correctness of Variable Avoidance**

If $\Gamma \vdash S \Updownarrow^x T_1 \mathrel{..} T_2$ then

1. $x \notin \mathsf{fv}(T_1)$ and $\Gamma \vdash T_1 <: S$
2. $x \notin \mathsf{fv}(T_2)$ and $\Gamma \vdash S <: T_2$

*Proof.* By induction on the derivation of $\Gamma \vdash S \Updownarrow^x T_1 \mathrel{..} T_2$. We only show a few cases.

**Case**
$$\frac{\Gamma \vdash \overline{S \Updownarrow^x S' \mathrel{..} S''} \qquad \Gamma \Vdash \overline{S <: S'}}{\Gamma \vdash C[\overline{S}] \Updownarrow^x C[\overline{S'}] \mathrel{..} C[\overline{S'}]}\ (\text{A-DEALIAS})$$

By part 1 of the IH, $\overline{x \notin \mathsf{fv}(S')}$, so by definition $x \notin \mathsf{fv}(C[\overline{S'}])$ which proves part 1 of the case. By Theorem 7.5.6, $\Gamma \vdash \overline{S <: S'}$ and by part 2 of the IH, $\Gamma \vdash \overline{S' <: S}$, therefore PS-INV finishes part 2 of the case.

$$\textbf{Case}\quad \frac{\begin{array}{c}\textsf{class}\ C[\overline{X <: N}] \lhd B[\overline{U}]\\ \sigma = [\overline{S/X}] \quad \Gamma \vdash B[\overline{\sigma U}] \Uparrow^{x} T\end{array}}{\Gamma \vdash C[\overline{S}] \Updownarrow^{x} \textsf{Nothing} .. T}\ (\text{A-Super})$$

> Part 1 is easy. For Part 2, by inversion of $\Gamma \vdash B[\overline{\sigma U}] \Uparrow^{x} T$ and by part 2. of the IH we must have $x \notin \textsf{fv}(B[\overline{\sigma U}])$ and $\Gamma \vdash B[\overline{\sigma U}] <: T$. By PS-Class, $\Gamma \vdash C[\overline{S}] <: B[\overline{\sigma U}]$ and GS-Trans finishes the case.

■

## 7.6 Translation

The new cases in the translation are defined in Figure 7.12. We translate DS type members as DOT type members and therefore DS type selections as DOT type selections. Local blocks are easily represented using our let-binding syntactic sugar. Translating a type member $A$ into a type $[\![A]\!]$ is straightforward, but for proper classes we also need a declaration $(\!| A |\!)$ which forces us to arbitrarily pick one of the bound of the type member. Class typing ensures that this is a valid choice as discussed in subsection 7.4.2.

### 7.6.1 Meta-theory

We only show the most interesting changes compared to subsection 6.5.1.

We cannot directly carry over Lemma 4.3.10 because "this" may be free in the method types and type member bounds declared in $C$ or one of its base class which prevents us from performing a rewriting step critical to the proof. Instead, we replace it by two less powerful lemmas which will be good enough for our purposes.

**Lemma 7.6.1**

Given $\textsf{tparams}(C) = \overline{X <: N}$, $\Gamma \dashv\vdash \Delta$, then $\Delta \vdash |C[\overline{T}]| <: |\sigma|([\![\textsf{vparams}(C)]\!] \wedge \textsf{baseArgs}(C))\}$ where $\sigma = [\overline{T/X}]$.

*Proof.* We follow the same reasoning as in Lemma 4.3.10 but with occurences of $\{\textsf{this} \Rightarrow [\![C]\!]\}$ replaced by $\{\textsf{this} \Rightarrow [\![\textsf{vparams}(C)]\!],\ \textsf{baseArgs}(C)\}$. By inversion of $\vdash C$ ok via DT-Class only $\overline{X}$ may be free in the type and value parameters of $C$ which allows us to perform the following renaming:

$$\{\textsf{this} \Rightarrow [\![\textsf{vparams}(C)]\!],\ \textsf{baseArgs}(C)\} = \{z \Rightarrow \tau([\![\textsf{vparams}(C)]\!] \bigwedge \textsf{baseArgs}(C))\}$$

where $\tau = [\overline{z.X/|X|}]$. ■

**Lemma 7.6.2**

Given $\textsf{tparams}(C) = \overline{X <: N}$, then $\Gamma \vdash x : C[\overline{T}]$ implies $|\Gamma|_{[x]} \vdash x :_{(!)} \theta[\![C]\!]$ where $\theta = [x/\textsf{this}]$.

*Proof.* We can distinguish two cases.

**Figure 7.12: Translating DS types, expressions and definitions to DOT**

All definitions from Figure 6.8 are carried over.

**Type Translation** $\boxed{|T| \coloneqq T_{\text{DOT}}}$

$$|x.L| \coloneqq x.L$$

**Expression Translation** $\boxed{|e|_\Gamma \coloneqq t_{\text{DOT}}}$

$$|\{\textbf{val}\ x = e_1;\ e_2\}|_\Gamma \coloneqq \textbf{let}\ x = |e_1|_\Gamma\ \textbf{in}\ |e_2|_\Gamma$$

**Method Translation** $\boxed{(\!| m |\!)_C \coloneqq d_{\text{DOT}}}$

$$\frac{\begin{array}{c}\textbf{class}\ C[\overline{X <: N}]\ \ldots \quad \Gamma = \overline{X <: N}, \mathsf{this} : C[\overline{X}] \\ \mathsf{mtype}(\mathsf{this}.m, C[\overline{X}]) = [\overline{Y <: P}] \to (\overline{x : U}) \to U_0 \\ \mathsf{mbody}(m, C[\overline{X}]) = e_0\end{array}}{(\!| m |\!)_C \coloneqq m(\mathsf{mtag} : |\overline{Y <: P}|, \overline{x : |U|}) : |U_0| = |e_0|_{\Gamma, \overline{Y <: P}, \overline{x:U}}}$$

**Type Declaration Translation** $\boxed{(\!| TD |\!) \coloneqq d_{\text{DOT}}}$

$$\frac{\begin{array}{c}\mathsf{tparams}(C) = \overline{X <: N} \\ \mathsf{ttype}(\mathsf{this}.A, C[\overline{X}]) = S\ ..\ T\end{array}}{\begin{array}{c}(\!| A |\!) \coloneqq (A = |T|) \\ [\![ A ]\!] \coloneqq (A : |S|\ ..\ |T|)\end{array}}$$

**Class Translation** $\boxed{(\!| C |\!) \coloneqq \overline{d_{\text{DOT}}}}$

$$\frac{\textbf{class}\ C[\overline{X <: N}]\ \ldots \quad \mathsf{baseArgs}(C) = \bigwedge \overline{Z = S}}{\begin{array}{l}(\!| C |\!) \coloneqq (\!| \mathsf{vparams}(C[\overline{X}]) |\!), (\!| \mathsf{mnames}(C) |\!)_C, \overline{Z = |S|}, (\!| \mathsf{tnames}(C) |\!)_C \\ (\!| C |\!)^T \coloneqq (\!| C |\!), \overline{X = T}\end{array}}$$

$$\frac{\textbf{trait}\ C[\overline{X <: N}]\ \ldots\ \{\overline{\mathsf{type}\ L >: S <: T};\ \ldots\}}{[\![ C ]\!] \coloneqq [\![ \mathsf{mnames}(C) ]\!]_C \wedge \mathsf{baseArgs}(C) \wedge \bigwedge \overline{(A : |S|\ ..\ |T|)}}$$

**Case** $x = \mathsf{this}$

In this case, $|\Gamma|(x) = [\![C]\!]^{\overline{|X|}}$ and $\theta[\![C]\!] = [\![C]\!]$, hence

$$\dfrac{\dfrac{}{|\Gamma|_{[x]} \vdash \mathsf{this} :_{(!)} [\![C]\!]^{\overline{|X|}}}\ (\text{VAR})}{|\Gamma|_{[x]} \vdash \mathsf{this} :_{(!)} [\![C]\!]}\ (\text{SUB, AND11})$$

**Case** $x \neq \mathsf{this}$

In this case, $|\Gamma|(x) = |C[\overline{T}]|$ and it is easy to see that $|\Gamma|_{[x]} \vdash |C[\overline{T}]| <: \{\mathsf{this} \Rightarrow [\![C]\!]\}$, hence

$$\dfrac{\dfrac{\dfrac{}{|\Gamma|_{[x]} \vdash x :_{(!)} |C[\overline{T}]|}\ (\text{VAR})}{|\Gamma|_{[x]} \vdash x :_{(!)} \{x \Rightarrow \theta[\![C]\!]\}}\ (\text{SUB})}{|\Gamma|_{[x]} \vdash x :_{(!)} \theta[\![C]\!]}\ (\text{VARUNPACK})$$

■

**Theorem 7.6.3: Translation preserves substitution**

$|\sigma S| = |\sigma||S|$

*Proof.* By structural induction on $S$ as in Theorem 4.3.7.

**Case** $S = x.L$

By definition, $\sigma = [\overline{T/X}]$ and $|\sigma| = [\overline{|T|/|X|}]$ for some $\overline{T}$, $\overline{X}$. Therefore, $|\sigma(x.L)| = |x.L|$ since $x.L \notin \mathsf{dom}(\sigma)$ and $|\sigma||x.L| = |x.L|$ since $|x.L| = x.L \notin \mathsf{dom}(|\sigma|)$.

■

**Lemma 7.6.4**

Given $\Gamma \dashv\vdash \Delta$, $\Gamma \vdash \sigma(\theta S)$ wf and $(\overline{X <: N}, \mathsf{this} : C[\overline{X}]) \vdash S$ pwf, then if either $\Gamma \vdash x : C[\overline{T}]$ or $\Delta \vdash_{[x]} x :_! |C[\overline{T}]|$, we must have $\Delta \vdash \theta|S| =:= |\sigma(\theta S)|$ where $\theta = [x/\mathsf{this}]$ and $\sigma = [\overline{T/X}]$.

*Proof.* By structural induction on $S$. Cases $S = T_1$ **&** $T_2$ and $S = T_1$ **|** $T_2$ easily follow by the IH.

**Case** $S = Z$

By inversion of $(\overline{X <: N}, \mathsf{this} : C[\overline{X}]) \vdash Z$ pwf, we must have $Z = X_i \in \overline{X}$. We find $\theta|X_i| = x.X_i$ and $|\sigma(\theta X_i)| = |\sigma X_i| = |T_i|$.

**Subcase** $x = \mathsf{this}$

We must have $C[\overline{T}] = C[\overline{X}]$ and $\theta|S| = |S| = |\sigma(\theta S)|$ which finishes the subcase.

**Subcase** $x \neq \mathsf{this}$

Either $\Gamma \vdash x : C[\overline{T}]$ which implies $\Delta \vdash_{[x]} x :_! |C[\overline{T}]|$ or we already have $\Delta \vdash_{[x]} x :_! |C[\overline{T}]|$. Hence,

$$\cfrac{\cfrac{\Delta \vdash x :_! |C[\overline{T}]|}{\Delta \vdash x :_! (X_i = T_i)}\ (\textsc{Sub})}{\Delta \vdash x.X_i =:= |T_i|}\ (\textsc{Sel1}, \textsc{Sel2})$$

**Case** $S = C[\overline{T}]$

By definition,

$$|\sigma(\theta C[\overline{T}])| = |(C[\overline{\sigma(\theta T)}])| = \mathsf{ct}.C \wedge \bigwedge \overline{X = |\sigma(\theta T)|}$$
$$\theta|C[\overline{T}]| = \mathsf{ct}.C \wedge \bigwedge \overline{X = \theta|T|}$$

By the IH, $\Delta \vdash \overline{|\sigma(\theta T)| =:= \theta|T|}$ and finishing the case is easy.

**Case** $S = \mathsf{this}.L$

We have $\theta|S| = x.L$ and $|\sigma(\theta S)| = |x.L| = x.L$ so GS-Refl finishes the case.

**Case** $S = y.L$ where $y \neq \mathsf{this}$

We have $\theta|S| = |S| = y.L$ and $|\sigma(\theta S)| = |y.L| = y.L$ so GS-Refl finishes the case once again. ■

**Lemma 7.6.5: Context truncation preserves environment entailment**

If $\Gamma \dashv\vdash \Delta$ and $x \in \mathsf{dom}(\Gamma)$, then $\Gamma_{[x]} \dashv\vdash \Delta_{[x]}$

*Proof.* By induction on $\Gamma \dashv\vdash \Delta$. Case EE-Empty is trivial.

**Case** $$\cfrac{\Gamma' \dashv\vdash \Delta \qquad \overline{\Delta \vdash |X| <: |N|}}{\Gamma', \overline{X <: N} \dashv\vdash \Delta}\ (\textsc{EE-Typs})$$

$(\Gamma', \overline{X <: N})_{[x]} = \Gamma'_{[x]}$ and the IH finishes the case.

$$\textbf{Case}\quad \frac{\begin{array}{c}\mathsf{tparams}(C) = \overline{X <: N} \\ \overline{X <: N} \dashv\Vdash \Delta', \mathsf{this} : T \\ \Delta', \mathsf{this} : T \vdash \mathsf{this} :_! [\![C]\!] \wedge \overline{X : \bot .. |N|}\end{array}}{\overline{X <: N}, \mathsf{this} : C[\overline{X}] \dashv\Vdash \Delta', \mathsf{this} : T, \Delta''}\ (\text{EE-This})$$

If $x = \mathsf{this}$ then $\Gamma_{[x]} = \Gamma$ and $\Delta_{[x]} = \Delta', \mathit{this} : T$ so EE-This finishes the case. Otherwise $x \notin \mathsf{dom}(\Gamma)$ and the case is trivially true.

$$\textbf{Case}\quad \frac{\Gamma' \dashv\Vdash \Delta' \quad y \neq \mathsf{this}}{\Gamma', y : T \dashv\Vdash \Delta', y : |T|, \Delta''}\ (\text{EE-Var})$$

If $x = y$ then $\Gamma_{[x]} = \Gamma$ and $\Delta_{[x]} = \Delta', x : |T|$ so EE-Var finishes the case. Otherwise $\Gamma_{[x]} = \Gamma'_{[x]}$ and $\Delta_{[x]} = \Delta'_{[x]}$ so the IH finishes the case.

■

**Theorem 7.6.6: Partial well-formedness preservation**

If $\Gamma$ pwf and $\Gamma \dashv\Vdash \Delta$, then $\Gamma \vdash S$ pwf implies $\Delta \vdash |S|$ wf.

*Proof.* By induction on the derivation of $\mathsf{fv}(S)$. We only show the additional case compared to Theorem 6.5.1.

**Case** $\mathsf{fv}(x.L) := \{x\}$

$|x.L| = x.L$ and since $\Gamma \dashv\Vdash \Delta$, we have $x \in \mathsf{dom}(\Delta)$ so $\Delta \vdash x.L$ wf.

■

**Theorem 7.6.7: Subtyping preservation**

If $\Gamma$ pwf and $\Gamma \dashv\Vdash \Delta$, then $\Gamma \vdash S <: T$ implies $\Delta \vdash |S| <: |T|$.

*Proof.* By induction on the derivation of $\Gamma \vdash S <: T$ like in Theorem 6.5.2. We only show the additional cases DS-SelThis1 and DS-SelOther1 since DS-SelThis2 and DS-SelOther2 are similar.

$$\textbf{Case}\quad \frac{\Gamma \vdash \mathsf{this} : C[\overline{X}] \quad \mathsf{ttype}(\mathsf{this}.L, C[\overline{X}]) = S_1 .. S_2}{\Gamma \vdash \mathsf{this}.L <: S_2}\ (\text{DS-SelThis1})$$

By inversion of EE-This, $\Delta_{[\mathsf{this}]} \vdash [\![C]\!]^{\overline{|X|}}$ where $[\![C]\!] = (... \wedge [\![L]\!]_C \wedge ...)$ and $[\![L]\!]_C = (L : |S_1| .. |S_2|)$ by definition. Hence,

$$\dfrac{\dfrac{\dfrac{\dfrac{}{\Delta_{[\mathsf{this}]} \vdash \mathsf{this} :_! \llbracket C \rrbracket^{\overline{|X|}}}\,(\textsc{Var})}{\Delta_{[\mathsf{this}]} \vdash \mathsf{this} :_! (L : |S_1| \mathbin{..} |S_2|)}\,(\textsc{Sub}, 2.4.5)}{\Delta_{[\mathsf{this}]} \vdash \mathsf{this} :_! (L : \bot \mathbin{..} |S_2|)}\,(\textsc{Trans}, \textsc{Typ})}{\Delta \vdash \mathsf{this}.L <: |S_2|}\,(\textsc{Sel1})$$

**Case** $$\dfrac{\Gamma \vdash x : T \quad \Gamma_{[x]} \vdash T <: C[\overline{U}] \quad x \neq \mathsf{this} \qquad (\mathsf{type}\, L >: S_1 <: S_2) \in \mathsf{tdecls}(C) \quad \sigma = [\overline{U/X}] \quad \theta = [x/\mathsf{this}]}{\Gamma \vdash x.L <: \sigma(\theta S_2)}\,(\textsc{DS-SelOther1})$$

By inversion of EE-Var, $\Delta(x) = |T|$. Hence,

$$\dfrac{\dfrac{\dfrac{\dfrac{\dfrac{}{\Delta_{[x]} :_! |T|}\,(\textsc{Var}) \quad \dfrac{}{\Delta_{[x]} \vdash |T| <: |C[\overline{U}]|}\,(\text{IH})}{\Delta_{[x]} :_! |C[\overline{U}]|}\,(\textsc{Sub})}{\Delta_{[x]} :_! \{x \Rightarrow \theta\llbracket C \rrbracket\}}\,(\textsc{Sub})}{\Delta_{[x]} \vdash x :_{(!)} \theta\llbracket C \rrbracket}\,(\textsc{VarUnpack}) \qquad \llbracket C \rrbracket = \ldots \wedge (L : |S_1| \mathbin{..} |S_2|) \wedge \ldots}{\dfrac{\dfrac{\Delta_{[x]} \vdash x :_{(!)} (L : \theta|S_1| \mathbin{..} \theta|S_2|)}{\Delta_{[x]} \vdash x :_{(!)} (L : |\sigma(\theta S_1)| \mathbin{..} |\sigma(\theta S_2)|)}\,(\textsc{Sub}, 7.6.4)}{\Delta \vdash x.L <: |\sigma(\theta S_2)|}\,(\textsc{Sel1})}\,(\textsc{Sub}, 2.4.5)$$

] ∎

**Lemma 7.6.8: Class translation preserves methods**

Given $\Delta$ wf, $\Delta_{[\mathsf{ct}]} \vdash \mathsf{ct} :_! \llbracket CT \rrbracket$, $\Delta \vdash \llbracket CT \rrbracket$, $\Gamma \vdash x_0 : T$ and $\Delta \vdash \overline{y : |\sigma U|}$, $|V| <: |\sigma P|$ where $\sigma = [\overline{V/Y}]$ then $\mathsf{mtype}(x_0.m, T) = [\overline{Y <: P}] \to (\overline{y : U}) \to U_0$ implies $\Delta, x_{\mathsf{mtag}} : \{_ \Rightarrow \overline{Y = |V|}\} \vdash x_0.m(x_{\mathsf{mtag}}, \overline{y}) : |\sigma U_0|$.

*Proof.* By induction on the derivation of $\mathsf{mtype}(m, T)$. Cases DM-Super,DM-AndLR,DM-AndL and DM-AndR are respectively similar to cases PM-Super, PM-AndLR, PM-AndL and PM-AndR of Lemma 5.5.8.

**Case** $$\frac{\theta = [x/\mathsf{this}] \quad \tau = [\overline{T/X}] \qquad (\mathbf{def}\ m[\overline{Y <: P'}](\overline{x : U'}) : U_0' = \underset{\sim}{e_0}) \in \mathsf{mdecls}(C)}{\mathsf{mtype}(x.m, C[\overline{T}]) \coloneqq [\overline{Y <: \tau(\theta P')}] \to (\overline{y : \tau(\theta U')}) \to \tau(\theta U_0')} \text{(DM-Impl)}$$

$$\dfrac{\dfrac{\dfrac{\Gamma \vdash x_0 : C[\overline{T}]}{\Delta \vdash x_0 : \theta[\![C]\!]}\text{(7.6.2)} \qquad [\![C]\!] = \ldots \wedge [\![m]\!]_C \wedge \ldots}{\Delta \vdash x_0 : (m(\mathsf{mtag} : \theta|\overline{Y <: P'}|, \overline{y : \theta|U|}) : \theta|U_0|)}\text{(Sub, 2.4.5)}}{\Delta \vdash x_0 : (m(\mathsf{mtag} : |\overline{Y <: \tau(\theta P')}|, \overline{y : |\tau(\theta U)|}) : |\tau(\theta U_0)|)}\text{(Sub, 7.6.4)}$$

By a similar argument than in case GT-Invk of Theorem 4.3.18 we find

$$\Delta \vdash x_0.m(x_{\mathsf{mtag}}, \overline{y}) : |\tau(\theta U_0)|$$

No special handling is required for dependent parameters since TApp' is already generic enough to handle them.

■

**Lemma 7.6.9: Type member translation preserves overriding relationship**

Given $\mathsf{tparams}(C) = \overline{X_C <: N_C}$, $B[\overline{U}] \in \mathsf{parents}(C[\overline{X_C}])$, $\mathsf{tparams}(B) = \overline{X_B <: \ldots}$, $\Gamma = (\overline{X_C <: N_C}, \mathsf{this} : C[\overline{X_C}])$ and $\Gamma \dashv\!\!\!\mid \Delta$, then $L \in \mathsf{tnames}(B)$ implies $\Delta \vdash [\![L]\!]_C <: [\![L]\!]_B$.

*Proof.* Let

$$\mathsf{ttype}(\mathsf{this}.L, B[\overline{X_B}]) = T_1 \mathrel{..} T_2$$
$$\mathsf{ttype}(\mathsf{this}.L, C[\overline{X_C}]) = S_1 \mathrel{..} S_2$$

then $\mathsf{ttype}(\mathsf{this}.L, B[\overline{U}]) = \sigma T_1 \mathrel{..} \sigma T_2$ by observation and we have

$$\dfrac{\dfrac{\dfrac{\Delta \vdash |T_1| <: |\sigma S_1|, |\sigma S_2| <: |T_1|}{\Delta \vdash |T_1| <: |\sigma||S_1|, |\sigma||S_2| <: |T_1|}\text{(7.6.3)} \qquad \dfrac{}{\Delta \vdash \overline{U =:= X_B}}\text{(4.3.8)}}{\Delta \vdash |T_1| <: |S_1|, |S_2| <: |T_1|}\text{(Trans, 2.4.6)}}{\Delta \vdash (L : |S_1| \mathrel{..} |S_2|) <: (L : |T_1| \mathrel{..} |T_2|)}\text{(Typ)}$$

Hence, we only need to prove that $\Delta \vdash |T_1| <: |\sigma S_1|, |\sigma S_2| <: |T_1|$. We proceed by inversion of $\mathsf{ttype}(\mathsf{this}.L, C[\overline{X_C}])$.

**Case** $$\frac{(\mathsf{type}\ L >: S_1 <: S_2) \in \mathsf{tdecls}(C)}{\mathsf{ttype}(\mathsf{this}.L, C[\overline{X_C}]) \coloneqq S_1 \mathrel{..} S_2}\text{(TT-Member)}$$

By inversion of $\vdash C$ ok via either DT-Class or DT-Trait we must have $\mathsf{override}_\Gamma(L, C[\overline{X_C}], B[\overline{U}])$ and therefore $\Gamma \vdash \sigma T_1 <: S_1, S_2 <: \sigma T_2$. Theorem 7.6.7 finishes the case.

**Case** $$\frac{\mathsf{parents}(C[\overline{X_C}]) = \overline{P} \quad (\mathsf{type}\, L \,...) \notin \mathsf{tdecls}(N)}{\mathsf{ttype}(\mathsf{this}.L,\, C[\overline{X_C}]) \coloneqq \mathsf{ttype}(\mathsf{this}.L,\, \& \overline{P})} \;(\text{TT-Super})$$

By inversion of $\mathsf{ttype}(\mathsf{this}.L,\, \& \overline{P})$ via one of TT-AndLR, TT-AndL and TT-AndR we must have

$$\mathsf{ttype}(\mathsf{this}.L,\, \& \overline{P}) = \left(\bigvee \overline{S_1'}\right) .. \left(\& \overline{S_2'}\right) \qquad \text{where } \sigma T_1 \in \overline{S_1} \text{ and } \sigma T_2 \in \overline{S_2}$$

By LS-Or21 and LS-Or22, $\Delta \vdash |\sigma T_1| <: \bigvee \overline{|S_1|}$ and by PS-And11 and PS-And12, $\bigwedge \overline{|S_2|} <: |\sigma T_2|$. ■

**Theorem 7.6.10: Typing translation is type-preserving**

If $\Gamma \dashv\vdash \Delta$ and $\Gamma \vdash e : T$, then $\Delta \vdash |e|_\Gamma : |T|$.

*Proof.* By induction on the derivation of $\Gamma \vdash e : T$ as in Theorem 6.5.3. Case DT-Invk proceeds like case GT-Invk of Theorem 5.5.9.

**Case** $$\frac{\Gamma \vdash e_1 : S \quad \Gamma, x : S \vdash e_2 : T \quad \Gamma, x : S \vdash T \Uparrow^x T'}{\Gamma \vdash \{\{\mathbf{val}\, x = e_1;\, e_2\}\} : T'} \;(\text{DT-Block})$$

We have $|\{\mathbf{val}\, x = e_1;\, e_2\}|_\Gamma = (\mathbf{let}\ x = |e_1|_\Gamma\ \mathbf{in}\ |e_2|_\Gamma)$.

$$\dfrac{\dfrac{}{\Delta \vdash |e_1|_\Gamma : |S|}\,(\text{IH}) \quad \dfrac{\dfrac{}{\Delta, x : |S| \vdash |e_2|_\Gamma : |T|}\,(\text{IH}) \quad \dfrac{}{\Delta, x : |S| \vdash |T| <: |T'|}\,(7.6.7)}{\Delta, x : |S| \vdash |e_2|_\Gamma : |T'|}\,(\text{Sub})}{\Delta \vdash \mathbf{let}\ x = |e_1|_\Gamma\ \mathbf{in}\ |e_2|_\Gamma : |T'|}\,(\text{Let},\, 7.5.7)$$

■

**Theorem 7.6.11: Program translation is type-preserving**

If $\varnothing \vdash_{\text{DS}} T$ wf and $\varnothing \vdash_{\text{DS}} e : T$ then $\varnothing \vdash_{\text{DOT}} \mathbf{let}\ \mathsf{ct} = \{\mathsf{ct} \Rightarrow (\!| CT |\!)\}\ \mathbf{in}\ |e|_\varnothing : |T|$.

*Proof.* As in Theorem 6.5.5 but using Theorem 7.6.10. ■

# 8 Conclusion

In this thesis, we rigorously bridged the gap between Scala and DOT for the first time via type-preserving compilation. This involved specifying a significant subset of Scala, as well as extending DOT itself with new rules and a generalized type safety theorem.

## 8.1 Future work

### 8.1.1 Extending DOT

This work served as a real-world benchmark for the two main flavors of DOT: wfDOT [Amin, Grütter, et al. 2016] and oopslaDOT [Rompf and Amin 2016]. We demonstrated that the limitations imposed by wfDOT are not mere inconvenience but real showstoppers for modeling Scala. We therefore believe that existing extensions of wfDOT such as pDOT [Rapoport and Lhoták 2019] should be "rebased" on oopslaDOT, although we have not investigated how much effort this would require.

### 8.1.2 Specifying Scala

The road ahead is clear: the Scala language has a large surface which still needs to be formalized. We believe that the meta-theoretical techniques we developed in this thesis should let us encode many more Scala features. Below, we present a non-exhaustive list of such features.

#### Inner classes

FJI [Igarashi and Pierce 2002] extends Featherweight Java with *inner* classes: classes defined inside other classes. The paper defines both operational semantics for FJI and a translation from FJI into FJ whose semantics are proven to be equivalent to the operational definition.

While one could implement a translation from FJI into DOT by composing the existing FJI-into-FJ and FJ-into-DOT translations, it would be more interesting to define a simpler translation from FJI into pDOT that does not involve flattening the class hierarchy. To reuse our type-preserving compilation proofs, this would require a version of pDOT built on top of oopslaDOT as mentioned in subsection 8.1.1.

It seems that no attempt has been made so far to combine FJI with FGJ. Once this work is done, combining FJI with Dependent Scala should be straightforward.

### Local classes

Local classes are classes defined in a local block, usually in a method. They can capture variables from their environment and therefore can be used to implement closures. Despite being a long-standing feature of Java [Gosling et al. 2015, § 14.3], there are no "FJ with local classes" calculus in the literature. However, FJ&$\lambda$ [Bettini et al. 2018] does extend FJ with Java 8 lambdas [Gosling et al. 2015, § 15.27] which can express an important subset of the semantics of local classes.

Properly supporting local classes in our source calculus would require some amount of rethinking since our formalization relies heavily on the presence of a single global class table known ahead of time. As an intermediate step, one could instead just support lambdas as in FJ&$\lambda$.

Unlike inner classes, local classes are not reachable via a path, and therefore should be translatable into oopslaDOT without having to combine it with pDOT first.

### Definition-site variance

Scala lets us write variance annotations on class type parameters, for example given **`trait`** `List[+X]`, then $\Gamma \vdash S <: T$ implies $\Gamma \vdash \texttt{List}[S] <: \texttt{List}[T]$.

It should be easy to extend Dependent Scala to support such annotations by taking inspiration from existing FJ-like calculi with definition-site variances [Emir et al. 2006; Kennedy and Pierce 2007].

A possible DOT representation is sketched out in [Rompf and Amin 2016, §§ 5.2, 7]. In our case, for subtyping preservation to hold, we would translate $\texttt{List}[T]$ to $\mathsf{ct}.\texttt{List} \wedge \{_ \Rightarrow \texttt{X} <: |T|\}$. Similarly, baseArgs must take variance into account, but the interaction of inheritance and variance in Scala is somewhat complex and still under active discussion (see https://github.com/lampepfl/dotty/issues/11834).

### Use-site variance (also known as "wildcards")

Java wildcards are a long-running topic of studies due to their complex interactions with other type system features [Cameron, Drossopoulou, and Ernst 2008; Igarashi and Viroli 2006; Daniel Smith and Cartwright 2008; Tate, Leung, and Lerner 2011]. Supporting them is important for expressiveness since they would let us return more precise types in baseArgs (Section 6.3) and variable avoidance (subsection 7.4.1).

Note that wildcard capture is more expressive in Scala than in Java. Consider,

```
class Box[T] {
  def push(x: T): Unit = ???
  def pop(): T = ???
}
class Test {
  def pushPop(x: Box[?]): Unit =
    x.push(x.pop())
}
```

The corresponding Java code does not typecheck, but it works in Scala 3 where the type of both `x.pop()` and the argument of `x.push` is `x.T` (users cannot write this type, but internally type parameters are handled as if they were type members). We anticipate that our formalization could support this after allowing type selections on type parameters and adding an extra typing rule of the form,

$$\frac{\Gamma \vdash x : C[\overline{? >: S <: T}] \quad \mathsf{tparams}(C) = \overline{X <: \ldots}}{\Gamma \vdash x : C[\overline{x.X}]} \qquad \text{(T-CAPTURE)}$$

Since Dependent Scala already desugars method calls to only involve variables as receivers and arguments, this should be enough to support all possible wildcard captures. If this works, it would make our formalization of wildcards significantly simpler than the usual one based on existential types [Cameron, Drossopoulou, and Ernst 2008].

The type translation of wildcards into DOT is straightforward: the type `Box[?` $<: T$`]` should be translated as $\mathsf{ct.Box} \wedge \{_ \Rightarrow \mathsf{X} <: |T|\}$. For type-preservation to hold, an extra DOT rule corresponding to T-CAPTURE is likely to be necessary:

$$\frac{\Gamma \vdash x : (L : S \,..\, T)}{\Gamma \vdash x : (L = x.L)} \qquad \text{(CAPTURE)}$$

### Pattern matching

Pattern matching in Scala has a large surface syntax [Odersky et al. 2021a, ch. 8; Liu et al. 2022], but the core semantics (including GADT-like inferred local constraints) have been formalized in cDOT [Boruch-Gruszecki et al. 2022]. Because cDOT extends pDOT which itself extends wfDOT, extending our type-preserving translation proofs to use cDOT as a target calculus will first require rebasing pDOT on top of oopslaDOT as mentioned in subsection 8.1.1.

### Higher-kinded types

[Odersky, Martres, and Petrashko 2016] explores how to model higher-kinded types in a DOT-like setting but concludes that a direct representation is a better approach, at least for a compiler implementation. As a stepping stone towards a higher-kinded DOT, [Stucki and Giarrusso 2021] defines $F^{\omega}_{..}$, an extension of $F^{\omega}_{<:}$ with (possibly higher-kinded) *type intervals*, but without type members. A sketch of $F^{\omega}_{..}$ extended with type members is discussed in [Stucki 2017, ch. 6].

#### Distributivity of intersections and unions in subtyping

As mentioned in subsection 5.5.1, we were unable to extend DOT with a rule of the form,

$$\Gamma \vdash (m(x : S) : T_1) \wedge (m(x : S) : T_2) <: (m(x : S) : T_1 \wedge T_2) \qquad \text{(AND-FUN)}$$

This is unfortunate since Scala subtyping does rely on this rule in practice. In fact, to be faithful to Scala we would need a more general rule of the form,

$$\Gamma \vdash (m(x : S_1) : T_1) \wedge (m(x : S_1) : T_2) <: (m(x : S_1 \vee S_2) : T_1 \wedge T_2) \quad \text{(AND-FUN')}$$

While AND-FUN is standard [Barendregt, Coppo, and Dezani-Ciancaglini 1983], AND-FUN' seems more controversial.[1] It is consistent with the system presented in [Pottier 1998], but that system only allows intersection types in negative (contravariant) positions and union types in positive (covariant) ones.

As remarked in [Giarrusso et al. 2020, § 4.4], DOT also lacks a rule for distributivity of intersections over unions which Scala assumes:

$$\Gamma \vdash (S \vee T) \wedge U <: (S \wedge U) \vee (T \wedge U) \qquad \text{(DISTR-}\wedge\text{-}\vee\text{-<:)}$$

#### Type inference

Scala source code is more flexible than the calculi we've developed so far: type arguments and method result types can be omitted and inferred by the typechecker. [Daniel Smith and Cartwright 2008] specifies how constraints are accumulated given a set of subtyping rules based on Java (augmented with first-class intersection types, union types, and wildcards with both lower-bounds and upper-bounds), but it leaves out the actual typing procedure. While Scala type inference is local (in particular, mutually recursive methods cannot all omit their result types), the approach used in the Scala 3 compiler is broadly similar to [Parreaux 2020] which describes a sound and complete global type inference algorithm for a structural type system with unions and intersections.

### 8.1.3 Mechanization

In this work, we did not attempt to mechanize our type-preserving translation proofs. It is not clear to us if this is something that could be on top of the existing oopslaDOT mechanization. For example, the existing mechanization auto-assigns a numerical label to type declarations based on the order they appear in a given object initialization, but we really need to be able to distinguish the type declarations corresponding to different class type parameters. Ideally, a mechanization would be presented much like this thesis as a series of calculi without duplicating the same proofs every time, but proof reuse seems to still be an active area of research [Delaware, Cook, and Batory 2011; Delaware, S. Oliveira, and Schrijvers 2013; Forster and Stark 2020].

[1] See https://github.com/lampepfl/dotty-feature-requests/issues/51.

## 8.2 Related work

### 8.2.1 Type-preserving compilation

The original presentation of FGJ [Igarashi, Pierce, and Wadler 2001] already included a proof of type-preserving translation into FJ, but compensating for type erasure requires introducing casts in the translation, and the type safety theorem of FJ does not apply to program with downcasts. So the translation in itself did not establish soundness.

[League, Shao, and Trifonov 2002] describes a type-preserving compilation scheme of FJ into System $F_\omega$ with multiple extensions including recursive types. They include support for casts and separate compilation as their goal is to develop a practical Java compiler.

### 8.2.2 Other works on DOT

For completeness sake, we mention [Amin and Rompf 2017] which recasts oopslaDOT with big-step semantics and [Rapoport, Kabir, et al. 2017] which provides an alternative proof of soundness for wfDOT including a full mechanization.

### 8.2.3 Multiple Inheritance and the Diamond Problem

What should happen when multiple matching methods from unrelated classes are inherited? There is no standard solution here but languages usually pick one of the following approaches:

- In Java and C++ with virtual inheritance, the class definition is considered invalid and an error is emitted.
- In C++ with non-virtual inheritance, the ambiguity resolution is delayed until the method call site, where the user can "upcast" the receiver to manually resolve the ambiguity. See [Wasserrab et al. 2006] for a precise treatment of inheritance in C++ including a soundness proof (but make sure to prepare a pot of coffee first). A similar solution is implemented on top of Featherweight Java by [Wang et al. 2018] which also lets the implementer of a method manually specify which method they are overriding in case of ambiguity.
- Like Scala, several languages will attempt to determine a linearization order for the parent classes and use that to resolve the ambiguity. The **C3 linearization algorithm** [Barrett et al. 1996] originally defined for Dylan is especially popular, being notably used by Python and Raku. This form of linearization is guaranteed to be monotonic: two classes will always appear in the same order in any given linearization. This isn't true in Scala when traits are involved which lets us define class hierarchies more freely at the cost of making linearization harder to reason about.

### 8.2.4 Intersection types

Featherweight Java was first extended with interfaces and intersection types faithful to Java semantics in FJ&$\lambda$[2] [Bettini et al. 2018]. In Java, intersection types are not first class types: the operands of the intersection cannot be type variables and the intersection itself can only appear in casts and upper-bounds of type parameters. FJP&$\lambda$ [Dezani-Ciancaglini, Giannini, and Venneri 2019] generalized FJ&$\lambda$ to allow intersections in any position (as in Scala) and [Dezani-Ciancaglini, Giannini, and Venneri 2020] presented a type-preserving translation FJP&$\lambda$, into FJ&$\lambda$.

Pathless Scala can be seen as a generalization of FJP&$\lambda$, but we found it easier to extend FGJ with traits and intersections rather than to extend FJP&$\lambda$ with polymorphism and generalize its notion of interfaces to traits. We make use of a fragment of FJ&$\lambda$ stripped of intersections and lambdas to model Java bytecode as a calculus in Appendix A.

### 8.2.5 Union types

[Igarashi and Nagira 2006] first extended FJ with union types as well as a *case analysis* expression complete with exhaustiveness checks which resembles pattern matching with type tests in Scala. Unlike Scala, they allow selecting a method on a union if a method with the given name exists on each side of the union, even if it is not defined in a common base type.

[Rehman et al. 2022] develops a calculus with both unions and *disjoint switches* inspired by Ceylon which requires the cases of a switch to correspond to non-overlapping type tests. Interestingly, Scala 3's match types construct also relies on type disjointness to define its reduction algorithm as described in [Blanvillain et al. 2022, § 2.2].

[2] FJ&$\lambda$ doesn't allow an abstract method to override a concrete one so it is slightly less expressive than Java.

# A Type erasure for Pathless Scala

This chapter is adapted from [Martres 2021].

While DOT has been very useful as a reasoning tool for various aspects of the Scala type system, it is not really suitable for answering questions such as "How do I compile this Scala program to Java bytecode?".[1]

To answer this question our main source of inspiration will be [Igarashi, Pierce, and Wadler 2001] which defines two calculi: Featherweight Java (FJ) which models single-class inheritance and Featherweight Generic Java (FGJ) which adds type parameters to the language, and then proceeds to define a way to compile FGJ to FJ via *erasure*.

Real Scala compilers erase traits to Java interfaces, but FJ does not model interfaces so cannot be directly used as a target for our erasure. Instead our target calculus is a fragment of FJ&$\lambda$ [Bettini et al. 2018] which extends FJ with interfaces. FJ&$\lambda$ also supports intersections and lambdas, but because these features are not present in Java bytecode, they are not useful for our purpose and we do not use them in our erasure mapping.

FJ&$\lambda$ stripped of intersections and lambdas makes for a great target calculus as it closely models most of the important aspects of Java bytecode, although we would really need to extend it with overloading to describe Scala's erasure faithfully.

Our target calculus is a fragment of FJ&$\lambda$ [Bettini et al. 2018] which extends FJ with interfaces. FJ&$\lambda$ also supports intersections and lambdas, but because these features are not present in Java bytecode, they are not useful for our purpose and we do not use them in our erasure mapping. We name the resulting fragment Featherweight Java with Default methods (FJD).[2]

---

[1] The answer to this question matters even when compiling Scala to a different backend such as JavaScript, because alternative backends strive to preserve the semantics of the JVM to ease cross-compilation [Doeraene 2018, § 2.1].

[2] FJI was already taken by Featherweight Java with Inner classes [Igarashi and Pierce 2002].

**Figure A.1: FJD: Syntax**

| | | | |
|---|---|---|---|
| $B, C, D, E$ | Class name | $L$ ::= | Class declaration |
| $f, g$ | Class field | $\textbf{class}\, C \lhd B, \overline{D}\, \{\overline{E\, f};\ K;\ \overline{M}\}$ | proper class |
| $m$ | Method name | $\textbf{interface}\ C \lhd \overline{B}\, \{\overline{H};\ \overline{M}\}$ | interface |
| $\Gamma$ ::= | Context | $H$ ::= | Abstract method |
| $\varnothing \mid \Gamma, x : C$ | | $C\, m(\overline{C\, x})$ | |
| | | $M$ ::= | Concrete method |
| | | $H = e_0$ | |
| | | $e$ ::= | Expression |
| | | $x$ | variable |
| | | $e.f$ | field access |
| | | $e_0.m(\overline{e})$ | method call |
| | | $\textbf{new}\, C(\overline{e})$ | object |
| | | $(C)e$ | cast |

## A.1 Type Erasure

Given a type environment $\Gamma$, we write $|T|_\Gamma$ for the type erasure of $T$ which is defined in FGJ as:

$$|X|_\Gamma := |\Gamma(X)|_\Gamma$$
$$|C[...]|_\Gamma := C$$

In general, we strive to have erasure preserve as much of the structure of the original program as possible to keep the translation simple and to allow interoperability between programs written in the source and target language. In particular, the mapping above preserves subtyping in FGJ: if $\Gamma \vdash S <:_{FGJ} T$ then $|S|_\Gamma <:_{FJ} |T|_\Gamma$ (see [Igarashi, Pierce, and Wadler 2001, Lemma A.3.5]) which reduces the amount of casts that need to be inserted when erasing expressions to a minimum (see [Igarashi, Pierce, and Wadler 2001, Theorem 4.5.3]).

Unfortunately, no matter how we erase intersection types, we cannot preserve subtyping in general because although $T_1$ **&** $T_2$ is the greatest lower bound of $T_1$ and $T_2$, there might not exist a specific type in FJD representing the greatest lower bound of $|T_1|_\Gamma$ and $|T_2|_\Gamma$.[3] Nevertheless, since we're trying to preserve as much structure as possible, it seems logical to define:

$$|T_1 \,\&\, T_2|_\Gamma := \mathsf{erasedGlb}(|T_1|_\Gamma, |T_2|_\Gamma)$$

where `erasedGlb` always returns one of its arguments. In fact this is what both Java and Scala do, but they differ on the implementation of `erasedGlb`:

- Java simply defines $\mathsf{erasedGlb}(T_1, T_2) := T_1$ [Gosling et al. 2015, § 4.6]. This means that

[3] Technically, subtyping would be preserved if we erased all types to Object, but this wouldn't be practical since it would require many more casts in expression erasure and impede interoperability between Scala and Java.

the user can tweak the erasure by reordering types which can be useful for evolving code in a binary-compatible way.

- Scala 2 defines `erasedGlb` to prefer subtypes over supertypes (thus actually returning the greatest lower bound of the erased types) and proper classes over traits (because both casting and method call are usually faster on classes than on interfaces [Click and Rose 2002; Shipilëv 2020]). Unfortunately, completely specifying the behavior of Scala 2 here is extremely hard because it inadvertently depends on implementation details of the compiler[4]
- Scala 3 preserves the two properties from Scala 2 mentioned above and additionally ensures that erasure preserves commutativity of intersection ($|T_1 \mathbin{\&} T_2|_\Gamma = |T_2 \mathbin{\&} T_1|_\Gamma$) by applying a tie-break based on the lexographical order of the names of the compared types. The following pseudo-code accurately specifies its behavior[5]:

```
1 def erasedGlb(tp1: Type, tp2: Type): Type =
2   if tp1.isProperClass && !tp2.isProperClass then
3     return tp1
4   if tp2.isProperClass && !tp1.isProperClass then
5     return tp2
6   if tp1 <: tp2 then return tp1
7   if tp2 <: tp1 then return tp2
8   if tp1.name <= tp2.name then tp1 else tp2
```

The Scala 3 algorithm preserves most interesting properties of intersections but has one non-obvious shortcoming: it does not preserve associativity, consider:

```
trait X; trait Y; trait Z extends X
```

Then `|(X&Y)&Z|` = `Z` but `|X&(Y&Z)|` = `X`. The problem is that while the lexicographic ordering by itself is total, it is applied inconsistently because *incomparability of subtyping is not transitive*: in our example neither `X <: Y` nor `Y <: X` making `X` and `Y` incomparable, but even though `Y` and `Z` are also incomparable it is not true that `X` and `Z` are incomparable.

To rectify this we propose[6] ordering classes by *the number of base types they have*. In other words, we replace the subtyping checks on lines 6 and 7 in the listing above by:

[4] For the unsavory details, see https://github.com/lampepfl/dotty/blob/3.2.0/compiler/src/dotty/tools/dotc/core/unpickleScala2/Scala2Erasure.scala.

[5] The complete implementation also special-cases value types and array types which we do not model in our calculus, see `erasedGlb` in https://github.com/lampepfl/dotty/blob/3.2.0/compiler/src/dotty/tools/dotc/core/TypeErasure.scala.

[6] Since this change would break binary compatibility, it will have to wait until the next major version of Scala.

```
val relativeLength = ℒ(tp1).length - ℒ(tp2).length
if relativeLength > 0 then return tp1
if relativeLength < 0 then return tp2
```

This means that we still prefer subtypes over supertypes since a subclass necessarily has more base types than any of its parent, but incomparability is now transitive which is enough to make `erasedGlb` itself transitive.

In the rest of this section, we will assume `erasedGlb` prefers classes over traits as well as subtypes over supertypes but otherwise will stay independent of any particular implementation.

## A.2 Expression Erasure

Because type erasure does not preserve subtyping we might need to insert casts both on prefixes of calls as well as on method arguments. To keep the typing rules in Figure A.2 readable, we delegate casting $|e|_\Gamma$ to $T$ to an auxiliary judgment $|e|_\Gamma^T$ which is mutually recursive with the main judgment:

$$\frac{e' = |e|_\Gamma \qquad \Gamma \vdash_{FJD} e' : S}{|e|_\Gamma^T \coloneqq \begin{cases} e' & \text{if } S <:_{FJD} T \\ (T)e' & \text{otherwise} \end{cases}}$$

**Figure A.2: PS: Expression Erasure**

$$\boxed{|e_{\text{PS}}|_\Gamma = e_{\text{FJD}}}$$

$$|x|_\Gamma \coloneqq x \qquad \text{(ER-Var)}$$

$$\frac{\Gamma \vdash e_0 : T_0 \qquad |T_0|_\Gamma = C}{|e_0.f|_\Gamma \coloneqq |e_0|_\Gamma^C.f} \qquad \text{(ER-Field)}$$

$$\frac{\Gamma \vdash e_0 : T_0 \quad \text{erasedReceiver}_\Delta(m, T_0) = C \qquad \text{mtype}_{\text{FJD}}(m_C, C) = (\overline{x : D}) \Rightarrow D_0 \quad e_i' = |e_i|_\Gamma^{D_i}}{|e_0.m[\overline{V}](\overline{e})|_\Gamma \coloneqq |e_0|_\Gamma^C.m_C(\overline{e'})} \qquad \text{(ER-Invk)}$$

$$\frac{|N|_\Gamma = C \quad \text{vparams}_\Delta([\,]_{\text{FJD}}](C) = \overline{f : D} \qquad e_i' = |e_i|_\Gamma^{D_i}}{|\textbf{new}\, N(\overline{e})|_\Gamma \coloneqq \textbf{new}\, C(\overline{e'})} \qquad \text{(ER-New)}$$

Casting the prefix of a getter call to the appropriate type is easy: we know that `erasedGlb` will always return the most specific class type in an intersection and that traits do not contain getters, therefore if $\mathsf{vparams}_\Gamma(T_0) = \overline{f : T}$ then $\mathsf{vparams}_{\Gamma_{\mathrm{FJD}}}(|T_0|_\Gamma) = \overline{f : |T|_\Gamma}$ and ER-FIELD is straight-forward, but finding the right cast for the receiver of a method call is more involved.

Given x : L & R and the class table:

```
trait L { def l(): Object }
trait R { def r(): Object }
```

Then the type of $|\mathsf{x}|_\Gamma$ will be either $L$ or $R$ (depending on the definition of `erasedGlb`), but that means that one of `x.l()` and `x.r()` will require casting the receiver, therefore ER-INVK relies on the following auxiliary function:

$$\mathsf{erasedReceiver}_\Delta(m, X) \mathrel{\mathbf{:=}} \mathsf{erasedReceiver}_\Delta(m, \Gamma(X))$$

$$\mathsf{erasedReceiver}_\Delta(m, C[\ldots]) \mathrel{\mathbf{:=}} C$$

$$\mathsf{erasedReceiver}_\Delta(m, T_1 \mathbin{\&} T_2) := \begin{cases} \mathsf{erasedReceiver}_\Delta(m, T_1) & \text{if } \mathsf{mtype}(m, T_1) \text{ is defined} \\ \mathsf{erasedReceiver}_\Delta(m, T_2) & \text{otherwise} \end{cases}$$

Additionally, erasure does not preserve method names: $m$ is erased to $m_C$ where $C$ is the type of the receiver, this is justified in the following section.

## A.3 Class Table Erasure

Given the class table:

```
trait X; class Y extends X
trait L[T] { def foo(): T }
trait R[T <: X] { def foo(): T }
class A ◁ Object, L[Y], R[Y] {
  def foo(): Y = new Y
}
```

One might hope we could erase it just by erasing each type and expression appearing in it:

```
interface L { Object foo() }
interface R { X foo() }
class A ◁ Object, L, R {
  Y foo() { return new Y(); }
}
```

But that would be incorrect: a method in FJD must have exactly the same type as the methods it overrides (just like in Java bytecode). Compilers normally handle this by generating synthetic *bridge methods* [Bracha et al. 2003]:

```
interface L { Object foo() }
interface R { X foo() }
class A ◁ Object, L, R {
  Y foo() { return new Y(); }
  Object foo() { return <overload of foo returning Y>(); }
  X foo() { return <overload of foo returning Y>(); }
}
```

Notice that the types of the new methods added in A match the types of the overridden methods in L and R and simply forward to the actual implementation of foo in A, thus restoring the semantics present in the source program. But we cannot directly reuse this technique since our target calculus does not support overloading, faced with the same problem FGJ adopted the following strategy:

> In [Generic Java], the actual erasure is somewhat more complex, involving the introduction of bridge methods [...] instead, the rule E-Method merges two methods into one by inline-expanding the body of the actual method into the body of the bridge method.

But this works because FGJ only supports single-class inheritance, whereas in the example above we need two bridges in A corresponding to the two traits containing an overridden foo. Like FGJ, we shy away from introducing overloading in our target calculus and instead employ the following scheme:

- When erasing a call to $m$, we replace it by a call to $m_C$ where $C$ is the erased receiver of $m$ (see the previous section).
- When erasing the declaration of $m$ in $C$, we rename it to $m_C$.
- When erasing a class $C$, we add enough bridge methods so that erased calls to $m$ always end up being forwarded to the implementer of $m$ in $C$.

For our example this means we get:

```
interface L { Object foo_L() }
interface R { X foo_R() }
class A ◁ Object, L, R {
  Y foo_A { return new Y(); }
  Object foo_L { return this.foo_A(); }
  X foo_R { return this.foo_A(); }
}
```

This scheme wouldn't be practical in a real compiler since it would make it much harder for Java and Scala code to interoperate, but as a model we believe it's close enough to the real thing to be useful. The exact rules are described in Figure A.3 which makes use of the following judgments:

$$\frac{\begin{array}{c}\mathsf{mtype}_{\mathsf{FJD}}(m_E, E) = (\overline{x : T}) \rightarrow T_0 \\ \mathsf{mtype}_{\mathsf{FJD}}(m_D, D) = (\overline{x : U}) \rightarrow U_0 \\ x_0 = \mathsf{this}.m_D(\overline{e}) \\ e_i = \begin{cases} x_i & \text{if } T_i = U_i \\ (U_i)x_i & \text{otherwise}\end{cases}\end{array}}{\mathsf{bridge}(m_E, m_D) \coloneqq T_0\, m_E(\overline{T\, x})\ \{\mathsf{return}\ e_0;\ \}}$$

$$\frac{\begin{array}{c}\mathsf{mimpl}(m, N) = D[\overline{T}] \\ \overline{E[...]} = \left\{\mathbf{n} \in \mathcal{L}(N) \setminus D[\overline{T}] \;\middle|\; \mathbf{def}\ m\ ... \in \mathsf{mdecls}(\mathbf{n})\right\}\end{array}}{\mathsf{bridges}(m, N) \coloneqq \overline{\mathsf{bridge}(m_E, m_D)}}$$

Note that this definition of `bridges` can generate unnecessary bridges since it does not take into account that a parent class might already have defined an equivalent bridge.

**Figure A.3: PS: Class Table Erasure**

**Method erasure** $\boxed{|M_{\text{PS}}|_C = M_{\text{FJD}}}$

$$\frac{\begin{array}{c}\mathbf{class}\, C[\overline{X <: N}]\ ... \\ \Gamma = \overline{X <: N}, \mathsf{this} : C[\overline{X}], \overline{Y <: P}, \overline{x : T} \quad D = |T_0|_\Gamma\end{array}}{\begin{array}{c}|\mathbf{def}\ m[\overline{Y <: P}](\overline{x : T}) : T_0\ \underset{\sim\sim}{=}\ \underset{\sim\sim}{e_0}|_C \coloneqq \\ D\ m_C(\overline{|T|_\Gamma\, x})\ \underset{\sim\sim\sim\sim\sim\sim\sim\sim\sim\sim\sim\sim}{\{\ \mathbf{return}\ |e_0|_\Gamma^D;\ \}}\end{array}} \quad \text{(ER-Method)}$$

**Class erasure** $\boxed{|L_{\text{PS}}| = L_{\text{FJD}}}$

$$\frac{\begin{array}{c}\Gamma = \overline{X <: N} \quad K = C(\overline{|U|_\Gamma\, g}, \overline{|T|_\Gamma\, f})\ \{\ \mathsf{super}(\overline{g});\ \mathsf{this}.\overline{f} = \overline{f};\ \} \\ \overline{M'} = \overline{|M|_C} \cup \left\{\mathsf{bridges}(m, C) \;\middle|\; m \in \mathsf{mnames}(C)\right\}\end{array}}{\begin{array}{c}|\mathbf{class}\, C[\overline{X <: N}](\overline{g : U}, \overline{f : T}) \triangleleft P(\overline{g}), \overline{Q}\ \{\overline{M}\}| \coloneqq \\ \mathbf{class}\, C \triangleleft |P|_\Gamma, \overline{|Q|_\Gamma}\ \{\overline{|T|_\Gamma\, f};\ K;\ \overline{M'}\}\end{array}} \quad \text{(ER-Class)}$$

$$\frac{\Gamma = \overline{X <: N} \quad \overline{M'} = \overline{|M|_C}}{|\mathbf{trait}\ C[\overline{X <: N}] \triangleleft \overline{Q}\ \{\overline{M}\}| \coloneqq \mathbf{interface}\ C \triangleleft \overline{|Q|_\Gamma}\ \{\overline{M'}\}} \quad \text{(ER-Trait)}$$

## A.4 Future work

In this work we've focused on erasing Scala types into "bytecode Java" types, but in practice we also need to worry about erasing Scala types into "source Java" types: the bytecode format defines a `Signature` attribute [Lindholm et al. 2015, § 4.7.8] which lets us specify a polymorphic Java method signature that will be ignored by the JVM at runtime but used by the Java compiler for typechecking, thus improving the interoperability between Scala and Java. It would be useful to specify an erasure from PS into full FJ&$\lambda$ as a way to model this process. The Java compiler will also use this attribute if it is available to compute the erased signature it will emit when invoking the method, therefore we should also define an erasure of FJ&$\lambda$ into FJD based on the semantics of Java erasure and verify that the composition of these two mapping are equivalents to the erasure mapping of PS into FJD to avoid issues such as https://github.com/scala/bug/issues/4214.

# Bibliography

Amin, Nada (2016). "Dependent Object Types". PhD thesis. EPFL. DOI: 10.5075/epfl-thesis-7156. URL: http://infoscience.epfl.ch/record/223518.

Amin, Nada, Samuel Grütter, Martin Odersky, Tiark Rompf, and Sandro Stucki (2016). "The Essence of Dependent Object Types". In: *A List of Successes That Can Change the World: Essays Dedicated to Philip Wadler on the Occasion of His 60th Birthday*. Cham, Switzerland: Springer, pp. 249–272. ISBN: 978-3-319-30935-4. DOI: 10.1007/978-3-319-30936-1_14.

Amin, Nada, Adriaan Moors, and Martin Odersky (2012). "Dependent object types". In: *19th International Workshop on Foundations of Object-Oriented Languages*. CONF. New York, NY, USA: Association for Computing Machinery.

Amin, Nada and Tiark Rompf (2017). "Type Soundness Proofs with Definitional Interpreters". In: *SIGPLAN Not.* 52.1, pp. 666–679. ISSN: 0362-1340. DOI: 10.1145/3093333.3009866.

Amin, Nada, Tiark Rompf, and Martin Odersky (2014). "Foundations of path-dependent types". In: *ACM SIGPLAN Notices*. Vol. 49. 10. New York, NY, USA: Association for Computing Machinery, pp. 233–249. DOI: 10.1145/2660193.2660216.

Amin, Nada and Ross Tate (2016). "Java and Scala's Type Systems Are Unsound: The Existential Crisis of Null Pointers". In: *Proceedings of the 2016 ACM SIGPLAN International Conference on Object-Oriented Programming, Systems, Languages, and Applications*. OOPSLA 2016. Amsterdam, Netherlands: Association for Computing Machinery, pp. 838–848. ISBN: 9781450344449. DOI: 10.1145/2983990.2984004.

Barendregt, Henk, Mario Coppo, and Mariangiola Dezani-Ciancaglini (1983). "A filter lambda model and the completeness of type assignment". In: *Journal of Symbolic Logic* 48.4, pp. 931–940. DOI: 10.2307/2273659.

Barrett, Kim, Bob Cassels, Paul Haahr, David A. Moon, Keith Playford, and P. Tucker Withington (1996). "A monotonic superclass linearization for Dylan". In: *ACM SIGPLAN Notices*. Vol. 31. New York, NY, USA: Association for Computing Machinery, pp. 69–82. DOI: 10.1145/236337.236343.

Bettini, Lorenzo, Viviana Bono, Mariangiola Dezani-Ciancaglini, and Betti Venneri (2018). "Java & Lambda: a Featherweight Story". In: *Logical Methods in Computer Science* Volume 14, Issue 3. DOI: 10.23638/LMCS-14(3:17)2018. arXiv: 1801.05052.

Blanvillain, Olivier, Jonathan Immanuel Brachthäuser, Maxime Kjaer, and Martin Odersky (2022). "Type-Level Programming with Match Types". In: *Proc. ACM Program. Lang.* 6.POPL. DOI: 10.1145/3498698.

Boruch-Gruszecki, Aleksander, Radosław Waśko, Yichen Xu, and Lionel Parreaux (2022). "A case for DOT: Theoretical Foundations for Objects With Pattern Matching and GADT-style Reasoning". In: *Proc. ACM Program. Lang.* 6.OOPSLA2. DOI: 10.1145/3563342.

Bracha, Gilad, Norman Cohen, Christian Kemper, Martin Odersky, David Stoutamire, Kresten Thorup, and Philip Wadler (2003). *Adding Generics to the Java Programming Language: Public Draft Specification Version 2.0.* URL: http://www.javainthebox.net/laboratory/J2SE1.5/LangSpec/Generics/materials/adding_generics-2_2-ea/spec10.pdf (visited on July 30, 2022).

Cameron, Nicholas, Sophia Drossopoulou, and Erik Ernst (2008). "A Model for Java with Wildcards". In: *ECOOP 2008 – Object-Oriented Programming*. Berlin, Germany: Springer, pp. 2–26. ISBN: 978-3-540-70592-5. DOI: 10.1007/978-3-540-70592-5_2.

Canning, Peter, William Cook, Walter Hill, Walter Olthoff, and John C. Mitchell (1989). "F-Bounded Polymorphism for Object-Oriented Programming". In: *Proceedings of the Fourth International Conference on Functional Programming Languages and Computer Architecture.* FPCA '89. Imperial College, London, United Kingdom: Association for Computing Machinery, pp. 273–280. ISBN: 0897913280. DOI: 10.1145/99370.99392.

Click, Cliff and John Rose (2002). "Fast subtype checking in the HotSpot JVM". In: *JGI '02: Proceedings of the 2002 joint ACM-ISCOPE conference on Java Grande.* New York, NY, USA: Association for Computing Machinery, pp. 96–107. DOI: 10.1145/583810.583821.

Delaware, Benjamin, William Cook, and Don Batory (2011). "Product Lines of Theorems". In: *SIGPLAN Not.* 46.10, pp. 595–608. ISSN: 0362-1340. DOI: 10.1145/2076021.2048113.

Delaware, Benjamin, Bruno C. d. S. Oliveira, and Tom Schrijvers (2013). "Meta-Theory à La Carte". In: *SIGPLAN Not.* 48.1, pp. 207–218. ISSN: 0362-1340. DOI: 10.1145/2480359.2429094.

Dezani-Ciancaglini, Mariangiola, Paola Giannini, and Betti Venneri (2019). "Intersection Types in Java: Back to the Future". In: *Models, Mindsets, Meta: The What, the How, and the Why Not? Essays Dedicated to Bernhard Steffen on the Occasion of His 60th Birthday.* Cham, Switzerland: Springer, pp. 68–86. ISBN: 978-3-030-22347-2. DOI: 10.1007/978-3-030-22348-9_6.

– (2020). "Deconfined Intersection Types in Java". In: *Recent Developments in the Design and Implementation of Programming Languages.* Ed. by Frank S. de Boer and Jacopo Mauro. Vol. 86. OpenAccess Series in Informatics (OASIcs). Dagstuhl, Germany: Schloss Dagstuhl–Leibniz-Zentrum für Informatik, 3:1–3:25. ISBN: 978-3-95977-171-9. DOI: 10.4230/OASIcs.Gabbrielli.3. URL: https://drops.dagstuhl.de/opus/volltexte/2020/13225.

Doeraene, Sébastien Jean R. (2018). "Cross-Platform Language Design". PhD thesis. EPFL. DOI: 10.5075/epfl-thesis-8733.

Dunfield, Jana and Neel Krishnaswami (2021). "Bidirectional Typing". In: *ACM Comput. Surv.* 54.5. ISSN: 0360-0300. DOI: 10.1145/3450952.

Emir, Burak, Andrew Kennedy, Claudio Russo, and Dachuan Yu (2006). "Variance and Generalized Constraints for C♯ Generics". In: *ECOOP 2006 – Object-Oriented Programming*. Berlin, Germany: Springer, pp. 279–303. ISBN: 978-3-540-35727-8. DOI: 10.1007/11785477_18.

Forster, Yannick and Kathrin Stark (2020). "Coq à La Carte: A Practical Approach to Modular Syntax with Binders". In: *Proceedings of the 9th ACM SIGPLAN International Conference on Certified Programs and Proofs.* CPP 2020. New Orleans, LA, USA: Association for Computing Machinery, pp. 186–200. ISBN: 9781450370974. DOI: 10.1145/3372885.3373817.

Giarrusso, Paolo G., Léo Stefanesco, Amin Timany, Lars Birkedal, and Robbert Krebbers (2020). "Scala step-by-step: soundness for DOT with step-indexed logical relations in Iris". In: *Proc. ACM Program. Lang.* 4.ICFP, pp. 1–29. ISSN: 2475-1421. DOI: 10.1145/3408996.

Gosling, James, Bill Joy, Guy Steele, and Gilad Bracha (2015). *The Java Language Specification, Java SE 8 Edition*. Oracle. URL: https://docs.oracle.com/javase/specs/jls/se8/jls8.pdf.

Greenman, Ben, Fabian Muehlboeck, and Ross Tate (2014). "Getting F-Bounded Polymorphism into Shape". In: *Proceedings of the 35th ACM SIGPLAN Conference on Programming Language Design and Implementation*. PLDI '14. Edinburgh, United Kingdom: Association for Computing Machinery, pp. 89–99. ISBN: 9781450327848. DOI: 10.1145/2594291.2594308.

Hu, Jason (2019). *Comparison Between Different DOTs*. URL: https://hustmphrrr.github.io/blog/2019/compare-dots.html (visited on July 30, 2022).

Igarashi, Atsushi and Hideshi Nagira (2006). "Union Types for Object-Oriented Programming". In: *Proceedings of the 2006 ACM Symposium on Applied Computing*. SAC '06. Dijon, France: Association for Computing Machinery, pp. 1435–1441. ISBN: 1595931082. DOI: 10.1145/1141277.1141610.

Igarashi, Atsushi and Benjamin C. Pierce (2002). "On Inner Classes". In: *Inform. And Comput.* 177.1, pp. 56–89. ISSN: 0890-5401. DOI: 10.1006/inco.2002.3092.

Igarashi, Atsushi, Benjamin C. Pierce, and Philip Wadler (2001). "Featherweight Java: a minimal core calculus for Java and GJ". In: *ACM Trans. Program. Lang. Syst.* 23.3, pp. 396–450. ISSN: 0164-0925. DOI: 10.1145/503502.503505.

Igarashi, Atsushi and Mirko Viroli (2006). "Variant parametric types: A flexible subtyping scheme for generics". In: *ACM Trans. Program. Lang. Syst.* 28.5, pp. 795–847. ISSN: 0164-0925. DOI: 10.1145/1152649.1152650.

Jeffery, Alex (2019). "Dependent Object Types with Implicit Functions". In: *Proceedings of the Tenth ACM SIGPLAN Symposium on Scala*. Scala '19. London, United Kingdom: Association for Computing Machinery, pp. 1–11. ISBN: 9781450368247. DOI: 10.1145/3337932.3338811.

Jung, Ralf, Robbert Krebbers, Jacques-Henri Jourdan, Aleš Bizjak, Lars Birkedal, and Derek Dreyer (2018). "Iris from the ground up: A modular foundation for higher-order concurrent separation logic". In: *J. Funct. Program.* 28. ISSN: 0956-7968. DOI: 10.1017/S0956796818000151.

Kabir, Ifaz and Ondřej Lhoták (2018). "$\kappa$DOT: Scaling DOT with Mutation and Constructors". In: *Proceedings of the 9th ACM SIGPLAN International Symposium on Scala*. Scala 2018. St. Louis, MO, USA: Association for Computing Machinery, pp. 40–50. ISBN: 9781450358361. DOI: 10.1145/3241653.3241659.

Kabir, Ifaz, Yufeng Li, and Ondřej Lhoták (2020). "$\iota$DOT: A DOT Calculus with Object Initialization". In: *Proc. ACM Program. Lang.* 4.OOPSLA. DOI: 10.1145/3428276.

Kennedy, Andrew J and Benjamin C. Pierce (2007). "On decidability of nominal subtyping with variance". In: CONF.

League, Christopher, Zhong Shao, and Valery Trifonov (2002). "Type-Preserving Compilation of Featherweight Java". In: *ACM Trans. Program. Lang. Syst.* 24.2, pp. 112–152. ISSN: 0164-0925. DOI: 10.1145/514952.514954.

Lindholm, Tim, Frank Yellin, Gilad Bracha, and Alex Buckley (2015). *The Java Virtual Machine Specification, Java SE 8 Edition*. Oracle. URL: https://docs.oracle.com/javase/specs/jvms/se8/jvms8.pdf.

Liu, Fengyun et al. (2022). *Option-less pattern matching*. EPFL. URL: https://docs.scala-lang.org/scala3/reference/changed-features/pattern-matching.html (visited on July 30, 2022).

Martres, Guillaume (2021). "Pathless Scala: a calculus for the rest of Scala". In: *SCALA 2021: Proceedings of the 12th ACM SIGPLAN International Symposium on Scala*. New York, NY, USA: Association for Computing Machinery, pp. 12–21. ISBN: 978-1-45039113-9. DOI: 10.1145/3486610.3486894.

Meyer, Bertrand (1992). "Applying 'design by contract'". In: *Computer* 25.10, pp. 40–51.

Nieto, Abel (2017). "Towards Algorithmic Typing for DOT". In: DOI: 10.48550/arXiv.1708.05437. arXiv: 1708.05437.

Odersky, Martin et al. (2021a). *The Scala Language Specification, Scala 2.13 Edition*. EPFL. URL: https://www.scala-lang.org/files/archive/spec/2.13/.

– (2021b). *Wildcard Arguments in Types | Scala 3 Language Reference*. URL: https://docs.scala-lang.org/scala3/reference/changed-features/wildcards.html (visited on July 30, 2022).

– (2022). *Trait Parameters | Scala 3 Language Reference*. EPFL. URL: https://docs.scala-lang.org/scala3/reference/other-new-features/trait-parameters.html (visited on July 30, 2022).

Odersky, Martin, Guillaume Martres, and Dmitry Petrashko (2016). "Implementing higher-kinded types in Dotty". In: *SCALA 2016: Proceedings of the 2016 7th ACM SIGPLAN Symposium on Scala*. New York, NY, USA: Association for Computing Machinery, pp. 51–60. ISBN: 978-1-45034648-1. DOI: 10.1145/2998392.2998400.

Odersky, Martin and Matthias Zenger (2005). "Scalable component abstractions". In: *SIGPLAN Not.* 40.10, pp. 41–57. ISSN: 0362-1340. DOI: 10.1145/1103845.1094815.

Parreaux, Lionel (2020). "The Simple Essence of Algebraic Subtyping: Principal Type Inference with Subtyping Made Easy (Functional Pearl)". In: *Proc. ACM Program. Lang.* 4.ICFP. DOI: 10.1145/3409006.

Pierce, Benjamin C. and David N. Turner (2000). "Local Type Inference". In: *ACM Trans. Program. Lang. Syst.* 22.1, pp. 1–44. ISSN: 0164-0925. DOI: 10.1145/345099.345100.

Pottier, François (1998). "Type inference in the presence of subtyping: from theory to practice". PhD thesis. INRIA.

Rapoport, Marianna, Ifaz Kabir, Paul He, and Ondřej Lhoták (2017). "A Simple Soundness Proof for Dependent Object Types". In: *Proc. ACM Program. Lang.* 1.OOPSLA. DOI: 10.1145/3133870.

Rapoport, Marianna and Ondřej Lhoták (2017). "Mutable WadlerFest DOT". In: *Proceedings of the 19th Workshop on Formal Techniques for Java-like Programs*. FTFJP'17. Barcelona, Spain: Association for Computing Machinery. ISBN: 9781450350983. DOI: 10.1145/3103111.3104036.

– (2019). "A path to DOT: formalizing fully path-dependent types". In: *Proceedings of the ACM on Programming Languages* 3, pp. 1–29. ISSN: 2475-1421. DOI: 10.1145/3360571.

Rehman, Baber, Xuejing Huang, Ningning Xie, and Bruno C. d. S. Oliveira (2022). "Union Types with Disjoint Switches". In: *36th European Conference on Object-Oriented Programming (ECOOP 2022)*. Ed. by Karim Ali and Jan Vitek. Vol. 222. Leibniz International Proceedings in Informatics (LIPIcs). Dagstuhl, Germany: Schloss Dagstuhl – Leibniz-Zentrum für Informatik,

25:1–25:31. ISBN: 978-3-95977-225-9. DOI: 10.4230/LIPIcs.ECOOP.2022.25. URL: https://drops.dagstuhl.de/opus/volltexte/2022/16253.

Rompf, Tiark and Nada Amin (2016). "Type soundness for dependent object types (DOT)". In: *SIGPLAN Not.* 51.10, pp. 624–641. ISSN: 0362-1340. DOI: 10.1145/3022671.2984008.

Shipilëv, Aleksey (2020). *The Black Magic of (Java) Method Dispatch.* URL: https://shipilev.net/blog/2015/black-magic-method-dispatch (visited on July 30, 2022).

Smith, Dan (2022). *JEP 402: Classes for the Basic Primitives (Preview).* URL: https://openjdk.org/jeps/402 (visited on July 30, 2022).

Smith, Daniel and Robert Cartwright (2008). "Java Type Inference is Broken: Can We Fix It?" In: *SIGPLAN Not.* 43.10, pp. 505–524. ISSN: 0362-1340. DOI: 10.1145/1449955.1449804.

Stucki, Sandro (2017). "Higher-Order Subtyping with Type Intervals". PhD thesis. Lausanne: IINFCOM, p. 141. DOI: 10.5075/epfl-thesis-8014. URL: http://infoscience.epfl.ch/record/232408.

Stucki, Sandro and Paolo G. Giarrusso (2021). "A theory of higher-order subtyping with type intervals". In: *Proc. ACM Program. Lang.* 5.ICFP, pp. 1–30. ISSN: 2475-1421. DOI: 10.1145/3473574.

Sulzmann, Martin, Manuel M. T. Chakravarty, Simon Peyton Jones, and Kevin Donnelly (2007). "System F with Type Equality Coercions". In: *Proceedings of the 2007 ACM SIGPLAN International Workshop on Types in Languages Design and Implementation.* TLDI '07. Nice, Nice, France: Association for Computing Machinery, pp. 53–66. ISBN: 159593393X. DOI: 10.1145/1190315.1190324.

Tate, Ross, Alan Leung, and Sorin Lerner (2011). "Taming Wildcards in Java's Type System". In: *SIGPLAN Not.* 46.6, pp. 614–627. ISSN: 0362-1340. DOI: 10.1145/1993316.1993570. URL: https://doi.org/10.1145/1993316.1993570.

Wang, Yanlin, Haoyuan Zhang, Bruno C. d. S. Oliveira, and Marco Servetto (2018). "FHJ: A Formal Model for Hierarchical Dispatching and Overriding". In: *32nd European Conference on Object-Oriented Programming (ECOOP 2018).* Ed. by Todd Millstein. Vol. 109. Leibniz International Proceedings in Informatics (LIPIcs). Dagstuhl, Germany: Schloss Dagstuhl–Leibniz-Zentrum fuer Informatik, 20:1–20:30. ISBN: 978-3-95977-079-8. DOI: 10.4230/LIPIcs.ECOOP.2018.20. URL: http://drops.dagstuhl.de/opus/volltexte/2018/9225.

Wasserrab, Daniel, Tobias Nipkow, Gregor Snelting, and Frank Tip (2006). "An operational semantics and type safety proof for multiple inheritance in C++". In: *SIGPLAN Not.* 41.10, pp. 345–362. ISSN: 0362-1340. DOI: 10.1145/1167515.1167503.

Wright, A. K. and M. Felleisen (1994). "A Syntactic Approach to Type Soundness". In: *Inform. And Comput.* 115.1, pp. 38–94. ISSN: 0890-5401. DOI: 10.1006/inco.1994.1093.

smarter@ubuntu.com
guillaume.martres.me
Birthdate: 19/05/1993
Nationalities: French and Tunisian

# Guillaume Martres

## Education

2016–Present **Ph.D. in Computer Science**, *EPFL*, Lausanne, Switzerland
Working on the Scala 3 compiler and language specification. The subject of my PhD thesis is: *Type-Preserving Compilation of Class-Based Languages* [draft].

2013–2015 **Master in Computer Science**, *EPFL*, Lausanne, Switzerland
The subject of my Master thesis was:
*Implementing value classes in Dotty, a compiler for Scala.*

2010–2013 **Bachelor in Computer Science**, *EPFL*, Lausanne, Switzerland

## Employment History

06/2015–08/2016 **Research Intern**, *Mozilla*, Mountain View, California
I participated in the development of the AV1 video codec, notably by integrating features from Daala.

10/2015–05/2016 **Scientific Assistant**, *EPFL*, Lausanne, Switzerland
I worked on the Dotty research compiler that eventually became Scala 3.

07/2014–09/2014 **Software Engineering Contractor**, *Mozilla*, Remote
I worked on the research Daala video codec.

07/2013–10/2013 **Software Engineering Intern**, *Google*, Mountain View, California
I worked on the reference encoder for the VP9 video codec.

05/2012–08/2012 **Student Developer**, *Google*
I took part in the Google Summer of Code by writing an HEVC decoder for Libav (this decoder was subsequently completed with the help of many contributors and also merged in FFmpeg).

07/2011–10/2011 **Student Developer**, *European Space Agency*
I participated in the Summer of Code in Space organized by the European Space Agency and contributed to the Marble virtual globe and atlas by adding support for satellites display.

## Skills

Languages I have experience with: Scala, Rust, Haskell, C, C++ (especially with Qt), Java

Languages I'm actively using: Scala

Operating Systems: Linux (especially Debian-based distributions).

## Notable Open Source contributions

Scala: Besides my work on the compiler, I'm also a member of the Scala Improvement Process committee where we review and vote on proposed changes to the language.

Libav/FFmpeg: Work on the HEVC decoder.

KDE: I maintained the Gluon game engine audio subsystem, ported the Kvkbd virtual keyboard from KDE 3 to KDE 4, contributed to several projects including the Muon package manager.

Kubuntu: I did packaging work.

Miscellaneous: I contribute to several projects on Github.